\documentclass[11pt,a4paper]{article}
\pdfoutput=1
\usepackage{jheppub}

\usepackage{wasysym}
\usepackage{slashed}
\usepackage{amsmath}
\usepackage{hyperref}
\usepackage[utf8]{inputenc}
\usepackage[T1]{fontenc}
\usepackage{lmodern} % load
\usepackage{amssymb}
\usepackage{mathtools}
\usepackage{mathrsfs}
\usepackage{amsfonts}
\usepackage{bbm}
\usepackage{relsize}
\usepackage{enumerate}
\usepackage{array}
\usepackage{booktabs}
\usepackage{textcomp}
\usepackage{caption}
\usepackage{cite}
\usepackage{cancel}
\usepackage{units}
\usepackage{multirow}
\usepackage{pst-all}
\usepackage{longtable}
\usepackage{relsize}
\usepackage[]{natbib}
\usepackage{hyperref}
\usepackage{amsfonts}
\usepackage{amssymb}
\usepackage{mathrsfs}
\usepackage{graphicx}
\usepackage{amsmath,amsthm,amssymb,amscd}
\usepackage{subfig}
\usepackage{hyperref}
\usepackage{multirow}
\usepackage{color}
\usepackage{amsfonts}
\usepackage{float}
\usepackage{mathtools,slashed}
\usepackage{bbold}
\usepackage{amsmath}
\usepackage{bbm}
\usepackage{slashed}

\newcommand{\cd}{{\cal D}}

\def\eqa{\begin{eqnarray}}
\def\eqae{\end{eqnarray}}
\def\eq{\begin{equation}}
\def\eqe{\end{equation}}
\def\be{\begin{equation}}
\def\ee{\end{equation}}
\def\bea{\begin{eqnarray}}
\def\eea{\end{eqnarray}}
\def\ba{\begin{array}}
\def\ea{\end{array}}
\def\bd{\begin{displaymath}}
\def\ed{\end{displaymath}}

\def\Tr{{\rm Tr}}

\def\>{\rangle}
\def\<{\langle}

\def\e{\epsilon}

\def\m{\mu}
\def\n{\nu}

\def\r{\rho}
\def\s{\sigma}

\def\F{\Phi}

\def\SS{\mathbb{S}_2\times\mathbb{S}_1}

\def\SIN{\mathbb{S}_2\times I_N}
\def\SIS{\mathbb{S}_2\times I_S}

\title{Deriving the 3D Topologically Twisted Index}

\author[]{Alejandro Cabo-Bizet}

\affiliation[]{Instituto de Astronom\'{i}a y F\'{i}sica del Espacio (CONICET-UBA) Ciudad Universitaria, C.P. 1428
Buenos Aires, Argentina}

\begin{document}
\title{Factorising the 3D Topologically Twisted Index}

%\input author_list.tex       % D0 authors (remove the first 3 lines
                             % of this file prior to submission, they
                             % contain a time stamp for the authorlist)
                             % (includes institutions and visitors)
\date{\today}
\abstract{
We explore the path integration -- upon the contour of hermitian (non-auxliary) field configurations -- of topologically twisted  $\mathcal{N}=2$ Chern-Simons-matter theory (TTCSM) on $\mathbb{S}_2$ times a segment. In this way, we obtain the formula for the 3D topologically twisted index, first as a convolution of TTCSM on $\mathbb{S}_2$ times halves of $\mathbb{S}_1$,  second as TTCSM on $\mathbb{S}_2$ times $\mathbb{S}_1$ -- with a puncture --, and third as  TTCSM on $\mathbb{S}_2 \times \mathbb{S}_1$. In contradistinction to the first two cases, in the third case, the vector multiplet auxiliary field $D$ is constrained to be anti-hermitian.

  }

\arxivnumber{}

\maketitle

%\flushbottom

\section{Introduction}

Recently, it has been pointed out a relation between 3D topologically twisted (TT) index of \citep{BZ}  and the entropy of supersymmetric $AdS_4$ black holes  \citep{BZ2,BZ3} via the gauge/gravity correspondence. This is an interesting observation, that could potentially show a way to new insights into the microscopic structure of black holes, via careful analysis of relevant supersymmetric localisation results \citep{Chung,Nedelin} and through holographic principles \citep{Strominger, Sen}.

%This note aims to provide an alternative explanation to the origin of the sum over fluxes $\mathfrak{m}$ of the 3D TT index of \citep{BZ}. In \citep{BZ} the fluxes $\mathfrak{m}$ were identified as saddle points over a complex path of integration, namely away from the real path. However, it was argued, that due to the integration over the supermultiplet of zero modes $(u_0, \bar{u}_0, \lambda_0, \bar{\lambda}_0, D_0)$ this path can be deformed into the real path without affecting the value of the Jeffrey Kirwan (JK) contour integrals that define the topologically twisted index. In this way %and in contradistinction with the results in \citep{},
%the authors predicted that even while integrating with the real conditions, the sum over fluxes $\mathfrak{m}$ must be present.  The puzzle then arises, as to how could the sum over fluxes $\mathfrak{m}$ emerge should one had localised the integral over the real path of fields in the beginning, without complexifying $D$.

 The 3D TT index of \citep{BZ}\footnote{Strictly speaking is a semi topological A-twist on $\mathbb{S_2}$ \citep{Witten0}. We will see what this means in a while.} is a summation over magnetic fluxes $\mathfrak{m}$ --and integration over the Coulomb branch--
\begin{eqnarray}
Z=\sum_{\mathfrak{m}} \int_{JK} d u Z_\mathfrak{m}[u], \label{sumfluxes}
\end{eqnarray}
where  $Z_\mathfrak{m}$, is the partition function of the $\mathcal{N}=2$ $\mathbb{S}_2\times \mathbb{S}_1$ Chern-Simons- SYM- matter theory (with semi-topological A twisting \citep{Witten0} over $\mathbb{S}_2$), about a supersymmetric saddle point that includes, among non trivial values for other fields in the vector multiplet, the following background gauge potential  \begin{eqnarray}
A_{(0)}&=& -\frac{\mathfrak{m}}{2 R^2}\cos{\theta} %-1)
d\phi + u dt,  \label{flux}
\end{eqnarray}
with $\mathfrak{m}$ being a generator having weight vectors in the co-root lattice of the corresponding gauge group $\mathcal{G}$.
The contour of integration in \eqref{sumfluxes} -- denoted as JK -- encloses the poles selected by a Jeffrey-Kirwan (JK) recipe (See \citep{JK0, BZ,  Closset2, Honda, ClossetNew}).

Consistency with Aharony and Giveon-Kutasov dualities suggests the presence of the sum over 1 loop contributions about the fluxes $\mathfrak{m}$. However, -- at least to our knowledge -- the origin of the latter sum, is not completely clear when integration over real path of fields is performed, by following the supersymmetric localisation method of \citep{Pestun}. 

Let us explain the point made in the last sentence of the previous paragraph in more detail.  Once we have defined a localising term for the vector multiplet sector, $\tau Q$ -( or $\tau \tilde{Q}$-),  we must specify reality conditions on the fields in the gauge multiplet:  $(A_\mu, \sigma, \lambda, \bar{\lambda}, D)$. The bosonic part of the aforementioned localising terms must be semi-positive definite and should vanish at specific set of classical configurations: the zero locus. The zero locus is the space of supersymmetric configurations that obey the specific set of reality conditions. The final result for the localised quantity is a sum over the zero locus.  Potentials of the form \eqref{flux} are not part of any zero locus -- by themselves --,
% \begin{eqnarray}
%A_{(0)}&=& -\frac{\mathfrak{m}}{2 R^2}\cos{\theta} %-1)
%d\phi.
%\end{eqnarray}
because they do not preserve any of the supercharges at disposal $(Q,\tilde{Q})$. In fact, the $\tau Q(\tilde{Q})V$ terms when evaluated at \eqref{flux}, are proportional to $\mathfrak{m}^2$, independent on $\tau$ and consequently suppressed -- with respect to the $\mathfrak{m}=0$ vacuum -- when the naive "semiclassical" limit $\tau\rightarrow \infty$ is taken \footnote{Configurations in the zero locus are the ones surviving the large $\tau$ limit. Notice that the gauge connections \eqref{flux} can not depend on $\tau$, due to the quantisation condition (GNO) for the fluxes $ \frac{1}{2 \pi}\int_{\mathbb{S}_2} F_{\theta \phi}=\mathfrak{m}$. Hence the norm of the classical configuration \eqref{flux}, $e^{-\int_{\mathbb{S}_2 \times \mathbb{S}_1} \tau QV |_{\eqref{flux}}}\sim O(e^{-\tau})$, is exponentially suppressed in any possible $\tau\rightarrow \infty$ limit.}. The issue then arises, as to how is it that is possible to get non trivial results from the backgrounds \eqref{flux}, in the "semiclassical" limit $\tau\rightarrow \infty$?

Potentials \eqref{flux} are supersymmetric when complemented with non trivial profiles of other fields in the vector multiplet.  For instance, \eqref{flux} preserve both $Q$ and $\tilde{Q}$ \citep{BZ} when
\begin{eqnarray}
D_{(0)}&=&i \frac{ \mathfrak{m} }{2 R^2} \label{D}.
\end{eqnarray}
Consequently, \eqref{flux} + \eqref{D} is part of the zero locus of any combination $(\alpha_1 Q+ \alpha_2 \tilde{Q})V$ with $\alpha_1$ and $\alpha_2$ arbitrary c-numbers% provided the vector multiplet auxiliary field, $D$ to be non-hermitian
. In \citep{BZ} it was argued that the saddle points in question do contribute to the integration along the real path, even though they are not along the real line of integration, as $D_{(0)}$ in \eqref{D} is complex,. To show that they must contribute, the authors noticed the consequences of integration over the super manifold of zero modes $(u_0, \bar{u}_0, \lambda_0, \bar{\lambda}_0, D_0)$.  Specifically, they claimed that integration of $D_0^i$ along the line $\mathbb{R}+i \eta$ with $\eta>0$(<0) was independent on $\eta$. Given that the integrand is a function of $D^{i}_{(0)}+D^{i}_{0}$ then by defining $\eta^i:=-i ~Im[D_{(0)}^i]=-i \frac{\mathfrak{m}^i}{2 R^2}$ the original "complex contour" gets shifted to the "real" one \footnote{Proper analysis \citep{BZ, Tachikawa, Hori} reduces the integration over the supermultiplet of zero modes to the JK  middle dimensional contour in the complex plane of bosonic expectation values $(u,\bar{u}):=(A_t+i \sigma,A_t-i \sigma)$.}.

%In this manuscript we will explore other two possibilities.
The supersymmetrisation of \eqref{flux} by mean of \eqref{D} is not possible, should we had chosen $D$ to be hermitian. In the latter case there are two choices that supersymmetrise \eqref{flux}. Namely, there are two specific real but globally ill defined  -- they are non periodic -- profiles for $\sigma$ that do the work
 \begin{eqnarray}
 \sigma^\pm_{(0)}(t,t^\pm_0):= \frac{\mathfrak{m}}{2 R^2} (\pm t-t_0^{\pm})+\, C.C.G.\, \label{sigma}
 \end{eqnarray}

Where $C.C.G.$ stands for constant commuting generator and $t^\pm_0$ are arbitrary constants. For $\sigma^-_{(0)}$, the magnetic flux preserves $Q$ and for $\sigma^+_{(0)}$, $\tilde{Q}$. Thence, naively speaking, one would say that \eqref{flux}+\eqref{sigma}$^-(^+)$ is part of the zero locus of the $Q(\tilde{Q})V$ term with vector multiplet fields being real (hermitian). In summary, we have counted three possibilities, that putting aside global issues, could justify the sum over fluxes \eqref{sumfluxes}. %However, should we impose the auxiliary field $D$ to be hermitian, \eqref{D} gets a priori excluded.

 The reader could wonder whether there exists a constant value for $\sigma$ to source the fluxes \eqref{flux}. In the case of 3D non TT index, like for instance the superconformal index, there is a  constant profile for $\sigma$ sourcing the fluxes $\eqref{flux}$ \citep{Imamura, Borokhov, Cremonesi1}  (See equation (21) of \citep{Imamura}). This potential contribution to the BPS conditions comes from $O(\frac{1}{R})$ terms in the superalgebra. However, these terms are proportional to $\mathcal{D}_\mu \epsilon$ and hence are not present in the case of the type $A$ semi-topological twisting. In conclusion, the only possible supersymmetriser for the the fluxes \eqref{flux} in the topologically twisted analysis and without considering a complexification of $D$, is something like \eqref{sigma}.

Notice that \eqref{sigma} has explicit dependence on the time coordinate $t$ and is non periodic under $t\rightarrow t+2\pi$. % We will see how to fix this issue on due time, but  at first, and only because of simplicity of presentation, let us simply introduce the general idea of this note by using \eqref{sigma}.% due to the fact there is no other way to make \eqref{flux} supersymmetric without considering a complexification of $D$.
% However we stress
One way to solve this problem is to exclude a point out of the $\mathbb{S}_1$. In the latter case, the fluxes will be present and one could ask whether %one can blend the results at $I_N$ and $I_S$ in such a way
the TT index formula of \citep{BZ} is recovered or not in this case. That will be one of the scopes of this work.

We will also explore whether the aforementioned formula for the TT index can be obtained out of dividing the $\mathbb{S}_1$ into two open patches $I_N=(0,t_0)$ and $I_S=(t_0,2\pi)$. %On the track, we will check that it does not matter which supercharge is used on any of the patches, the final result for the partition function will be the same. 
As naive intuition suggests, the result for the index will come after "glueing together", in the sense of \citep{Pasquetti,Pasquetti2}, the theories on $\SIN$ and $\SIS$. We will start by attacking the problem in this way, because then is easier to take the limit $t_0 \rightarrow 2\pi$ on one of the blocks to recover the procedure mentioned in the previous paragraph. For that, necessarily, we will need to compute the partition function of 3D TTCSM on $\mathbb{S}_2$ times an open segment $I$.

 Before entering in the bulk of the paper, let us explain in more detail one of the technical motivations that drove us to understand these issues. As the supercharges $Q_\epsilon$ and $\tilde{Q}_\epsilon$  are nilpotent (the $Q_\epsilon$ and $\tilde{Q}_\epsilon$  are our version of the supercharges $Q$ and $\tilde{Q}$ defined in equations (2.6) and (2.7) of \citep{BZ}), the  localising term used in \citep{BZ}, which in our conventions is
\begin{eqnarray}\nonumber\\
\label{eq:veclag1}
 Q_\epsilon \tilde{Q}_\epsilon \Tr(\bar{\lambda}^{\dagger}\lambda-4 D\s)&=&\mathcal{L}_{SYM}%+\Tr \sigma \mathcal{D}_\mu \left(\star F\right)^\mu
 + \Tr  ~\mathcal{D}_\mu \mathcal{J}^\mu \label{Qexact} \\
&=&(-\tilde{Q}_\epsilon Q_\epsilon+ i v^\mu \mathcal{D}_\mu) \Tr(\bar{\lambda}^{\dagger}\lambda-4 D\s)%+ 2 \Tr \sigma \mathcal{D}_\mu \left(\star F\right)^\mu
,\label{Qexact2} \\ \nonumber
\end{eqnarray}
with
\begin{eqnarray}
\mathcal{J}^{\mu} &:=& (i v_\n F^{\m \n}\s-i v^\m D\s-%\z^{\dagger}\e
 \cd^\m
\s \s - i\e^{\m\n\r}v_\r \cd_\n \s \s + ~fermionic~ %\frac{i}{2}%\z^{\dagger}\e
%\bar{\l}^{\dagger}\g^\m\l
),\quad\quad\quad \label{current}
\end{eqnarray}
is not only $Q_\epsilon$ but $(Q+\tilde{Q})_\epsilon$-exact. %In these expressions we have used the Bianchi identity $\mathcal{D}_\mu \star F\right^\mu=0$.

 In virtue of \eqref{Qexact2}, the localising action \eqref{Qexact} is also $\tilde{Q}_\epsilon$-exact up to a total derivative. In fact one could have guessed this fact, from the known property $\{Q_\epsilon,\tilde{Q}_\epsilon\}=i v^\mu \mathcal{D}_\mu$ -- acting on gauge invariant objects --. Strictly speaking, the latter definition of $\mathcal{L}_{SYM}$ is not $Q_\epsilon$ ( resp. $\tilde{Q}_\epsilon$) supersymmetric. In fact, from the nilpotency of $Q_\epsilon$ and the RHS of \eqref{Qexact} (resp. \eqref{Qexact2}) it follows that $\mathcal{L}_{SYM}$ is $Q_\epsilon$ (resp. $\tilde{Q}_\epsilon$) supersymmetric only up to a total derivative.

The flux \eqref{flux} when sourced by a complex value of $D$, \eqref{flux} + \eqref{D}, belongs to the zero locus of $Q_\epsilon$ \citep{BZ}.
%In \citep{BZ} it was used the fact that by using the complex value for $D$ \eqref{flux} + \eqref{D}, the magnetic flux can be made part of the zero locus of $Q_\epsilon$. 
However, notice that the localising action \eqref{Qexact} is not positive definite when $D$ is assumed to be complex. Nevertheless, this is not a problem to perform localisation. One of our conclusions, is that one can always define a $Q_\epsilon$ exact term whose bosonic part is semi-positive definite under the reality conditions that define a given $Q_\epsilon$ BPS configuration. In fact, we checked that \eqref{flux} + \eqref{D} is not an exception to that rule.

The localisation term  \eqref{eq:veclag1} is both, $Q_\epsilon$-exact (precisely) and $\tilde{Q}_\epsilon$-exact (up to total derivatives). Thenceforth, is also possible to use the two possible supporting real values of $\sigma$ \eqref{flux}+\newline \eqref{sigma}$^-(^+)$ in such a way the magnetic fluxes become $Q_\epsilon(\tilde{Q}_\epsilon)$-BPS. There seems plausible to expect that with this real BPS configurations, there is not need to relax the reality condition for the auxiliary field $D$ in order to recover the sum over fluxes. What we are set to explore with our approach, is whether the final result for the path integral along the real contour will be the same or not obtained in \citep{BZ} -- they integrate over a complex contour--. In the least of the ambitions, our scope is to understand what are the assumptions or considerations that allow to recover their result.

Let us argue our point from a different perspective. From the reality condition for the bosonic fields in the gauge multiplet, it follows, that the bosonic part of $\mathcal{L}_{SYM}$
\begin{eqnarray}
\frac{1}{2}F^2+ \left(\mathcal{D}_\mu \sigma\right)^2+D^2 \label{suppression}
\end{eqnarray}
is positive definite.  How could be possible that field configurations with finite energy density $\mathcal{L}_{SYM}>0$ belong to the zero locus of the RHS of \eqref{eq:veclag1}?
\footnote{In \citep{BZ} (See also \citep{ClossetNew}) the authors define a "semiclassical limit" in which the one loop determinant contribution cancels out the classical suppression. In the presence of the supermultiplet of zero modes $(D_0,\ldots)$. I would like to thank Francesco Benini for drawing my attention to this point. In this way, they provide an argument in favour of the presence of the sum over fluxes even when the usual localisation arguments seems to say the contrary when integrating over real fields \citep{Honda}. In this work an alternative way to explain the presence of the aforementioned fluxes -- following the lines of the usual localisation program -- was pursued.}

 From the point of view of the real saddle points \eqref{flux}+\eqref{sigma} the answer to this question comes from the integration of the total derivative term $\Tr~ \mathcal{D}_\mu \mathcal{J}^\mu$, which is non trivial in this case. Evaluated on a magnetic flux and the corresponding scalar configuration $\sigma$ that is needed to preserve either $Q$ or $\tilde{Q}$, %the component along $\mathbb{S}_1$ of the current $\mathcal{J}^{\mu}$
%\begin{eqnarray}
%\mathcal{J}^{\mu} := (i v_\n F^{\m \n}\s-i v^\m D\s-%\z^{\dagger}\e
% \cd^\m
%\s \s- i\e^{\m\n\r}v_\r \cd_\n \s \s+\frac{i}{2}%\z^{\dagger}\e
%\bar{\l}^{\dagger}\g^\m\l),\quad\quad\quad
%\end{eqnarray}
% that we denote as $\mathcal{J}^{t}$, will inherit a branch point on $\mathbb{S}_1$ from the non trivial profile of $\sigma$ \eqref{sigma}.
the integration of $\mathcal{D}_\mu \mathcal{J}^\mu$ will provide a non trivial contribution that cancels the suppression contribution coming from \eqref{suppression}. This contribution comes from the term $-\int_{\mathbb{S}_2} \mathcal{D}_3 \sigma \sigma \bigg|_{\partial_{I_{N(S)}}}=-\int_{\mathbb{S}_2\times I_{N(S)}}\mathcal{L}_{SYM}$.

 This paper is organised as follows. In subsection \ref{recast} we recast the $Q_\epsilon$-superalgebra in terms of a redefined gauge potential, denoted as $\hat{A}$. This step, together with a proper selection of $Q_\epsilon$ exact localising term, is essential to deal with the explicit time dependence of the zero locus profile for $\sigma$. It will also allow to absorb some of the effects of such a dependence with the help of gauge degeneracy. However, we stress that $\sigma$ can not be absorbed completely $alla$ Stuckelberg. What we meant to transmit in the previous sentence, is that we can eliminate {\bf only} the explicit dependence on $t$ out of the quadratic operators to compute determinants upon, by a convenient choice of gauge and $Q_\epsilon$ localising term. In other words, we will fix the imaginary part of $A_t$ in a convenient way, that allows to eliminate the aforementioned $t$-dependent terms in a specific $Q_\epsilon$ exact localising term \footnote{Semi-positive definiteness of the bosonic part of the $Q_\epsilon$ exact localising term will not be affected at all by this choice.}. It must be noticed that quadratic terms of the form $\bar{\phi}^\dagger \mathcal{D}_t \sigma \phi$, which are proportional to the magnetic flux, remain inert under such a gauge choice freedom. Such terms will play the role of $i \,\bar{\phi}^\dagger D \phi$ in \citep{BZ}.

  A point could be made on whether to relax the reality condition for $A_t$ would mean to diverse or not from the "real" contour of integration. However, the definition of "real" contour comes after gauge fixing is imposed, namely, functional integration is performed over physical degrees of freedom. As the imaginary part of $A_t$ is going to be fixed to a value, there will not be physical fluctuations along it and hence functional integration is performed over hermitian degrees of freedom. But this subjective point aside, the important point in favour of the approach to be followed here, is that semi-positive definiteness of the bosonic part localising term is not sacrificed along the path of integration. Next, we go on to define our localising $Q_\epsilon V$ terms for vector and matter multiplets, as well as Chern-Simons terms. We obtain the $Q_\epsilon$-localisation locus for our reality conditions and repeat the same analysis from the $\tilde{Q}_\epsilon$ supercharge perspective. %and show under specific matching rules at the junction points $\partial I_N=\{0,\pi\}$ the glueing of the two blocks coincide with the 3D TT index of \citep{BZ}.

 Thereafter, we move on to compute the 3D TT partition functions on $\mathbb{S}_2\times I_{N,S}$.  %We will illustrate, how the 3D TT index can be seen as a  "glueing" of the partition functions on $I_N$ and $I_S$ as naive intuition predicts.
 We will show, that glueing together the partition functions on $I_N=(0,\pi)$ and  $I_S=(\pi,2\pi)$ results in the TT Index formula of \citep{BZ}.

  Our reality conditions for the temporal component of $\hat{A}_t$ imply that the moduli $u_i$ (the eigenvalues of its expectation value in the direction of a Cartan generator $h_i$) must be integrated along the real line; it also implies the annihilation of the VEV of $\sigma$ which as will be shown, is the imaginary part of $u$. The selection of the segment of integration for $u$ is not fixed a priori. Keeping this issue in mind, we move forward and using the periodicity properties of $Z_m[u]$ we define $(0,2\pi)$ as the segment of integration. In fact the poles in $Z_m[u]$ due to the presence of a single pair of chiral-antichiral multiplets, appear with periodicity $2\pi \mathbb{Z}$ in the variable $u$. Thence it is plausible to restrict integration to a single cell, as done in \citep{Shlomo}  (in their case $\mathbb{S}_2\times T_2$, only poles in the fundamental domain of the corresponding elliptic function are selected). In this way we take into consideration residues coming from a single pole in representation of a matter multiplet and not the full tale of images. This is, the remaining images of the selected pole do not signal the presence of additional matter content.  In Appendix \ref{Inte}, we illustrate how the result coincides with the JK prescription given in \citep{BZ}. To see that, one can see how to map the integral over the segment $(0,2\pi)$ to the integration over a contour closed over the upper half of the complex plane and it is thenceforth reduced to a computation of residues.
  From this perspective it is clear to see the origin of the so called "boundary contribution" \citep{BZ}. We focus on the rank one case.

  For completeness, we analyse the index of a massive deformation of $U(N)_{k=0}$ with $N$ chiral-antichiral multiplets in the fundamental and $N$ such other pairs in the anti-fundamental. %Following \citep{BZ}, in this example we commute the summation over $\mathfrak{m}$ with the integration over the moduli $\int_{JK} d u_i$. %This commutation is justified provided $\sum_{\mathfrak{m}} Z_{\mathfrak{m}}[u]$ to be uniformly convergent inside the JK path of integration.
Even though we will only flavour a $U(1)$, the result of the integration is not vanishing in this case \footnote{The technical reason behind this suspicion is that the $U(N)$ Vandermonde determinant $\prod_{i < j=1}^N \left(\sin{\frac{u_i -u_j}{2}}\right)^2$ would vanish evaluated at a point in the moduli space $\overrightarrow{u}_{(0)}$, should any pair of components $i-th$ and $j-th$ coincide ${u_{(0)}}_i={u_{(0)}}_j$ with $1\leq i,j\leq N$. }. The summation over $\mathfrak{m}$ will spread the $N$ poles at $x_{(0)}=e^{ i v}$ of the fundamental matter multiplets in $Z_{\mathfrak{m}}[x:=e^{i u}]$ over the circumference of unit radius in such a way a non trivial result is obtained for the topologically twisted index which in our toy example goes like $N^2$ for large $N$.\footnote{ The segment of integration $(0,2\pi)$ crosses the poles in this case. Thenceforth, we define the integration as its principal value (PV).}

\section{ The supercharges $Q_\epsilon$ and $\tilde{Q}_\epsilon$ }
In this section we reconstruct the supercharges $Q_\epsilon$ and $\tilde{Q}_\epsilon$ in our conventions, with special emphasis in total derivatives. Because the zero locus background we are about to work with, does not preserve time homogeneity, it will be useful to redefine the gauge potential in a specific way, such that the trouble caused by time dependence of the aforementioned background gets absorbed by gauge degeneracy and a particular selection of $Q_\epsilon$ exact localising terms. We call those terms  "unorthodox". They have a positive definite bosonic action along the path of integration of interest. %We compute every total derivative to be needed in the following section. %These localising terms simplify the analysis of one loop determinants in the presence of the aforementioned backgrounds.
The $Q_\epsilon$ exact terms proved to be a more convenient choice than the orthodox ones. In sake of completeness, we compare these "unorthodox" localising terms of vector and matter multiplet with the orthodox ones. All along this section we perform consistency checks that show that the redefinition mentioned above works perfectly. We define also "unorthodox" Chern-Simons terms as well as usual Chern-Simons terms. Finally we write down the $\mathbb{Z}_2$ that relates $Q_\epsilon$-results to $\tilde{Q}_\epsilon$-results and $vice\, versa$.

 The first two subsections are intended to a reader interested in technical insights. If the interest is to get the general idea of our analysis, then we recommend to jump to Subsection \ref{subsecSum} where the summary and scope of the  analysis presented in the first two sections is given.

\subsection{The supercharge in $I_N$: $Q_\epsilon$} \label{recast}

In this subsection we rewrite the $Q_\epsilon$-supersymmetry algebra in terms of a redefined gauge section that we denote as $\hat{A}$. Secondly, we define the corresponding $Q_\epsilon$-exact localisation actions for the vector and matter sectors, as well as Chern-Simons term to be used in $\mathbb{S}_2\times I_N$. Complementarily, in appendix \ref{conventions} we show  that the bosonic and fermionic parts of the localising $Q_\epsilon V$ actions, with the original form of the $Q_\epsilon$ algebra \eqref{algebra}-\eqref{algebraChiral0} \footnote{Before recasting it in terms of the redefined potential $\hat{A}$. }, coincide with the corresponding ones in \citep{BZ}.

%From now on in this section we will not use the su(supra-)ffix $N$ except for when necessary. Implicilty, the discussion in this subsection suppose the half $\mathbb{S}_2\times I_N$ as the domain chart of study.

Let us start by a complexified gauge multiplet $(A_\mu,\sigma, \lambda, \bar{\lambda},D)$ endowed with the following $Q_\epsilon$ supersymmetry algebra
\begin{eqnarray}
Q_{\epsilon} A_\mu&=& %Q_{\epsilon} A^*_\mu =
-\frac{i}{2}\left(%\bar{\epsilon}^\dagger\gamma_\mu \lambda
 -\bar{\lambda}^\dagger \gamma_\mu \epsilon\right), ~~~~
Q_{\epsilon} \sigma =%Q_{\epsilon} \sigma^*=
\frac{1}{2}\left(%\bar{\epsilon}^\dagger\lambda
-\bar{\lambda}^\dagger \epsilon\right),\nonumber\\
Q_{\epsilon} \lambda&=&- \frac{1}{2} \gamma^{\mu\nu} \epsilon F_{\mu \nu}+D \epsilon-i \gamma^\mu \epsilon \,\mathcal{D}_\mu \sigma ~~~~ Q_{\epsilon} \bar{\lambda}^\dagger=0% - \frac{1}{2}\bar{\epsilon}^\dagger \gamma^{\mu\nu} F_{\mu \nu}+D \bar{\epsilon}^\dagger+i \bar{\epsilon}^\dagger\gamma^\mu \mathcal{D}_\mu \sigma \nonumber
\nonumber
 \\
 Q_{\epsilon} D&=&%Q_{\epsilon} D^*= %-\frac{i}{2} \bar{\epsilon}^\dagger \gamma^\mu \mathcal{D}_\mu \lambda
 -\frac{i}{2}  (\mathcal{D}_\mu \bar{\lambda})^\dagger \gamma^\mu \epsilon
 %+\frac{i}{2} [\bar{\epsilon}^\dagger\lambda, \sigma]
 +\frac{i}{2}[\bar{\lambda}^\dagger \epsilon,\sigma].\label{algebra}
 \end{eqnarray}
  To us, the symbol $\dagger$ will be exclusive to the operation of complex conjugation. What in \citep{BZ} has been called $\lambda^\dagger$ we will call $\bar{\lambda}^\dagger$, namely
 %We stress that $Q_{\epsilon}$ is not a "Real" charge in the sense
% \begin{eqnarray}
% (\delta_\epsilon F)^* \neq \delta_\epsilon F^*.
% \end{eqnarray}
% We will comment on this point in a while.
 \begin{eqnarray}
\lambda^\dagger_{there} \equiv \bar{\lambda}^\dagger_{here}.
\end{eqnarray}
To us, $\lambda^\dagger$ is not an independent field but the complex conjugated of $\lambda$. The covariant derivative $\mathcal{D}_\mu$ is absolute, namely it includes the gauge connection, the $R-$Symmetry background gauge connection, the spin connection, flavour background connection (See \eqref{CovDer}). The killing spinor $\epsilon$ is commuting and hence $Q_\epsilon:= \epsilon_ \alpha Q^\alpha$ is anti-commuting. More details on our conventions are given in Appendix \ref{conventions}.

 For convenience, let us show a property of the supersymmetry algebra \eqref{algebra} that will result useful later on when dealing with the zero locus that includes the $\sigma^-_{(0)}$ background \eqref{sigma}. Notice that \eqref{algebra} can be rewritten as
 \begin{eqnarray}
Q_{\epsilon} \hat{A}_{\theta, \phi }&=& -\frac{i}{2}\left(%\bar{\epsilon}^\dagger\gamma_\mu \lambda
 -\bar{\lambda}^\dagger \gamma_{\theta, \phi} \epsilon\right), ~~ Q_{\epsilon} \hat{A}_t=0 \nonumber\\
Q_{\epsilon} \sigma &=& \frac{1}{2}\left(%\bar{\epsilon}^\dagger\lambda
-\bar{\lambda}^\dagger \epsilon\right),~~~~~
Q_{\epsilon} \lambda = -\frac{1}{2} \gamma^{\mu\nu} \epsilon \hat{F}_{\mu \nu}+D \epsilon-i \gamma^3 \epsilon \hat{\mathcal{D}}_3 \sigma\nonumber\\
Q_{\epsilon} \bar{\lambda}^\dagger&=&0% - \frac{1}{2}\bar{\epsilon}^\dagger \gamma^{\mu\nu} F_{\mu \nu}+D \bar{\epsilon}^\dagger+i \bar{\epsilon}^\dagger\gamma^\mu \mathcal{D}_\mu \sigma \nonumber
~~~~~~~
 Q_{\epsilon} D= %-\frac{i}{2} \bar{\epsilon}^\dagger \gamma^\mu \mathcal{D}_\mu \lambda
 -\frac{i}{2}  (\hat{\mathcal{D}}_\mu \bar{\lambda})^\dagger \gamma^\mu \epsilon \label{algebra1}
 \end{eqnarray}
where $\hat{\mathcal{D}}_\mu$ is the fully covariant derivative but with a redefined gauge potential
$\hat{A}_\mu$ constructed out of $A_\mu$ and $\sigma$. The redefined potential $\hat{A}_\mu$ coincides with $A_\mu$ except for the temporal component
\begin{eqnarray}
\hat{A}_3 := A_3 +i \sigma. \label{effPot}
\end{eqnarray}
$\hat{F}_{\mu \nu}$ is the field strength of $\hat{A}$.

The explicit form of the killing spinor $\epsilon$  \eqref{ksform}, is cardinal in achieving the form \eqref{algebra1}. This is because in order to rearrange the spinor structure and being able to absorb the $\mathcal{D}_{\theta} \sigma$ and $\mathcal{D}_{\phi} \sigma$ derivative terms into the field strength term $\gamma^{\mu \nu} F_{\mu \nu}$ of the gaugino variation in \eqref{algebra}  one must use that
\begin{eqnarray}
\gamma^2\epsilon=-i \gamma^1 \epsilon,
\end{eqnarray}
together with the relations
\begin{eqnarray}
 e^\theta_1 (\star F_\theta)
&=&e^\theta_1 (\star \hat{F}_\theta)- i e^\phi_2 \ \mathcal{D}_\phi \sigma \nonumber\\
 e^\phi_2 (\star F_\phi)
&=&e^\phi_2 (\star \hat{F}_\phi)+ i e^\theta_1 \ \mathcal{D}_\theta \sigma. \label{Identidad}
\end{eqnarray}

 In this paper the numerical indices $1,2,3$ will refer to flat space indices, so the reader should not confuse quantities such as $F_{12}$ with $F_{\theta \phi}$, the latter being obtained from the former by mean of the driebeins. %Also it should be stressed that the derivative $\mathcal{D}_\mu$ is covariant under every local symmetry (even space time symmetries) the object it acts upon is charged of.

 As for the complexified vector multiplet, we are going to be interested in the following reality conditions % as our middle dimensional contour of functional integration
\begin{eqnarray}
A_\theta= A_\theta^*, ~ A_\phi= A_\phi^*, ~  \hat{A}_t=\hat{A}_t^*,%Im[A_t]=  -\sigma  ,
~ \sigma=\sigma^*, ~ D=D^*.% ~ \lambda=\lambda^*, ~\bar{\lambda}=\bar{\lambda}^*.
\label{ComplexCond}
\end{eqnarray}
%In our path of integration, $\lambda$ and $\bar{\lambda}$ are not going to be considered as hermitian.
Notice that the reality condition is imposed upon $\hat{A}_3$ not upon $A_3$. In terms of $A_3$ the condition is $Im[A_3]=\sigma$. However, as will become clear when computing one-loop determinants we are going to gauge fix $A_3=i \sigma+\ldots$ and thence the imaginary part of $A_t$ will be non physical. The $\dots$ stand for a constant and hermitian generator, namely a real zero mode that we will call $u$.

As for the matter multiplets $(\phi, \psi, F)$ and $(\bar{\phi}, \bar{\psi}, \bar{F})$, the nilpotent algebra associated to $Q_\epsilon$ reads
%\begin{eqnarray}
%Q_{\epsilon} \phi &=&%\bar{\epsilon}^\dagger \psi
%0, ~~~~~~
%Q_{\epsilon} \bar{\phi}^\dagger =  \bar{\psi}^\dagger \epsilon, \nonumber\\
%Q_{\epsilon} \psi &=& i \gamma^\mu \epsilon \mathcal{D}_\mu \phi +i \epsilon \sigma \phi,%+\bar{\epsilon} F
%~~~~~~ Q_{\epsilon} \bar{\psi}^\dagger = %i \gamma^\mu \bar{\epsilon} \mathcal{D}_\mu \bar{\phi}+i \bar{\phi} \sigma \bar{\epsilon}+
%-\bar{F}^\dagger \epsilon^c^\dagger,~~~~~~\nonumber\\
%Q_{\epsilon} F&=& i (\epsilon^c)^\dagger \gamma^\mu \mathcal{D}_\mu \psi- i \sigma (\epsilon^c)^\dagger \psi-i (\epsilon^c)^\dagger \lambda \, \phi,~~~~~~
%Q_{\epsilon} \bar{F}^\dagger= 0.% i \bar{\epsilon}^\dagger \gamma^\mu \mathcal{D}_\mu \psi- i \bar{\epsilon}^\dagger \bar{\psi} \sigma+i \bar{\phi} \bar{\epsilon}^\dagger \bar{\lambda}
%\label{algebraChiral0}
%\end{eqnarray}\footnote{Where $\epsilon^C:=C \epsilon^*$ and $C=-i \sigma_2$. Noice, our $C$ conjugation matrix is real.}
\begin{eqnarray}
Q_{\epsilon} \phi &=&%\bar{\epsilon}^\dagger \psi
0, ~~~~~~
Q_{\epsilon} \bar{\phi}^\dagger =  -\bar{\psi}^\dagger \epsilon, \nonumber\\
Q_{\epsilon} \psi &=&- i \gamma^\mu \epsilon \mathcal{D}_\mu \phi -i \epsilon \sigma \phi,%+\bar{\epsilon} F
~~~~~~ Q_{\epsilon} \bar{\psi}^\dagger = %i \gamma^\mu \bar{\epsilon} \mathcal{D}_\mu \bar{\phi}+i \bar{\phi} \sigma \bar{\epsilon}-
+\bar{F}^\dagger {\epsilon^c}^\dagger,~~~~~~\nonumber\\
Q_{\epsilon} F&=& -i (\epsilon^c)^\dagger \gamma^\mu \mathcal{D}_\mu \psi-i \sigma (\epsilon^c)^\dagger \psi+i (\epsilon^c)^\dagger \lambda \, \phi,~~~~~~
Q_{\epsilon} \bar{F}^\dagger= 0.% i \bar{\epsilon}^\dagger \gamma^\mu \mathcal{D}_\mu \psi- i \bar{\epsilon}^\dagger \bar{\psi} \sigma+i \bar{\phi} \bar{\epsilon}^\dagger \bar{\lambda}
\label{algebraChiral0}
\nonumber
\end{eqnarray}\footnote{Where $\epsilon^C:=C \epsilon^*$ and $C=-i \sigma_2$. Notice, the $C$ conjugation matrix used in this paper is real.}
As already said for the gaugini,
\begin{eqnarray}
\phi_{there}^\dagger\equiv\bar{\phi}^\dagger_{here}, \, \,
\psi^\dagger_{there}\equiv\bar{\psi}^\dagger_{here},\end{eqnarray}
where $there$ refers to \citep{BZ}.

 For computational convenience we will rewrite \eqref{algebraChiral0} as
\begin{eqnarray}
Q_{\epsilon} \phi &=&%\bar{\epsilon}^\dagger \psi
0, ~~~~~~ Q_{\epsilon} \bar{\phi}^\dagger = - \bar{\psi}^\dagger \epsilon\nonumber \\
Q_{\epsilon} \psi &=& -i \gamma^\mu \epsilon \hat{\mathcal{D}}_\mu \phi,~~~~~~
Q_{\epsilon} \bar{\psi}^\dagger= %i \gamma^\mu \bar{\epsilon} \mathcal{D}_\mu \bar{\phi}+i \bar{\phi} \sigma \bar{\epsilon}+
%-
\bar{F}^\dagger {\epsilon^c}^\dagger \nonumber\\
Q_{\epsilon} F&=& -i (\epsilon^c)^\dagger \gamma^\mu \hat{\mathcal{D}}_\mu \psi+i (\epsilon^c)^\dagger \lambda \, \phi, ~~~~~~
Q_{\epsilon} \bar{F}^\dagger= 0,% i \bar{\epsilon}^\dagger \gamma^\mu \mathcal{D}_\mu \psi- i \bar{\epsilon}^\dagger \bar{\psi} \sigma+i \bar{\phi} \bar{\epsilon}^\dagger \bar{\lambda}
\label{algebraChiral}
\end{eqnarray}
were $\hat{\mathcal{D}}_\mu$ is the fully covariant derivative but with the effective gauge potential
$\hat{A}_\mu$ defined in \eqref{effPot}. %constructed out of $A_\mu$ and $\sigma$. The effective potential $\hat{A}_\mu$ coincides with $A_\mu$ except for the temporal component
%\begin{eqnarray}
%\hat{A}_3 \equiv A_3 +i \sigma.
%\end{eqnarray}

We are going to integrate the matter multiplets along the path
\begin{eqnarray}
\phi%=\phi^*=\bar{\phi}
=\bar{\phi}, ~ F%=F^*= \bar{F}
=\bar{F}, \label{ComplexCondChiral}
\end{eqnarray}
which guaranties positive semi-definiteness of our $Q_\epsilon V_{Matter}$. As shall be argued upon in sections to come, we are interested in matter contents that do not create gauge/parity anomaly. This condition imposes a constraint on the gauge representation carried by the matter content \citep{Redlich1, Redlich2}.

%\begin{eqnarray}
%Q^*_{\epsilon} \phi &=&%\bar{\epsilon}^\dagger \psi
%0\\
%Q^*_{\epsilon} \bar{\phi} &=& \bar{\psi}^ \dagger \epsilon  \\
%Q^*_{\epsilon} \psi &=&- i \gamma^\mu^T \epsilon \mathcal{D}_\mu \phi +i \epsilon \sigma \phi%+\bar{\epsilon} F
%\\
%Q^*_{\epsilon} \bar{\psi}&=& %i \gamma^\mu \bar{\epsilon} \mathcal{D}_\mu \bar{\phi}+i \bar{\phi} \sigma \bar{\epsilon}+
%-\bar{F}^* \epsilon^c\\
%Q^*_{\epsilon} F&=&- i (\mathcal{D}_\mu \psi)^\dagger \gamma^\mu \epsilon^c- i \sigma \psi^\dagger \epsilon^c -i \lambda^\dagger \epsilon^c  \, \phi \\
%Q^*_{\epsilon} \bar{F}&=& 0% i \bar{\epsilon}^\dagger \gamma^\mu \mathcal{D}_\mu \psi- i \bar{\epsilon}^\dagger \bar{\psi} \sigma+i \bar{\phi} \bar{\epsilon}^\dagger \bar{\lambda}
%\label{algebraChiral2}
%\end{eqnarray}

%\paragraph{Super Yang Mills Action and boundary term}
\vspace{.2cm}

Let us start by writing the $Q_\epsilon V$ localising term for the gauge multiplet
\begin{eqnarray}
Q_\epsilon V_{vector}&:=&Q_{\epsilon} \left(~~(\overset{\bullet}{Q_{\epsilon} \lambda}) \lambda\right), \label{locV} \\
(\overset{\bullet}{Q_{\epsilon} \lambda})&:=& \left(Q_\epsilon \lambda\right)^* \bigg|_{\hat{A}^*\rightarrow \hat{A}, ~ \sigma^*\rightarrow \sigma, ~ D^* \rightarrow D},\label{eqbullet0}
\end{eqnarray}
In a while we will comment about \eqref{locV} and its relation with the term used in \citep{BZ}, that we have written in the LHS of \eqref{Qexact}. %As can be checked after comparing the RHS of \eqref{QVector} with the RHS of \eqref{Qexact}, they are equivalent up to a total derivative term. %that integrates trivially when evaluated on the non trivial background fields.
%\begin{eqnarray}
%\overset{\bullet}{(\delta_\epsilon \lambda)}\equiv \frac{1}{2} F_{\mu \nu} \gamma ^{\mu \nu}\epsilon - D^\dagger \epsilon +i \mathcal{D}_\mu\sigma \gamma^\mu \epsilon
%\end{eqnarray}

%By using the form of the algebra
%\eqref{algebra} we obtain
%\begin{eqnarray}
%Q V^{Vector}\equiv Tr \bigg[ \left | \star F _\mu+ %i
% \mathcal{D}_\mu \sigma\right |^2 +|D|^2+i (\hat{\mathcal{D}}_\mu\bar{\lambda})^\dagger \gamma^\mu \lambda\bigg] . \label{QVector2}
%\end{eqnarray}

 In subsections to come the non covariant form of the algebra \eqref{algebra1} and the localising term \eqref{locV} will be used to compute one loop determinants. This will prove to be useful at the technical level. Mainly because with the aforementioned form of the algebra and localising term, the explicit time dependence of the zero locus background $\sigma_{0}$ mentioned in the introduction, \eqref{sigma}, will be implicit.

 From the term $(\overset{\bullet}{Q_{\epsilon} \lambda}) Q_\epsilon \lambda$ in the expansion of \eqref{locV}, and with the help of \eqref{algebra1} is straightforward to get the bosonic part
\begin{eqnarray}
Q_{\epsilon}V^{Vector}_B&\equiv&% |D-i F_{1 2}-i \hat{\mathcal{D}}_3 \sigma |^2+|\hat{F}_{13}- i \hat{F}_{2 3}|^2. \label{QVvec2}
D^2+ \left(F_{1 2}+ \hat{\mathcal{D}}_3 \sigma \right)^2+\left(\hat{F}_{13}\right)^2+ \left(\hat{F}_{2 3}\right)^2. \label{QVvec2}
\end{eqnarray}
It should be stressed that all equalities in this subsection do take into consideration total derivatives. Notice that with the reality conditions \eqref{ComplexCond} this bosonic term is positive definite.

As for the fermionic part, one gets
\begin{eqnarray}
Q_{\epsilon} \overset{\bullet}{(Q_{\epsilon} \lambda)}=- i\bar{\lambda}^\dagger \gamma^\mu \overleftarrow{P}^+_\mu, \label{ddlambda}
\end{eqnarray}
with
\begin{eqnarray}
P^+_{\theta, \phi}:=\frac{1+\sigma_3}{2} \mathcal{D}_{\theta, \phi}~~~, ~~~~~ P^+_t:=\mathcal{D}_t.
\end{eqnarray}
From \eqref{ddlambda} we obtain the fermionic part of \eqref{locV}
\begin{eqnarray}
Q_\epsilon V^{Vector}_F&=&-( \bar{\lambda}_1^\dagger   ~~~ \bar{\lambda}_2^\dagger   )  \left(\begin{array}{c c}
 %-
 i \overleftarrow{\hat{\mathcal{D}}_t} &  0\\
 %-
 \frac{i}{R} ( \overleftarrow{\mathcal{D}_{\theta}}+i \csc \theta \overleftarrow{\mathcal{D}_\phi} ) & -i \overleftarrow{\hat{\mathcal{D}}_t}
 \end{array}\right) \left(\begin{array}{c} \lambda_1 \\  \lambda_2 \end{array}\right).  \label{QVectorStuckelberg}\end{eqnarray}
Notice that this expression is non covariant. The use of this non covariant form results convenient when computing 1-loop determinants. We stress that $\mathcal{D}_\phi$ should be understood as covariant derivative that includes the magnetic flux potential felt by the effective charge of the corresponding gaugino component. Specifically, the spin connection part of the covariant derivative is included in the latter mentioned magnetic flux contribution. See appendix \ref{1loopdets} for further details in this issue.

In Appendix \ref{loft} we show how to derive the localising action coming from \eqref{locV} but by using the covariant form of the algebra \eqref{algebra}. Is straightforward to check that the result for the action in the latter case is the same \eqref{QVvec2}+\eqref{QVectorStuckelberg} but after substitution of \eqref{effPot}. This coincidence, checks the consistency of using the algebra in terms of the redefined potential $\hat{A}$ \eqref{algebra1} instead of the usual one in terms of $A$ \eqref{algebra}.

 It is also possible to use the orthodox localising term, the one used in \citep{BZ} for instance, which in the conventions used in this paper is given by
\begin{eqnarray}
Q_\epsilon \tilde{Q}_\epsilon Tr \left(\bar{\lambda}^\dagger \lambda -4 \sigma D \right). \label{LocVCov}
\end{eqnarray}

After expanding \eqref{LocVCov} with the use of the non covariant form of the algebra \eqref{algebra1} one gets

\begin{eqnarray}
Q_\epsilon V^{vector}_{B}&:=&\frac{1}{2}\hat{F}_{\mu \nu}^2+ \left(\hat{\mathcal{D}}_3\sigma\right)^2+D^2 +2 \sigma i  \hat{\mathcal{D}}_\mu \hat{F}^\mu_{~3}  - 2\sigma i \hat{\mathcal{D}}_\mu \star \hat{F}^\mu-2 i \sigma \{\bar{\lambda}^\dagger,\lambda\}\nonumber \\ &-&2
i \hat{\mathcal{D}}_3 \left(  \sigma\left( D-  i \hat{\mathcal{D}}_3 \sigma\right) \right),\label{COVBOSVector}\\
&& \nonumber
\\
Q_\epsilon V^{vector}_{F}&:=&i \bar{\lambda}^\dagger \hat{\slashed{\mathcal{D}}} \lambda- i \hat{\mathcal{D}}_\mu \left(\bar{\lambda}^\dagger P^-\gamma^\mu  \lambda  \right),\label{COVFERVector} \\&&\nonumber
\end{eqnarray}
with $ \star F _\mu := \frac{1}{2} \epsilon_\mu^{~\nu \beta} F_{\nu \beta}$ and $P^{\mp}:=\frac{1\mp\sigma_3}{2} $. Trace over the gauge indices is assumed in \eqref{COVBOSVector} and \eqref{COVFERVector}. It is straightforward to show that after replacing the expression for the redefined gauge potential $\hat{A}$ in terms of $A$, \eqref{effPot}, in \eqref{COVFERVector}, one obtains the covariant form of the localising action. Namely, the localising term \eqref{LocVCov} worked out with the covariant form of the algebra \eqref{algebra}
\begin{eqnarray}
Q_\epsilon V^{vector}_B&:=&\frac{1}{2} F_{\mu \nu}^2+\left(\mathcal{D}_\mu\sigma\right)^2+D^2- i \sigma \{\bar{\lambda}^\dagger,\lambda\} -i\sigma \{\bar{\lambda}^\dagger, P^- \lambda\}\nonumber \\&+& 2\, i\, \mathcal{D}_\mu \bigg(\sigma v^\nu F^\mu_{~\nu}-v^\mu \sigma D + i \sigma D^\mu \sigma - \epsilon^{\mu \nu \beta} v_\beta \, \sigma \mathcal{D}_\nu \sigma  \bigg), \nonumber\\ && \nonumber\\
Q_\epsilon V^{vector}_F&:=& i \bar{\lambda}^\dagger \slashed{D} \lambda +i\sigma \{\bar{\lambda}^\dagger, P^- \lambda\}-i \mathcal{D}_\mu \left(\bar{\lambda}^\dagger P^-\gamma^\mu  \lambda\right).
\end{eqnarray}
Again, this coincidence checks the consistency of using the algebra in terms of the redefined potential $\hat{A}$ \eqref{algebra1} instead of the covariant one \eqref{algebra}.

As one could expect a priori, the difference between the localising term \eqref{locV} and \eqref{LocVCov} is the $Q_\epsilon$- exact term
\begin{eqnarray}
Q_\epsilon\left(  2 D \epsilon\lambda +\tilde{Q}_\epsilon \left(4 \sigma D\right)\right), \label{diffQV}
\end{eqnarray}
which, naively, in virtue of the localisation argument, means both localising terms are set to compute the same observables. However one must not forget that the for the reality conditions \eqref{ComplexCond}, the bosonic part of the localising term to be used here, \eqref{locV}, is positive definite, meanwhile the bosonic part of \eqref{LocVCov} is not.

% We will localise by using this term. This term will allow us to map the computation in the presence of a non trivial Bogomolnyi solution of the zero locus conditions to the approach used in \citep{BZ}.

%SUSY variation of SYM action is
%\begin{equation}
%Q_{\epsilon} S_{SYM} =  \int_{y=y_0} d\theta dx \sqrt{ g} Tr\left( - \frac{i y}{2} \delta_\epsilon F_{x \theta} \sigma-\frac{iy}{2} F_{x \theta} \delta_\epsilon \sigma-\frac{i}{2} \bar{\lambda}^\dagger \gamma^y \delta_\epsilon \lambda \right).
%\end{equation}

%Notice that for the boundary conditions \eqref{bcondition2} could be added to the SYM action in order to make it supersymetric is given by
%\begin{eqnarray}
%S_{bdry}=-sign  \int_{y=y_0} d\theta dx \sqrt{ g} Tr\left( \bar{\lambda} \gamma^y \lambda \right).
%\end{eqnarray}
%A possible path to choose is to impose boundary conditions onto $\lambda$ and $\bar{\lambda}$ in such a way $S_{bdry}$ vanish. In that case SYM would be by itself supersymmetric.

%Interesting: Note Aside:

%\begin{eqnarray}
%\gamma^{u x} \partial_{u}A_{x}|_{bdry}=\gamma^{u \theta} \partial_{u}A_{ \theta}|_{bdry}=0\\
%\gamma^{u}\partial_u \sigma|_{bdry}=0
%\\\gamma^u\partial_u \lambda|_{bdry}=\gamma^u\partial_u \bar{\lambda}|_{bdry}=0\label{eq7}
%\end{eqnarray}

%This set of boundary condition preserves $\mathcal{N}=(2,2)$  Superconformal algebra on the conformal boundary $T_2=S_1\times S_1$! Ref: Gomis. The boundary value of $A_u$ becomes a scalar. However we will need to break more supersymmetry

%The conformal "boundary" of $H_2$ is at $y=0$.
%\paragraph{Chern-Simons Term}

%\paragraph{Complexified chiral multiplet}

Next, the same analysis will be performed for the matter sector of the theory. Let us define the following localising term %. %We perform a computation analog to the one presented in appendix \ref{Matter} but by using the form of the algebra \eqref{algebraChiral}  and the following localising term
\begin{eqnarray}
Q_\epsilon V^{Matter}&:=&%Q_\epsilon V^{Matter}_{I}+Q_\epsilon V^{Matter}_{II}
Q_\epsilon \bigg( i \epsilon \gamma^\mu \psi \hat{\mathcal{D}}_\mu \bar{\phi}^\dagger+ F \bar{\psi}^\dagger\epsilon^c + i \bar{\phi}^\dagger \epsilon \lambda \phi \bigg).\label{MatterAc}
%Q_\epsilon V^{Matter}_I&:=&Q_{\epsilon}\left( \left(\overset{\bullet}{Q_{\epsilon} \psi}\right)^\dagger \psi + \bar{\psi}^\dagger  \left(\overset{\bullet}{Q_{\epsilon} \bar{\psi}}\right)\right)\\&=&
%Q_{\epsilon} \left(\left(\overset{\bullet}{Q_{\epsilon} \psi}\right)^\dagger\right) \psi+\left(\overset{\bullet}{Q_{\epsilon} \psi}\right)^\dagger  Q_{\epsilon} \, \psi+\left(Q_{\epsilon} \bar{\psi}\right)^\dagger  \overset{\bullet}{ Q_{\epsilon} \bar{\psi}}-\bar{\psi}^\dagger Q_{\epsilon}\left(\overset{\bullet}{Q_{\epsilon}{\bar{\psi}}}\right), \nonumber\\ Q V^{Matter}_{II}&:=&-2 iQ_{\epsilon} \left( \bar{\phi} \, \overset{\bullet}{Q_{\epsilon}\sigma} \, \phi \right)
 \end{eqnarray}
%
%with the over-dotted objects defined as follows

%\begin{eqnarray}
%\left(\overset{\bullet}{Q_{\epsilon} \psi}\right) ^\dagger&:=&\left(Q_{\epsilon} \psi\right)^\dagger\bigg|_{\phi\rightarrow \bar{\phi}}=- i \epsilon \gamma^\mu  \hat{\mathcal{D}}_\mu \bar{\phi}^\dagger. \\
%\overset{\bullet}{Q_{\epsilon} \bar{\psi}}&:=&%-
 %\left(Q_\epsilon \bar{\psi}^\dagger\right)^\dagger \bigg |_{\bar{F}\rightarrow F}= F \epsilon^c.\\
%\overset{\bullet}{Q_{\epsilon} \sigma }&:=& \left(Q_\epsilon \sigma\right)^\dagger\bigg|_{\bar{\lambda}\rightarrow \lambda}=-\frac{1}{2} \epsilon \lambda.
%\end{eqnarray}

In appendix \ref{Matter}, it will be seen how this localising term differs from the one used in \citep{BZ}, by using the covariant form of the algebra \eqref{algebra} and \eqref{algebraChiral0}%, namely the algebra in terms of the potential $A$, not the one in terms of the redefined potential $\hat{A}$
. The difference being a $Q_\epsilon$ exact term, as shall be seen below  and thenceforth both localising terms are set to compute the same observables.

The expansion of \eqref{MatterAc} under the redefined form of the $Q_\epsilon$ algebra, \eqref{algebra1} and \eqref{algebraChiral}, is given by the sum of the following bosonic and fermionic terms

\begin{eqnarray}
Q_\epsilon V^{Matter}_{B}&=& ({\hat{\mathcal{D}}}^{\mu}{\bar{\phi}})^\dagger  \, {\hat{\mathcal{D}}}_{\mu}{\phi} +  \, \bar{\phi}^\dagger\left(\hat{\mathcal{D}}_3\sigma+i D- \epsilon^{\mu \nu}_{~~\beta}v^{\beta} \left(q V_{\mu \nu}+W_{\mu \nu}\right) \right)\phi
\nonumber\\&&  +\bar{F}^\dagger F+\,\hat{\mathcal{D}}_{\mu}\left( i \, \epsilon^{\mu \nu}_{~~\beta}v^{\beta} \bar{\phi}^\dagger {\hat{\mathcal{D}}}_{\nu} \phi \right)
, \label{LBOSF0}
\\ && \nonumber \\
Q_\epsilon V^{Matter}_{F}&=&i \, \bar{\psi}^\dagger {\gamma}^{\mu} {\hat{\mathcal{D}}}_{\mu}{\psi}- i\,  \bar{\psi}^\dagger \lambda \,\phi- i \,
\bar{\phi}^\dagger\, \bar{\lambda}^\dagger P^- \psi- i \hat{\mathcal{D}}_\mu\left( \bar{\psi}^\dagger P^+ \gamma^\mu \psi\right).  \label{LFERF0} \\  && \nonumber
\end{eqnarray}

The saddle point expansion of these Lagrangians about the $Q_\epsilon$ localising locus will coincide with the one of \citep{BZ}.  Notice the presence of the term $\bar{\phi}^\dagger \mathcal{D}_3 \sigma \phi$. This term provides the flux dependence that in \citep{BZ} comes from the complex value of $D$. Specifically the $Q_\epsilon$ zero locus condition will fix $\mathcal{D}_3 \sigma$ to the density $-F_{12}$. %As shall be seen, from the global  perspective of $\mathbb{S}_1$ these configurations will correspond to 't Hooft insertions/defects. It is said global because from the point of view of a patch $I_N$($I_S$) these insertions are not evident. %Just to give a glimpse, these insertions will be domain walls that allow a jump in the topological charges of fluxes over $\mathbb{S}_2$ from $I_N$ to $I_S$ and $vice\, versa$.

To show the difference of the localising term \eqref{MatterAc} with the orthodox one, the one used in \citep{BZ}, let us write down the latter one in our conventions. The orthodox localising term for matter in the conventions of this paper is given by the super derivative
\begin{eqnarray}
Q_\epsilon \tilde{Q}_\epsilon\bigg( -\bar{\psi}^\dagger\psi -2 i \, \bar{\phi}^\dagger \sigma \phi  \bigg) \label{LOCMATORT},
\end{eqnarray}
where the $\tilde{Q}_\epsilon$ algebra has not been defined yet. We postpone the writing of $\tilde{Q}_\epsilon$ until next subsection, equation \eqref{algebra1South}, in order not to make too clumsy the presentation in this subsection.

After expanding \eqref{LOCMATORT} with the redefined form of the $Q_\epsilon$ algebra \eqref{algebra1} \eqref{algebraChiral} one obtains

\begin{eqnarray}
Q_\epsilon V^{Matter}_B&=&(\hat{\mathcal{D}}^{\mu} {\bar{\phi}})^\dagger  \,
 \hat{\mathcal{D}}_\mu \phi +  \, \bar{\phi}^\dagger\left(\hat{\mathcal{D}}_3\sigma+i D- \epsilon^{\mu \nu}_{~~\beta}v^{\beta} \left(q V_{\mu \nu}+W_{\mu\nu}\right) \right)\,\phi  \nonumber \\ && + {\bar{F}}^\dagger F +2 \bar{\phi}^\dagger \sigma \hat{\mathcal{D}}_3\phi+ \hat{\mathcal{D}}_{\mu}\left( i \, \epsilon^{\mu \nu}_{~~\beta}v^{\beta} \bar{\phi}^\dagger {\hat{\mathcal{D}}}_{\nu} \phi \right),\label{RefEqTot0} \\\nonumber &&\\
Q_\epsilon V^{Matter}_F &:=&i \, \bar{\psi}^\dagger {\gamma}^{\mu} {\hat{\mathcal{D}}}_{\mu}{\psi}-2 \,i \, \bar{\psi}^\dagger\, P^+\sigma \,\psi- i\,  \bar{\psi}^\dagger \lambda \,\phi- i \,
\bar{\phi}^\dagger\, \bar{\lambda}^\dagger \psi- i \hat{\mathcal{D}}_\mu\left( \bar{\psi}^\dagger P^+ \gamma^\mu \psi\right).
\end{eqnarray}

%\begin{eqnarray}&&\nonumber\\
%Q_\epsilon V^{Matter}_B&=&({\hat{\mathcal{D}}^{\mu} \bar{\phi})^\dagger  \,
 %\hat{\mathcal{D}}_\mu \phi +  \, \bar{\phi}^\dagger\left(\hat{\mathcal{D}}_3\sigma+i D- \epsilon^{\mu \nu}_{~~\beta}v^{\beta} \left(q V_{\mu \nu}+W_{\mu\nu}\right) \right)\,\phi  \nonumber \\ && + {\bar{F}}^\dagger F +2 \bar{\phi}^\dagger \sigma \hat{\mathcal{D}}_3\phi+ \hat{\mathcal{D}}_{\mu}\left( i \, \epsilon^{\mu \nu}_{~~\beta}v^{\beta} \bar{\phi}^\dagger {\hat{\mathcal{D}}}_{\nu} \phi \right),\label{RefEqTot0} \\\nonumber &&\\
%Q_\epsilon V^{Matter}_F &:=&i \, \bar{\psi}^\dagger {\gamma}^{\mu} {\hat{\mathcal{D}}}_{\mu}{\psi}-2 \,i \, \bar{\psi}^\dagger\, P^+\sigma \,\psi- i\,  \bar{\psi}^\dagger \lambda \,\phi- i \,
%\bar{\phi}^\dagger\, \bar{\lambda}^\dagger \psi- i \hat{\mathcal{D}}_\mu\left( \bar{\psi}^\dagger P^+ \gamma^\mu \psi\right). \\&&
%\nonumber
%\end{eqnarray}

With the use of the redefinition of $\hat{A}$ in terms of $A$ \eqref{effPot}, this expansion presented before coincides with the expansion in terms of the original form of the algebra \eqref{algebra} and \eqref{algebraChiral0} which is given by
\begin{eqnarray}
Q_\epsilon V^{Matter}_B&:=&D^{\mu}{\bar{\phi}}^\dagger  \, {{D}}_{\mu}{\phi} +\bar{\phi}^\dagger\,\left(%-\hat{\mathcal{D}}_3\sigma+
 i D\, + \sigma^2 \,- \epsilon^{\mu \nu}_{~~\beta}v^{\beta} \left(q V_{\mu \nu}+W_{\mu \nu}\right) \right) \phi \nonumber\\&+& \bar{F}^\dagger F%+\left(\hat{\mathcal{D}}_3 \bar{\phi}\right)^\dagger\sigma \phi
 +\mathcal{D}_{\mu}\left(% v^\mu \bar{\phi}^\dagger \sigma \phi +
v^\mu\, \bar{\phi}^\dagger \sigma \phi \,+\,i \, \epsilon^{\mu \nu}_{~~\beta}v^{\beta} \bar{\phi}^\dagger {D}_{\nu} \phi \right), \label{RefEqTot}\\ &&\nonumber\\
Q_\epsilon V^{Matter}_F&:=&i \, \bar{\psi}^\dagger {\gamma}^{\mu} {{D}}_{\mu}{\psi}-i \bar{\psi}^\dagger\, \sigma \,\psi- i\,  \bar{\psi}^\dagger \lambda \,\phi- i \,
\bar{\phi}^\dagger\, \bar{\lambda}^\dagger \psi- i {D}_\mu\left( \bar{\psi}^\dagger P^+ \gamma^\mu \psi\right),
\end{eqnarray}
Again, this coincidence checks the consistency of using the redefined form of the algebra \eqref{algebraChiral}. Let us take a moment aside to comment about the importance of total derivatives when dealing with BPS configurations of the kind \eqref{sigma}.  Notice that should one discard the total derivative term $\mathcal{D}_\mu\left(v^\mu\, \bar{\phi}^\dagger \sigma \phi\right)$ in \eqref{RefEqTot} the quadratic expansion about the zero loci of the form $\eqref{flux}$+$\eqref{sigma}$ would not depend on the fluxes $\mathfrak{m}$. Indeed, discarding such a term would be a mistake, because its integration over $\mathbb{S}_2\times I_N$ is non trivial. This is the reason why using the form \eqref{RefEqTot0} instead of \eqref{RefEqTot} is advisable.  However notice that \eqref{RefEqTot} has an explicit dependence on $\sigma$ that will give rise to an explicit dependence on $t$ in the quadratic expansion. Such a dependence can be absorbed by adding up a $Q_\epsilon$ exact term. The final and homogeneous in time quadratic operator expansion would be coming from the new $Q_\epsilon$- exact localising term \eqref{MatterAc} whose bosonic part has been written in \eqref{LBOSF0}.

The difference between the super-derivative localising term \eqref{LOCMATORT} and the one to be used to compute one loop determinants \eqref{MatterAc}, is the $Q_\epsilon$ exact term
\begin{eqnarray}
Q_\epsilon \bigg(2\, i \, \bar{\phi}^\dagger \sigma\, \epsilon\psi \bigg), \label{EqRefNonAP}
\end{eqnarray}
so in virtue of the localisation argument both localising terms are set to compute the same observables. However, as already argued \eqref{MatterAc} is the most convenient choice for our purposes. The $Q_\epsilon$ exact difference \eqref{EqRefNonAP} follows directly from equation \eqref{EqRefNew}.

We can also define the "unorthodox" supersymmetric Chern-Simons term
\begin{eqnarray}
\mathcal{L}_{CS}= -\frac{i k}{4 \pi}\left( \epsilon^{\mu \nu \beta}\left(\hat{A}_\mu \partial_\nu \hat{A_\beta} -\frac{2 i}{3}\hat{A}_\mu \hat{A}_\nu \hat{A}_\beta \right)  -\bar{\lambda}^\dagger\frac{ 1-\sigma_3}{2} \lambda \right). \label{ModifiedCS} 
\end{eqnarray}
$\mathcal{L}_{CS}$ is supersymmetric under \eqref{algebra1}. The variation of the bosonic CS part is a multiplication of $\lambda_2$ by a linear combination of $ \hat{F}_{13}$ and $\hat{F}_{23}$. The variation of the fermionic term, specially because  it involves only the product $\bar{\lambda}_2 \lambda_2$\footnote{From the algebra \eqref{algebra1} is easy to see that the variation of $\lambda_2$ involves only a linear combination of $ \hat{F}_{1 3}$ and $\hat{F}_{2 3}$.}, is again $\bar{\lambda}_2$ multiplied by the precise linear combination of $\hat{F}_{13}$ and $\hat{F}_{2 3}$ that cancels the variation of the bosonic term. As a classical contribution, the Chern-Simons term will not affect the 1-loop determinants, it will only provide an extra on shell value. This unorthodox CS term proved to be more natural to our analysis than the orthodox one. Its $Q_\epsilon$ variation gives the total derivative
\begin{eqnarray}
-i \frac{k}{4 \pi}  \hat{\mathcal{D}}_\mu\bigg( \epsilon^{\mu \nu \beta} \left(Q_\epsilon \hat{A}_\nu \right) \hat{A}_\beta \bigg).\label{TotDerChernN}
\end{eqnarray}
For completeness, we write down the orthodox Chern-Simons term which is symmetric under the covariant form of both algebras $Q_\epsilon$ and $\tilde{Q}_\epsilon$. In our conventions it is given by
\begin{eqnarray}
-\frac{i k}{4 \pi}\left( \epsilon^{\mu \nu \beta}\left(A_\mu \partial_\nu A_\beta -\frac{2 i}{3} A_\mu A_\nu A_\beta \right)  -\bar{\lambda}^\dagger \lambda+ 2 \, \sigma D \right). \label{ChernSimonsCovariant}
 \end{eqnarray}
 %for the analysis of global supersymmetry on $\mathbb{S}_1$ this term is also invariant up to a total derivative under the "antichiral" supercharge $\tilde{Q}_\epsilon$.
 Its total derivative $Q_\epsilon$ variation is
 \begin{eqnarray}
 -i \frac{k}{4 \pi}  \mathcal{D}_\mu\bigg( \epsilon^{\mu \nu \beta} \left(Q_\epsilon A_\nu \right) A_\beta - i\,  \bar{\lambda}^\dagger \gamma^\mu\epsilon \, \sigma \bigg).  
 \end{eqnarray}

We will also use the mixed CS term
\begin{eqnarray}
\mathcal{L}_T= - \frac{i}{2\pi} \hat{A}^T_3 tr F_{12}, \label{mCS}
\end{eqnarray}
with $\hat{A}^T_3:=A^T_3+ i \sigma^T$ being the component along $\mathbb{S}_1$ of a gauge background connection $\hat{A}^T$ associated to a $U(1)$ topological symmetry (See subsection 2.1.3 of \citep{BZ} for more details).  In \citep{BZ} other modified Chern-Simons terms are written. However \eqref{mCS} is supersymmetric by itself and it is enough to our purposes. For completeness we write down its total derivative $Q_\epsilon$ variation
\begin{eqnarray}
-\frac{i}{2 \pi}  \hat{\mathcal{D}}_\mu\bigg( \epsilon^{\mu \nu 3} \left(  \hat{A}^T_3\, Tr Q_\epsilon A_\nu \right) \bigg). \\
\nonumber
\end{eqnarray}

%\subsubsection{Supersymmetry on $D_2 \times S_1$}

%Before moving on, we comment on how to preserve supersymmetry on the hemisphere $D_2 \times S_1$, with metric \eqref{metric} and $0<\theta<\frac{\pi}{2}$.  The idea is to write down boundary conditions in such a way the restriction
%\begin{eqnarray}
%Q^{bdry}_\epsilon:=Q_\epsilon \bigg |_{\theta=\frac{\pi}{2}}
%\end{eqnarray}
%become also nilpotent algebra, namely $\left(Q^{bdry}_\epsilon \right)^2=0$. There is not a unique way to achieve that goal. For instance we could define boundary conditions in such a way that $Q^{bdry}_\epsilon:=0$. Here we follow a different logic and show a set of boundary conditions that preserve the $Q_\epsilon$- supersymmetry at the boundary $T_2$.

%It is straightforward to check that algebra \eqref{algebra} at $\theta=\frac{\pi}{2}$ reduces to
%\begin{eqnarray}
%\end{eqnarray}
%provided the following set of boundary conditions is imposed
%\begin{eqnarray}
%\partial_\theta A_\phi \bigg |_{\theta=\frac{\pi}{2}}=0, ~~ \partial_\theta A_t\bigg |_{\theta=\frac{\pi}{2}}=0, \\
%\partial_\theta \sigma \bigg |_{\theta=\frac{\pi}{2}}=0, ~~ \partial_\theta (\epsilon^C \bar{\lambda}) \bigg|_{\theta=\frac{\pi}{2}}=0 \label{bc}
%\end{eqnarray}
%From imposing the variations to obey \eqref{bc}, specifically $Q_\epsilon \sigma$, it comes out
%\begin{eqnarray}
%\partial_\theta (\epsilon \bar{\lambda}) \bigg|_{\theta=\frac{\pi}{2}}=0
%\end{eqnarray}

%\bigspace

\subsubsection*{The Localisation locus}

In this subsection the $Q_\epsilon$ locus consistent with the reality conditions \eqref{ComplexCond} and \eqref{ComplexCondChiral} is defined. This information plus gauge fixing conditions, specify the path of functional integration. At last, it is evaluated the localising term $Q_\epsilon V$ for vector and matter sectors around the aforementioned locus.

We start from the $Q_\epsilon$-BPS conditions
%The localisation term
%\begin{eqnarray}
%Tr[\delta_\epsilon\left((\overset{\bullet}{\delta_\epsilon\lambda})^\dagger \cdot \lambda\right)]&=&\mathcal{L}_{bosonic}+\mathcal{L}_{fermionic} \label{LocAction}\\
 %\mathcal{L}_{bosonic}&=&\bigg[ \left( \star F _\mu +  D_\mu \sigma\right)^2 +D^2\bigg] \\
% \mathcal{L}_{fermionic}&=& i (D_\mu \bar{\lambda})^\dagger \gamma^\mu \lambda-i [\sigma,\bar{\lambda}^\dagger]\lambda
%\end{eqnarray}
\begin{eqnarray}
\lambda=\bar{\lambda}&=&0,
\\
\epsilon^\dagger Q_{\epsilon}{\lambda}={\epsilon^C}^\dagger Q_{\epsilon}{\lambda}&=&0,\\ %\\
%(\overset{\bullet}{Q_{\epsilon}{\lambda}})^\dagger \epsilon=(\overset{\bullet}{Q_{\epsilon}{\lambda}})^\dagger\epsilon^C&=&0
\epsilon^\dagger Q_{\epsilon}{\psi}={\epsilon^C}^\dagger Q_{\epsilon}{\psi}&=&0,\\
\epsilon^\dagger Q_{\epsilon}{\bar{\psi}}={\epsilon^C}^\dagger Q_{\epsilon}{\bar{\psi}}&=&0,
 \end{eqnarray}
which reduce to
\begin{eqnarray}
\lambda = \bar{\lambda}= i F_{1 2}-D+i {\mathcal{D}}_3\sigma= \left(\mathcal{D}_1 \sigma +F_{2 3}\right)+ i \left( \mathcal{D}_2 \sigma+F_{3 1} \right)=0,\label{locusGauge}% \\
%-i F_{1 2}-D- \mathcal{D}_3 \sigma=\left(\mathcal{D}_1 \sigma +F_{3 1})+ i \left(\mathcal{D}_2 \sigma-F_{2 3}\right)=0
\\
\psi=\bar{\psi}={\hat{\mathcal{D}}}_3 \phi=\left(\mathcal{D}_1 + i\, \mathcal{D}_2 \right) \phi=\bar{F}=0.
\label{QeBPS} 
\end{eqnarray}
Before imposing reality conditions on the fields, this is the most general zero locus condition.
From reality conditions  and \eqref{locusGauge}
\begin{eqnarray}
D=0.
\end{eqnarray}
From \eqref{locusGauge}, $D=0$ and again the reality conditions \eqref{ComplexCond}, it follows the set of equations
\begin{eqnarray}
\hat{\mathcal{D}}_{3}\sigma =- F_{12}. \nonumber\\
F_{2 3}+i \mathcal{D}_2 \sigma= \hat{F}_{23} = 0, ~~~  F_{31}-i \mathcal{D}_1 \sigma=\hat{F}_{31}=0.
 \label{Bogomolnyi}
\end{eqnarray}
This is the same form of the Bogomolnyi equations reported in equation (9.7) of \citep{KapustinWitten}.  In the first equation we can substitute $\hat{\mathcal{D}}_3$ by $\mathcal{D}_3$ because $[\hat{A}_3,\sigma]=[A_3,\sigma]$.

In Appendix \eqref{AppLoop} we elaborate on how one arrives to the non vanishing part zero locus solution on the segment $I_N$
\begin{eqnarray}
F_{12}&=&\frac{ \mathfrak{m} }{2 R^2},~~ \hat{A}_3=u,\nonumber
\\
 \sigma^N&=& \sigma_0 - \,\frac{\mathfrak{m}}{2 R^2
 } \, t , ~~~  0<  t < t_0.\label{locusN}
\end{eqnarray}
We will use the following gauge potential parameterisation all along this paper
 \begin{eqnarray}
 \hat{A}_{(0)}= -\frac{
\mathfrak{m}}{2% R^2
}
\cos{\theta}  d\phi +u%{A_{(0)}}_3(t)
 dt  .%\nonumber\\
%\tilde{F}_{1 2 }_{(0)}=\frac{ \mathfrak{m} }{2 R^2}, ~~~ \sigma_{(0)}= \sigma_0- \frac{\mathfrak{m}}{2 R^2} t,
\label{monopoles0Bulk}  \label{'tHooftLineBulk}
\end{eqnarray}
At this point we can evaluate the classical Lagrangian densities on the zero locus. From equations \eqref{QVvec2} and  \eqref{ModifiedCS}
\begin{eqnarray}
(Q_\epsilon V_{Vector}) \bigg|_{\eqref{monopoles0},~ \lambda=0}&=&0, \\
\mathcal{L}^N_{CS} \bigg|_{\eqref{monopoles0}, ~ \lambda=0} &=& -\, \frac{i}{vol_{\mathbb{S}_2}}\,  k \,  u \cdot \mathfrak{m},  \label{ChernSimons}
\end{eqnarray}
where $u \cdot \mathfrak{m}:= \frac{1}{2}Tr(u~ \mathfrak{m})%\equiv u(\mathfrak{m})
$. In our conventions if $T_a$ and $T_b$ are Cartan generators then $Tr[T_a T_b]=2\delta_{a b}$. % associated to the roots, that in abuse of notation, we will denote as $u$ and $\mathfrak{m}$ too.
Finally we are ready to write down the classical contributions
\begin{eqnarray}
e^{ -\int_{\SIN} Q V_{Vector}}\bigg |_{\eqref{monopoles0}, ~\lambda=0}=1, ~~
e^{-S^N_{CS}}\bigg |_{\eqref{monopoles0},~\lambda=0}=e^{i\,\frac{t_0}{2\pi} \, k \,\left(2 \pi u\right) \cdot \mathfrak{m}}.\\\nonumber
\end{eqnarray}
%Even though $m^r$-monopoles have finite energy growing like $m^2$.
We will also use the contribution of the $U(1)$ topological factor \eqref{mCS}
\begin{eqnarray}
\xi ^{Tr[\mathfrak{m}]}, \label{TopSym}
\end{eqnarray}
to absorb inconvenient phase factors of the same form, like for instance $(-1)^{Tr[\mathfrak{m}]}$.

\subsection{The supercharge in $I_S$: $\tilde{Q}_\epsilon$}
In this subsection we first reconstruct the supercharge $\tilde{Q}_\epsilon$ that anti-commutes equivariantly with the chiral one $Q_\epsilon$. Namely
\begin{eqnarray}
\bigg\{\, Q_\epsilon, \, \tilde{Q}_\epsilon \, \bigg\}= i \hat{\mathcal{L}}_v, \label{AntiComm}
\end{eqnarray}
where the gauge covariant Lie derivative in the RHS is defined as
\begin{eqnarray}
\hat{\mathcal{L}}_v:= v^\mu \hat{\mathcal{D}}_\mu,
 \end{eqnarray}
 for every field except for the redefined gauge potential $\hat{A}$, for which the action is the following one
 \begin{eqnarray}
\hat{ \mathcal{L}}_v \hat{A}_\nu:= v^\mu \hat{F}_{\mu \nu}.
 \end{eqnarray}
 Notice that in the RHS of \eqref{AntiComm} there is not gauge transformation with $\sigma$ as parameter. This is because such a gauge transformation is implicit in the hatted potential $\hat{A}_3$ action upon the corresponding fields.  It is straightforward to find, that the $\tilde{Q}_\epsilon$ super-transformations for the gauge multiplet are given by
 \begin{eqnarray}
\tilde{Q}_{\epsilon} \hat{A}_{\theta, \phi }&=& -\frac{i}{2}\left(%\bar{\epsilon}^\dagger\gamma_\mu \lambda
 -\epsilon \gamma_{\theta, \phi} \lambda\right), ~~ \tilde{Q}_{\epsilon} \hat{A}_t=0 \nonumber\\
\tilde{Q}_{\epsilon} \sigma &=& \frac{1}{2}\left(%\bar{\epsilon}^\dagger\lambda
-\epsilon \lambda\right),~~~~~
%\tilde{Q}_{\epsilon} \bar{\lambda} = \frac{1}{2} \gamma^{\mu\nu} \epsilon \hat{F}_{\mu \nu}+D^\dagger \epsilon-i \gamma^3 \epsilon \hat{\mathcal{D}}_3 \sigma
\tilde{Q}_{\epsilon} \bar{\lambda}^\dagger =
 \frac{1}{2} \epsilon \gamma^{\mu\nu}  \hat{F}_{\mu \nu}-
 D \epsilon -
 i\epsilon \gamma^3  \hat{\mathcal{D}}_3 \sigma
\nonumber\\
\tilde{Q}_{\epsilon} \lambda&=&0% - \frac{1}{2}\bar{\epsilon}^\dagger \gamma^{\mu\nu} F_{\mu \nu}+D \bar{\epsilon}^\dagger+i \bar{\epsilon}^\dagger\gamma^\mu \mathcal{D}_\mu \sigma \nonumber
~~~~~~~
 \tilde{Q}_{\epsilon} D= %-\frac{i}{2} \bar{\epsilon}^\dagger \gamma^\mu \mathcal{D}_\mu \lambda
 \frac{i}{2}  \epsilon \gamma^\mu \hat{\mathcal{D}}_\mu \lambda. \label{algebra1South}
 \end{eqnarray}
In this subsection, we will only present the analysis in terms of the redefined potential $\hat{A}$. The covariant form of the $\tilde{Q}_\epsilon$ supercharge can be obtained by using the definition \eqref{effPot} and the identity \eqref{Identidad}. In Appendix \ref{OrtTermAnti} we compute the orthodox localising term in terms of the covariant form of $\tilde{Q}_\epsilon$.

The unorthodox localising term that will be used to compute one loop determinants for the vector multiplet is
\begin{eqnarray}
\tilde{Q}_\epsilon V_{vector}&:=&\tilde{Q}_\epsilon\left( \bar{\lambda}^\dagger \overset{\bullet}{\left( \tilde{Q}_\epsilon \bar{\lambda}^\dagger  \right)}  \right), \label{locV2}\\\nonumber
 \overset{\bullet}{\left( \tilde{Q}_\epsilon \bar{\lambda}^\dagger  \right)}&:=&  \left( \tilde{Q}_\epsilon \bar{\lambda}^\dagger \right)^* \bigg|_{\hat{A}^* \rightarrow \hat{A}, ~ \sigma^* \rightarrow \sigma, ~ D^* \rightarrow D},
\end{eqnarray}
which after expanded gives the following bosonic and fermionic terms
\begin{eqnarray}
\tilde{Q}_\epsilon V^{vector}_{B}&:=&D^2+ \left(F_{1 2} \, { \red-  }\, \hat{\mathcal{D}}_3 \sigma \right)^2+\left(\hat{F}_{13}\right)^2+ \left(\hat{F}_{2 3}\right)^2. \label{Qvector2South}\\
\tilde{Q}_\epsilon V^{vector}_{F}&:=&%+
i \bar{\lambda}^\dagger \gamma^\mu  \overrightarrow{P}^-_\mu \lambda \nonumber\\
&=&%+
( \bar{\lambda}_1^\dagger   ~~~ \bar{\lambda}_2^\dagger   )  \left(\begin{array}{c c}
 %-
 i \overrightarrow{\hat{\mathcal{D}}_t} &~~~   \frac{i}{R} ( \overrightarrow{\mathcal{D}_{\theta}}-i \csc \theta \overrightarrow{\mathcal{D}_\phi} )\\
 %-
0 & -i \overrightarrow{\hat{\mathcal{D}}_t}
 \end{array}\right) \left(\begin{array}{c} \lambda_1 \\  \lambda_2 \end{array}\right).
\end{eqnarray}

\begin{eqnarray}
P^-_{\theta, \phi}:=\frac{1-\sigma_3}{2} \mathcal{D}_{\theta, \phi}~~~, ~~~~~ P^-_t:=\mathcal{D}_t.
\end{eqnarray}
In Appendix \ref{OrtTermAnti} we present the analog results that come after use of the orthodox localising term.

As for the matter multiplets the $\tilde{Q}_\epsilon$ transformations are
\begin{eqnarray}
\tilde{Q}_{\epsilon} \phi &=&\epsilon \psi % \bar{\epsilon}^\dagger \psi
, ~~~~~~ \tilde{Q}_{\epsilon} \bar{\phi}^\dagger =0,  \nonumber \\
\tilde{Q}_{\epsilon} \bar{\psi}^\dagger &=& -i \epsilon \gamma^\mu \hat{\mathcal{D}}_\mu \bar{\phi}^\dagger,~~~~~~
\tilde{Q}_{\epsilon} \psi= %i \gamma^\mu \bar{\epsilon} \mathcal{D}_\mu \bar{\phi}+i \bar{\phi} \sigma \bar{\epsilon}+
F \epsilon^c \nonumber\\
\tilde{Q}_{\epsilon} \bar{F}^\dagger&=&-
 i \hat{\mathcal{D}}_\mu \bar{\psi}^\dagger \gamma^\mu \epsilon^c -i \bar{\phi}^\dagger \bar{\lambda}^\dagger \epsilon^c \, , ~~~~~~
\tilde{Q}_{\epsilon} F= 0.% i \bar{\epsilon}^\dagger \gamma^\mu \mathcal{D}_\mu \psi- i \bar{\epsilon}^\dagger \bar{\psi} \sigma+i \bar{\phi} \bar{\epsilon}^\dagger \bar{\lambda}
\label{algebraChiralSouth}
\end{eqnarray}
%It is straightforward to check that $\tilde{Q}_\epsilon^2 \bar{F}^\dagger=0$ with our convention $i [\mathcal{D}_\mu,\mathcal{D}_\nu]:=F_{\mu \nu}$.
The localising term to be used in the computation of one loop determinants for matter is
\begin{eqnarray}
\tilde{Q}_\epsilon V^{Matter}&:=& Q_\epsilon \bigg( i \, \bar{\psi}^\dagger \gamma^\mu \epsilon \, \hat{\mathcal{D}}_\mu \phi  -\bar{F}^\dagger \epsilon^c \psi -\bar{\phi}^\dagger \, \bar{\lambda}^\dagger\epsilon\, \phi \bigg),\label{QG2}%\tilde{Q}_\epsilon V^{Matter}_{I}+\tilde{Q}_\epsilon V^{Matter}_{II}
 \end{eqnarray}
whose expansion gives the following bosonic and fermionic parts
%\begin{eqnarray}
%\left(\overset{\bullet}{\tilde{Q}_{\epsilon} \bar{\psi}^\dagger}\right)^\dagger&:=&%\left(\tilde{Q}_\epsilon \bar{\psi}^\dagger\right)^\dagger \bigg |_{\bar{\phi}\rightarrow \phi}=
 %i \gamma^\mu \epsilon \hat{\mathcal{D}}_\mu \phi. \\
%\left(\overset{\bullet}{\tilde{Q}_{\epsilon} \psi}\right)^\dagger&:=&%\left(\tilde{Q}_{\epsilon} \psi\right)^\dagger\bigg|_{F\rightarrow \bar{F}}=%
%- \bar{F}^\dagger \epsilon^c.\\
%\overset{\bullet}{\tilde{Q}_{\epsilon} \sigma }&:=&%\left(\tilde{Q}_\epsilon \sigma\right)^\dagger\bigg|_{\lambda \rightarrow \bar{\lambda}}=
%\frac{1}{2} \bar{\lambda}^\dagger \epsilon.
%\end{eqnarray}
\begin{eqnarray}
\tilde{Q}_\epsilon V^{matter}_B&=&({\hat{\mathcal{D}}}^{\mu}{\bar{\phi}})^\dagger  \, {\hat{\mathcal{D}}}_{\mu}{\phi} +  \, \bar{\phi}^\dagger\left({\red-}\hat{\mathcal{D}}_3\sigma+i D- \epsilon^{\mu \nu}_{~~\beta}v^{\beta} \left(q V_{\mu \nu}+W_{\mu \nu}\right) \right)\phi
\nonumber\\&&  +\bar{F} F+\,\hat{\mathcal{D}}_{\mu}\left( i \, \epsilon^{\mu \nu}_{~~\beta}v^{\beta} \bar{\phi}^\dagger {\hat{\mathcal{D}}}_{\nu} \phi \right).\\ && \nonumber\\
\tilde{Q}_\epsilon V^{matter}_F&:=&  i \, \bar{\psi}^\dagger {\gamma}^{\mu} {\hat{\mathcal{D}}}_{\mu}{\psi}-i  \bar{\phi}^\dagger  \bar{\lambda}^\dagger\psi -i \bar{\psi}^\dagger P^- \lambda \phi \nonumber\\&&+i \hat{\mathcal{D}}_\mu\left( \bar{\psi}^\dagger \gamma^\mu P^- \psi\right)
\end{eqnarray}
One possible $\tilde{Q}_\epsilon$ invariant Chern-Simons term to use is
\begin{eqnarray}
\tilde{\mathcal{L}}_{CS}= -\frac{i k}{4 \pi}\left( \epsilon^{\mu \nu \beta}\left(\hat{A}_\mu \partial_\nu \hat{A_\beta} -\frac{2 i}{3}\hat{A}_\mu \hat{A}_\nu \hat{A}_\beta \right)\, {\red+} \,\, \bar{\lambda}^\dagger P^- \lambda \right). \label{ModifiedCSSouth}
\end{eqnarray}
whose $\tilde{Q}_\epsilon $variation gives the following total derivative
\begin{eqnarray}
-i \frac{k}{4 \pi}  \mathcal{D}_\mu\bigg( \epsilon^{\mu \nu \beta} \left(\tilde{Q}_\epsilon \hat{A}_\nu \right) \hat{A}_\beta \bigg). \label{TotDerChernSS}
\end{eqnarray}
The total derivative $\tilde{Q}_\epsilon$ variation of the covariant Chern-Simons term \eqref{ChernSimonsCovariant} is in this case
\begin{eqnarray}
 -i \frac{k}{4 \pi}  \mathcal{D}_\mu\bigg( \epsilon^{\mu \nu \beta} \left(\tilde{Q}_\epsilon A_\nu \right) A_\beta + i\, \epsilon \gamma^\mu \lambda \, \sigma \bigg).
\end{eqnarray}
The total derivative $\tilde{Q}_\epsilon$ variation of the modified Chern-Simons term \eqref{mCS}
\begin{eqnarray}
-\frac{i}{2 \pi}  \hat{\mathcal{D}}_\mu\bigg( \epsilon^{\mu \nu 3} \left(  \hat{A}^T_3\, Tr \tilde{Q}_\epsilon A_\nu \right) \bigg).
\end{eqnarray}
%\begin{eqnarray}
%\hat{A}_3 := A^S_3 -i \sigma. \label{effPotSouth}
%\end{eqnarray}

Following the procedure already detailed in the previous subsection one finds the following $\tilde{Q}_\epsilon$
 zero locus conditions
 \begin{eqnarray}
\lambda = \bar{\lambda}= i F_{1 2}-D-i \mathcal{D}_3\sigma%= \left(-\mathcal{D}_1 \sigma +F_{2 3})+ i \left(-\mathcal{D}_2 \sigma+F_{3 1}\right)
=\hat{F}_{23}+i \hat{F}_{31}=0,\label{locusGauge}% \\
%-i F_{1 2}-D- \mathcal{D}_3 \sigma=\left(\mathcal{D}_1 \sigma +F_{3 1})+ i \left(\mathcal{D}_2 \sigma-F_{2 3}\right)=0
\\
\psi=\bar{\psi}=\hat{\mathcal{D}}_3 \bar{\phi}^\dagger=(\mathcal{D}_1 -i \mathcal{D}_2) \bar{\phi}^\dagger=\bar{F}=0.
\label{QeBPSTilded}
\end{eqnarray}
For the matter multiplet one arrives to $\phi=\bar{\phi}^\dagger=0$ based on arguments used in the case of $\tilde{Q}_\epsilon$ in Appendix \ref{AppLoop}.
For the vector multiplet and for the reality conditions used in this work, the conditions for the bosons become $D=0$ together with the Bogomolnyi equations
\begin{eqnarray}
\hat{\mathcal{D}}_{3}\sigma ={\red + } F_{12}. \nonumber\\
F_{2 3}+i \mathcal{D}_2 \sigma= \hat{F}_{23} = 0, ~~~  F_{31}-i \mathcal{D}_1 \sigma=\hat{F}_{31}=0.
 \label{BogomolnyiTilded}
\end{eqnarray}
The red colour plus sign is written to highlight the difference with the $Q_\epsilon$ locus conditions \eqref{Bogomolnyi}.
Repeating the analysis of the previous subsection one obtains to the following form of the locus
\begin{eqnarray}
F_{12}&=&\frac{ \mathfrak{m}_S }{2 R^2},\nonumber
\\
 \sigma^S&=& \sigma_0 {\red+} \,\frac{\mathfrak{m}_S}{2 R^2
 } \left(t-2 \pi\right), ~~~  t_0 < t < 2\pi,\label{locusS}
\end{eqnarray}
where again the weights of the generator $\mathfrak{m}_S$ are GNO quantised and in the Cartan of $\sigma_0$. %The value of the flux $\mathfrak{m}_S$ is in principle decoupled from the value at $I_S$, that we have defined as $\mathfrak{m}$. However as shall be shown in the next section, we will select matching conditions that ensure continuity of these quantities, namely
%\begin{eqnarray}\nonumber\\
%\mathfrak{m}_S=\mathfrak{m}, ~~\sigma_0=\sigma^S_0,~~ \sigma^N(0)=\sigma^S(2\pi), ~ \sigma^N(\pi)=\sigma^S(\pi).\\
%\nonumber
%\end{eqnarray}

Next we are in conditions to write down the classical Lagrangian densities evaluated on the zero locus. From equations \eqref{Qvector2South} and  \eqref{ModifiedCSSouth} we get
\begin{eqnarray}
(\tilde{Q}_\epsilon V_{Vector}) \bigg|_{\eqref{monopoles0Bulk},~ \bar{\lambda}=0}&=&0, \\
\mathcal{L}^S_{CS} \bigg|_{\eqref{monopoles0Bulk}, ~ \bar{\lambda}=0} &=& -\, \frac{i}{vol_{\mathbb{S}_2}}\, k \,  u_S\cdot \mathfrak{m}_S.  \label{ChernSimonsSouth}
\end{eqnarray}
The classical on shell action values are
\begin{eqnarray}
e^{ -\int_{\SIS} Q V_{Vector}}\bigg |_{\eqref{monopoles0Bulk}, ~\bar{\lambda}=0}=1, ~~
e^{-S^S_{CS}}\bigg |_{\eqref{monopoles0Bulk},~\bar{\lambda}=0}=e^{i\,\frac{2 \pi -t_0}{2\pi} \, k\, \left(2\pi u_S\right) \cdot \mathfrak{m}_S}.\\\nonumber
\end{eqnarray}

\subsubsection{The $\mathbb{Z}_2$ transformation $\mathcal{ P}$: $Q_\epsilon \rightarrow \tilde{Q}_\epsilon$}
For the reality conditions \eqref{ComplexCond} the following $\mathbb{Z}_2$ transformation maps $Q_\epsilon$ to $\tilde{Q}_\epsilon$ and $vice\, versa$, in the following way
\begin{eqnarray}
\mathcal{P}:  \hat{A}_\mu  \rightarrow -\hat{A}^\dagger_\mu, ~~ \sigma\rightarrow \sigma^\dagger \\
 \bar{\lambda}^\dagger \rightarrow \lambda, ~ \lambda\rightarrow \bar{\lambda}^\dagger \\
 D\rightarrow -D^\dagger \\ \partial_\mu\rightarrow -\partial_\mu
\end{eqnarray}
and
\begin{eqnarray}
\mathcal{P}:  \phi\rightarrow \bar{\phi}^\dagger, ~~ \bar{\phi}^\dagger \rightarrow \phi , \\
 \bar{\psi}^\dagger\rightarrow \psi,~~ \psi\rightarrow \bar{\psi}^\dagger \\\nonumber
\end{eqnarray}
The complex conjugation plays a role even when supposing the reality conditions \eqref{ComplexCond}. This is because the supertransformations of real fields will be complex, as one can easily notice by observing carefully the RHS, for instance of \eqref{algebra1}. This $\mathbb{Z}_2$ transformation can be used as a tool to check the matching of $Q_\epsilon $ and $\tilde{Q}_\epsilon$ supersymmetric  of exact objects, like for instance the localising actions and their variations. Notice that $\mathcal{P}$ resembles a $CP$ transformation, but it is not because $\psi$ and $\bar{\psi}^\dagger$, for instance, are a priori independent degrees of freedom, not complex conjugated to each other.

\subsection{%Ensuring  $Q \cup \tilde{Q}$- SUSY on $\mathbb{S}_1$:
What is localisation computing?} \label{subsecSum}
This subsection is a summary of results presented in the previous ones. In particular we will stress what are the actions that $Q_\epsilon$ and $\tilde{Q}_\epsilon$ localisation integrate out. Both of these localising actions are equivalent up to total derivatives \footnote{The orthodox ones. The unorthodox ones are not related by a total derivative as we have already seen. The difference between the orthodox and unorthodox terms are $Q_\epsilon$ and $\tilde{Q}_\epsilon$ in both "chiral" and "antichiral" sectors, respectively. The addition of these exact terms spoils the total derivative relation between "chiral" and "antichiral" localising terms. %in passing from "orthodox" to "unorthodox" localising terms.
} as we have already seen. What we will further remark, is that when placed on a manifold with boundaries, topologically twisted Chern-Simons-matther theory (TTCSM) has two possible versions to be used. We will call these versions as $Q_\epsilon$ and $\tilde{Q}_\epsilon$- TTCSM respectively.

In the end of this subsection we explore how to supersymmetrise the modified Chern-Simons terms on segments, in a gauge covariant way. For that, an additional set of boundary conditions will need to be imposed upon the vector multiplet. As will be discussed in a while, this set of boundary conditions guaranties well definiteness of the variational problem on $\mathbb{S}_2$ times a segment in the vector multiplet sector. We will also identify and discuss about a set of boundary conditions for the matter sector that will guaranty both supersymmetry and well definiteness of the variational problem on $\mathbb{S}_2$ times a segment.

We can define a $Q-$exact term that differs from the Lagrangian of $\mathcal{N}=2$ SYM Lagrangian on $\mathbb{S}_2$ times a segment by a total derivative. Let us explain this last point in more detail. For us, SYM Lagrangian density is defined as
\begin{eqnarray}
\mathcal{L}^{Vector}_{SYM}&:=& \frac{1}{2} F_{\mu \nu}^2+\left(\mathcal{D}_\mu\sigma\right)^2+D^2+i \bar{\lambda}^\dagger \slashed{D} \lambda- i \sigma \{\bar{\lambda}^\dagger,\lambda\}, \\\nonumber \\\mathcal{L}^{Matter}_{SYM}&:=&\mathcal{D}^{\mu}{\bar{\phi}}^\dagger  \, {\mathcal{D}}_{\mu}{\phi} +\bar{\phi}^\dagger\,\left(%-\hat{\mathcal{D}}_3\sigma+
 i D\, + \sigma^2 \,- \epsilon^{\mu \nu}_{~~\beta}v^{\beta} \left(q V_{\mu \nu}+W_{\mu \nu}\right) \right) \phi \nonumber\\&+& \bar{F}^\dagger F+i \, \bar{\psi}^\dagger {\gamma}^{\mu} {{D}}_{\mu}{\psi}-i \bar{\psi}^\dagger\, \sigma \,\psi- i\,  \bar{\psi}^\dagger \lambda \,\phi- i \,
\bar{\phi}^\dagger\, \bar{\lambda}^\dagger \psi,
\end{eqnarray}
where trace over the gauge indices in the first line, is implicit. To study the variations of $S_{SYM}$ one can use the following identities
\begin{eqnarray}
\left(Q_\epsilon V \right)^N&-&\mathcal{L}_{SYM}=%Tr \mathcal{D}_\mu \sigma  \left(\star F\right)^\mu+
 \mathcal{D}_\mu J^\mu_N,\nonumber\\
\left(\tilde{Q}_\epsilon V \right)^S&-&\mathcal{L}_{SYM}=\mathcal{D}_\mu J^\mu_S,% -Tr \mathcal{D}_\mu \sigma  \left(\star F\right)^\mu- \mathcal{D}_\mu\left(v^\mu \bar{\phi}^\dagger\sigma \phi \right)+\ldots
\label{BoundaryNS}
\end{eqnarray}
where $\left(Q_\epsilon V\right)^N$ and $\left(\tilde{Q}_\epsilon V\right)^S$ are the orthodox localising terms, which for the vector multiplet are \eqref{LocVCov} and \eqref{locVCov2} respectively, and for the matter multiplet are \eqref{LOCMATORT} and \eqref{MatterAc2} respectively. The currents $J^\mu_{N(S)}$ can be divided into vector and matter part
\begin{eqnarray}
J^\mu_{N(S)}:={J^{Vector}_{N(S)}}^\mu+{J^{Matter}_{N(S)}}^\mu.
\end{eqnarray}
As for the vector currents, one gets
\begin{eqnarray}\nonumber\\
{J^{Vector}_N}^\mu&:=&Tr \left( 2 \, i \, \sigma\, v^\nu F^\mu_{~\nu}-2 \,i \,v^\mu \sigma D -2\, \sigma \mathcal{D}^\mu \sigma  -\,i \,\bar{\lambda}^\dagger P^-\gamma^\mu  \lambda \right),  \\ && \nonumber\\
{J^{Vector}_S}^\mu&:=& Tr \left( 2 \, i \, \sigma\, v^\nu F^\mu_{~\nu}+2 \,i \,v^\mu \sigma D -2\, \sigma \mathcal{D}^\mu \sigma  -\,i \,\bar{\lambda}^\dagger \gamma^\mu P^+  \lambda \right). \label{CurrentsNS} \\&&\nonumber
\end{eqnarray}
As for the matter currents one gets
\begin{eqnarray}
{J^{Matter}_N}^\mu&:=&~~ % v^\mu \bar{\phi}^\dagger \sigma \phi +
v^\mu\, \bar{\phi}^\dagger \sigma \phi \,+\,i \, \epsilon^{\mu \nu}_{~~\beta}v^{\beta} \bar{\phi}^\dagger {\mathcal{D}}_{\nu} \phi  - \,i \,\bar{\psi}^\dagger P^+ \gamma^\mu \psi, \\\nonumber &&\\
{J^{Matter}_S}^\mu&:=&-% v^\mu \bar{\phi}^\dagger \sigma \phi +
v^\mu\, \bar{\phi}^\dagger \sigma \phi \,+\,i \, \epsilon^{\mu \nu}_{~~\beta}v^{\beta} \bar{\phi}^\dagger {\mathcal{D}}_{\nu} \phi-\, i\,  \bar{\psi}^\dagger \gamma^\mu P^{ -} \psi.
\end{eqnarray}

 Relations \eqref{BoundaryNS} are of special interest, because the $N$(resp. $S$) $Q_\epsilon$( resp. $\tilde{Q}_\epsilon$)-exact term is nilpotent with respect to $Q_\epsilon$ (resp. $\tilde{Q}_{\epsilon}$).  Consequently, total derivatives do not come out in their SUSY variations. Thence from these equations one immediately reads out the $Q_\epsilon$ and $\tilde{Q}_\epsilon$ variations of the Lagrangian $\mathcal{L}_{SYM}$.

Thence the $Q_\epsilon$ and $\tilde{Q}_\epsilon$ variations of the $S^{N}_{SYM}$ and $S^{S}_{SYM}$ actions are
\begin{eqnarray}
Q_\epsilon S_{SYM}^N&=& \int_{\mathbb{S}_2}  \delta  J_{N}^3~ \bigg|^{t_0^-}_{0^-}, \nonumber\\ \tilde{Q}_\epsilon S^{S}_{SYM}&=&\int_{\mathbb{S}_2} -\delta  J_{S}^3~ \bigg|^{t_0^+}_{2\pi^+}.  \label{eqboundary}%&=&\,i\,\int_{S_2}\, Q_\epsilon \,\bigg(  tr \left(\bar{\lambda}^\dagger \lambda \, - \, 4 \sigma D\right) + \left(-\bar{\psi}^\dagger\psi\,-\,2 \, \bar{\phi}^\dagger \sigma \phi \,\right) \bigg) \bigg|^\pi_{0}%\nonumber\\
 %J_{N(\text{ resp. }S)}^3&:=&+(\text{ resp. }-)  ~ \left(Tr \sigma  \left(\star F\right)^3+\bar{\phi}^\dagger \sigma \phi \ldots\right)
\end{eqnarray}

% In \eqref{eqboundary} $\partial I_N:=\{0,\pi\}$ and $\partial I_S:=\{\pi,2\pi\}$. The second equality in the first line follows from the definitions \eqref{CurrentsNS} together with periodicity under $2\pi$.
%There are, boundary conditions at $\partial I_N$ respecting the zero locus \eqref{zerolocus} that could make this term vanish.
The $Q_\epsilon$ and $\tilde{Q}_\epsilon$ localisation are set to compute not  the partition function of $S^{N, S}_{SYM}$, but instead the path integral of the actions
\begin{eqnarray}
\tilde{S}^N&:=& S^N_{SYM}- \int_{\mathbb{S}_2}  J_{N}^3 \bigg|^{t_0^-}_{0^-},~~~\nonumber\\
 \tilde{S}^S&:=&S^{S}_{SYM}+ \int_{\mathbb{S}_2} J_{S}^3 \bigg|^{t_0^+}_{2\pi^+}, \label{SStilde}\\~~~Q_\epsilon \tilde{S}^N&=&\tilde{Q}_\epsilon \tilde{S}^S=0.\nonumber% \nonumber\\
%&:=&S_{SYM}-2 \int_{S2}J^3_S \bigg|_{\partial I_S} .
\end{eqnarray}
%After imposing the matching conditions for field and derivative values from one side to the other (this will be done in a while) $\tilde{S}$ can be interpreted as 3D SYM with the insertion of 2D defects at the boundaries of $I_N$ (Or equivalently of $I_S$). One of
The $\tilde{S}$ actions are precisely the orthodox localising terms. The consequence, at the technical level, of the total derivatives that define $\tilde{S}^{N,S}$ in terms of $S_{SYM}$ is to cancel the classical on shell value of $S_{SYM}$ evaluated on BPS configurations with finite flux and henceforth, the latter are not suppressed in the large $\tau$ limit (This cancellation follows through in the same way already explained in the introduction).

We can write down the localising Lagrangian densities associated to \eqref{SStilde} without explicitly writing the total derivative terms: they are simply the orthodox localising terms.

 Summarising, we write down the two different choices of localising terms -- from the results in the previous subsections--
\begin{eqnarray}
\left\{
	\begin{array}{cc}
		\tilde{\mathcal{L}}^N:=Q_\epsilon V_N=\bigg(\eqref{locV},\eqref{MatterAc}\bigg)\text{ or }\bigg(\eqref{LocVCov},\eqref{LOCMATORT}\bigg) &~~~~~ 0 < t < t_0 \\
		\tilde{\mathcal{L}}^S:=\tilde{Q}_\epsilon V_S=\bigg(\eqref{locV2}, \eqref{QG2}\bigg)\text{ or }\bigg(\eqref{locVCov2}, \eqref{MatterAc2}\bigg)&~~~~~t_0<t<2\pi .\end{array} \right\} \label{eqqq} \\\nonumber
\end{eqnarray}
The the supersymmetry variations of $\tilde{\mathcal{L}}^{N,S}$ are zero follows from the nilpotency property $Q_\epsilon^2 V_N=\tilde{Q}_\epsilon^2 V_S=0$. The option to the left (resp. right) in each parenthesis refers to the vector ( resp. matter) multiplet. We must stress that the first and second big parenthesis in each of the lines in \eqref{eqqq}, are not the same. The difference being a $Q_\epsilon$ (first line) and $\tilde{Q}_\epsilon$ (second line) exact terms as already shown in previous sections. In the first line for instance such a $Q_\epsilon$ exact difference is given by
\begin{eqnarray}
-\bigg(\eqref{diffQV},\eqref{EqRefNonAP}\bigg).
\end{eqnarray}
As we have already stressed several times in previous subsections, the first option that we have called before "unorthodox" terms,  is the convenient one for our purposes and due to technical issues. The currents $\mathcal{J}^\mu$ that define the difference between the Q-exact term and the SYM action that we wrote above, are coming from the "orthodox" terms, but the analog ones coming from the "unorthodox" terms can be easily deduced from the results presented in previous subsections.

 % In this way one is glueing two exactly supersymmetric actions each one of them localised at $\mathbb{S}_2\times I_N$ and $\mathbb{S}_2\times I_S$ respectively. The relation of the glued action $\tilde{S}$ with $S_{SYM}$ is given by equation \eqref{SStilde}.

 To stress differences, %that are only due to total derivative terms and the possible addition of extra $Q_\epsilon$ and $\tilde{Q}_\epsilon$ exact terms,
we will discern between the 3D TT $\mathcal{N}=2$ exact action coming from $Q_\epsilon$ and $\tilde{Q}_\epsilon$ by using the following terminology

\begin{itemize}

\item The theory {\bf $\tilde{S}^N$} we will call {\bf 3D $Q_\epsilon$-TT theory}.

\item The theory {\bf $\tilde{S}^S$} we will call {\bf 3D $\tilde{Q}_\epsilon$-TT theory}.

\end{itemize}

For completeness let us collect here the Chern-Simons terms that define two possible $TTCSM$, that were defined in the previous section as
 \begin{eqnarray}
 \left\{
	\begin{array}{clcl}
		 \mathcal{L}^N_{CS}~&:=&\eqref{ModifiedCS} &~~~~~ 0 < t < t_0 \\
		 \mathcal{L}^S_{CS}~&:=&\eqref{ModifiedCSSouth} &~~~~~t_0<t<2\pi  \end{array}, \, ~~~~\mathcal{L}_{T}~:= \eqref{mCS}\right\}, \label{CS}\\\nonumber
 \end{eqnarray}
and move on to analyse their variations. Is easy to see that the $\delta J^3$ component coming from the variation of the matter-Chern-Simons term \eqref{mCS} is zero (the super variation of these terms is again of the form $\mathcal{D}_\mu \delta J^\mu$) and thence this term does not suffer from supercharge leaking through the extrema of the corresponding patch $I_N$ or $I_S$.

As for the "unorthodox" Chern-Simons term variations \eqref{ModifiedCS} and \eqref{ModifiedCSSouth} there is a non trivial $\delta J^3$ component, that can be easily inferred out of the variations \eqref{TotDerChernN} and \eqref{TotDerChernSS} to be
 \begin{eqnarray}
 \delta J_N^3&=&-i \frac{k}{8 \pi} \bar{\lambda}^\dagger_2\, \left(A_1+i \,A_2\right) \nonumber \\
 \delta J_S^3&=&\,~~i \frac{k}{8 \pi} \lambda_2\, \left(A_1-i \,A_2\right) \label{CSBT}
 \end{eqnarray}

 These contributions can be cancelled with the addition of the following boundary action to the $N (S)$  modified CS terms
\begin{eqnarray}
S^{bdry}_{CS}=+(-)\frac{k}{8 \pi}\int_{S_2} tr \hat{A}_\mu {\hat{A}}^\mu \bigg|_{\partial I_{N(S)}}.
\end{eqnarray}
However this term breaks gauge invariance. In \citep{Yutaka} the reader can find two procedures to tackle this issue.
However, we will solve this problem in a different way.

First we remind that in taking the large $\tau$ limit in localisation we must redefine every field $\mathcal{X}$ in the theory in the following way
\begin{eqnarray}
\mathcal{X}_\tau=BPS_{\mathcal{X}}+\frac{1}{\sqrt{\tau}} \mathcal{X}^{(1)},  \label{FBPS}
\end{eqnarray}
where $BPS_\mathcal{X}$ denotes the profile of the field $\mathcal{X}$ in the BPS state of consideration. Before continuing we stress that apart from the presence of $CS$ boundary contributions coming from the variations \eqref{CSBT}, no constraint is needed over \eqref{FBPS} in order to preserve supersymmetry, because we have defined density Lagrangians whose supersymmetric variations vanish trivially.

However, as we want the Chern-Simons term to be supersymmetric, we will impose \eqref{CSBT} to vanish at $t=0$ and $t=t_0$ and in consequence we are forced to analyse the consistency of such constraint with the supersymmetry algebra of interest $Q_\epsilon$ or $\tilde{Q}_\epsilon$.  Specifically, when $\mathcal{X}$ is either $\lambda_2$( or $\bar{\lambda}^\dagger_{2}$) we will always consider
\begin{eqnarray}
BPS_{\lambda_2\left(\text{ or }\bar{\lambda}^\dagger_2\right)}=0.
\end{eqnarray}\footnote{We will exclude zero modes in our analysis.}
Thenceforth, should we impose at the boundaries of $I_N \,(\text{or } I_S)$ the following Dirichlet (D) conditions upon the $\mathcal{X}^{(1)}$ fluctuations
\begin{eqnarray}
\lambda^{(1)}_2\bigg |_{t=0\text{ and }t_0}=({\bar{\lambda}}^{(1)}_2)^\dagger\bigg |_{t=0 \text{ and }t_0}=0, \label{DCond}
\end{eqnarray}
the variation of the Chern-Simons terms \eqref{CSBT} vanish. After imposing \eqref{DCond} for $\lambda_2^{(1)}$ and $({\bar{\lambda}^{(1)}_2)^\dagger}$ and given the fact that the $BPS_\mathcal{X}$ are supersymmetric by definition,  in order not to break the superalgebra transformations \eqref{algebra1} and \eqref{algebra1South} at the boundaries $t=0$ and $t= t_0$ we must impose the $D$- conditions for every other $\mathcal{X}^{(1)}$ fluctuation in the vector multiplet. This guaranties that the full off shell fields $\mathcal{X}_\tau$ preserve super-transformations at the boundaries $t=0$ and $t=t_0$ once \eqref{DCond} is imposed. For the matter multiplet this is not necessary.
In a while, we will see what are the consequences of imposing the $D$-conditions at the boundaries.

So far we have only commented on the conditions that are needed to guaranty supersymmetry on $\mathbb{S}_2$ times an interval $I=(0,t_0)$. Let us pause to comment on how the aforementioned conditions also ensure a well defined variational problem on $\mathbb{S}_2\times I$. First, let us refer to the vector multiplet linear fluctuations that will be denoted as $\delta V$. The boundary term variations of the vector multiplet action \eqref{locV} and Chern Simons \eqref{ModifiedCS} actions once the EOM's are demanded, integrate to zero after imposing Dirichlet boundary conditions upon off shell fluctuations. Namely, should we demand  $\delta V(0) = \delta V(t_0) = 0$ the variational problem $\mathbb{S}_2\times I$ for the vector multiplet sector is well defined as it stands, because the boundary terms integrate to zero.

The variation of the matter multiplet action \eqref{MatterAc} is a total derivative once the EOM's are imposed. This total derivative variation integrates to zero on $\mathbb{S}\times I$, if we impose the following boundary condition
\begin{eqnarray}
\mathcal{X}(0)={\blue +} ({\red -}) \mathcal{X}(t_0), ~ \delta \mathcal{X}(0)={\blue +} ({\red -}) \delta \mathcal{X}(t_0), \label{BCfermions}
\end{eqnarray}
on fields in the matter multiplet $\mathcal{X}=\{\bar{\phi},\phi, \bar{\psi}, \psi,F, \bar{F}\}$ and their variations $\delta \mathcal{X}=\{\delta \bar{\phi},  \delta \phi, \delta \bar{\psi}, \delta \psi,F, \delta \bar{F}\}$. The choices {\blue +} and {\red -} are independent and in the next section they will be called {\blue even} and {\red odd} respectively. Clearly the {\blue +}  choice guaranties the boundary variations at both extrema to cancel each other, together with the $D$-conditions for the vector multiplet. The choice {\red -} does the work too, because the boundary terms are quadratic in matter background - fluctuations as one can directly infer out of \eqref{LBOSF0} and \eqref{LFERF0}. The variational problem on $\mathbb{S}_2 $ times the segment $(0,t_0)$ of length $t_0$ is well defined with the aforementioned conditions. Clearly \eqref{BCfermions} is
consistent with supersymmetry algebrae \eqref{algebraChiral} and \eqref{algebraChiralSouth} and in consequence they preserve supersymmetry. We finish this section by noticing that the D-conditions
\begin{eqnarray}
\mathcal{X}(0)= \mathcal{X}(t_0)=0, ~ \delta \mathcal{X}(0)= \delta \mathcal{X}(t_0)=0 \label{DMatter}
\end{eqnarray}
also guaranty the variational problem is well defined for matter. However, we will mainly use \eqref{BCfermions}, unless we explicitly say we use \eqref{DMatter}.

\paragraph{Two variational problems on $\SS$}
{\bf VP-I)}
To have a well defined variational problem on $\SS$, we take either the limit $t_0\rightarrow 2\pi$ of the action $\tilde{S}_N+{S_{N}}_{CS}$ 
coming from the Lagrangian density
\begin{eqnarray}
\mathcal{L}_{Q_\epsilon}:=\eqref{locV}+\eqref{MatterAc}+\eqref{ModifiedCS} \label{lQ}
\end{eqnarray}
or $t_0\rightarrow 0$ of $\tilde{S}_S+{S_S}_{CS}$ that comes from
\begin{eqnarray}
\mathcal{L}_{\tilde{Q}_\epsilon}:=\eqref{locV2}+\eqref{QG2}+\eqref{ModifiedCSSouth},
\end{eqnarray}
together with the D-conditions for the vector multiplet fluctuations and \eqref{BCfermions}  at $t=0$. The previous limit produces a well defined variational problem on $\SS$. However, an important fact we can not overview is that, for instance $\tilde{S}_{N}+{S_{N}}_{CS}$ with $t_0=2\pi$, has stationary points that  break the periodicity of $\mathbb{S}_1$ at $t=0$, as for instance the $Q_\epsilon$ BPS configuration \eqref{locusN}. The fact such solutions break the topology of $\mathbb{S}_1$ forces the exclusion of a point out of the $\mathbb{S}_1$ in order to have them into consideration. In fact \eqref{locusN} is a solution of the EOM's that come from the "unorthodox" $Q_\epsilon$ localising term \eqref{locV}(+\eqref{ModifiedCS}), specifically from \eqref{QVvec2}(+\eqref{ModifiedCS}) discarding back-reaction from the matter sector and gaugini which are zero in this background. It is easy to see that the linear differential of \eqref{QVvec2} vanishes identically upon evaluation on \eqref{locusN}. The variation of the CS \eqref{ModifiedCS} and interaction terms between the vector and matter multiplets, under a variation of the bosonic vector multiplet fields $\delta V$, integrates to $0$ if 
\begin{eqnarray}
\text{ D-condition  :   } \delta V(0):=0
\end{eqnarray}
is imposed at the point $t=0$ of $\mathbb{S}_1$. For instance, the total derivative $\mathcal{D}_3\left(\bar{\phi}^\dagger\delta \sigma \phi\right)$ that results (together with an interaction term in the EOM's times $ \delta \sigma$) from the variation of the term proportional to $\bar{\phi}^\dagger \mathcal{D}_3\sigma \phi$   in \eqref{RefEqTot0}, integrates to $0$ if $\delta \sigma(0)=0$. In conclusion, \eqref{locusN} is a minimum of the "unorthodox" $\tilde{S}_N+{S_{N}}_{CS}$ with $t_0\rightarrow 2\pi$, under variations $\delta V$ that obey the Dirichlet condition $\delta V(0)=0$ at the point $t=0$ of $\mathbb{S}_1$. However \eqref{locusN} with $t_0\rightarrow 2\pi$ is not periodic on $\mathbb{S}_1$ and in conclusion is {\bf singular} with $t_0\rightarrow2\pi$. On the 
\begin{itemize}
\item matter sector we can again impose Dirichlet boundary conditions \eqref{DMatter} 
\item or the less restrictive condition \eqref{BCfermions}.
\end{itemize}
Notice that the latter set of conditions allows for discontinuity of the fields at $t=0$.
 
 The weak point of this approach is that in order to keep non trivial fluxes one must sacrifice continuity at the point where the D-condition is imposed for the off shell fluctuations $\delta V$. In other words, in this case we must consider $\mathbb{S}_1$ with the point $t=0$ excluded.
 
 {\bf VP-II)} We can define a different variational problem on $\SS$ that allows for the presence of minima that are non trivial BPS configurations, with flux on $\mathbb{S}_2$ and respect the periodicity of $\mathbb{S}_1$. First we define the Lagrangian density
\begin{eqnarray}
\mathcal{L}_{mixed}=\left\{\begin{tabular}{cc} $\mathcal{L}_{Q_\epsilon}$:=\eqref{locV}+\eqref{MatterAc}+\eqref{ModifiedCS} & $0<t<t_0=\pi$ \\ $\mathcal{L}_{\tilde{Q}_\epsilon}$:=\eqref{locV2}+\eqref{QG2}+\eqref{ModifiedCSSouth}  & $t_0=\pi<t<2 \pi$  \end{tabular}\right., \label{refeq0}
\end{eqnarray}
where $\mathcal{L}_{Q_\epsilon}$( resp. $\mathcal{L}_{\tilde{Q}_\epsilon}$) is $Q_\epsilon$ ( resp. $\tilde{Q}_\epsilon$) supersymmetric up to a total derivative that integrates trivially with the forthcoming conditions. Second, we define matching (continuity) conditions from one side to the other of the junction points
\begin{eqnarray}
X(0^+)=X(0^-) \text{  and   }  X(\pi^+)=X(\pi^-). \label{MatchingCond}
\end{eqnarray}
The matching conditions \eqref{MatchingCond} select the following BPS configurations
  \begin{eqnarray}
\left\{ \begin{array}{cccccc} F_{12 }&=&\frac{\mathfrak{m}}{2 R^2}. & ~~~ \sigma&=&- \,\frac{\mathfrak{m}}{2 R^2
 }\, t , ~~~~~~ ~~~  0<  t < \pi.\\F_{12}&=&\frac{\mathfrak{m}}{2 R^2} &  ~~~\sigma&= & \,\frac{\mathfrak{m}}{2 R^2
 } \, (t-2 \pi) , ~~~  \pi<  t < 2\pi. \end{array}\right. \label{'t HooftContinuous}
 \end{eqnarray}\footnote{Notice that $D_3 \sigma$ is not continuously matched, but that is ok with the variational problem in question.}
Additionally, we demand the off shell variations to vanish at the junction points $t=0$ and $t=t_0=\pi$, namely
\begin{eqnarray}
\delta X(0)=\delta X(t_0=\pi)=0. \label{refeq}
\end{eqnarray} 
Out of the equations labeled in \eqref{refeq0}, it is straightforward to check that the D-conditions \eqref{refeq} guaranty that the total derivative part of the off shell variations of  $\mathcal{L}_{mixed}$ integrates to zero on ${\SS}$. Of course \eqref{FBPS} needs to be satisfied too and in the case of the matter multiplet we impose on top of that
\begin{eqnarray}
\mathcal{X}_{Matter}(0)= \mathcal{X}_{Matter}(t_0)=0.\label{refeq2}
\end{eqnarray}
When condition \eqref{refeq2} is imposed on top of \eqref{FBPS}, it kills any possible non trivial BPS configuration coming from the matter sector. In subsection \ref{OddEven} we will show how \eqref{refeq2} is consistent with the previously mentioned set of boundary conditions for the matter sector: \eqref{BCfermions}. In fact the results in subsection \ref{QuantiEven}, allow to quantise this last variational problem  by demanding {\blue even } (resp. {\red odd}) condition \eqref{BCfermions} on top of \eqref{refeq2} when $0<t<t_0=\pi$ ( resp. $\pi=t_0<t<2\pi$). In subsection \ref{FaCnes}  it will be shown that the quantisation of \eqref{refeq0}  mentioned in the previous sentence, results in the sum over fluxes that we have called 3D TT index on $\SS$, after localising to the BPS configurations \eqref{'t HooftContinuous}. 
  In the very end, supersymmetric localisation reduces the problem to quantum mechanics on $\mathbb{S}_1$ with Dirichlet conditions on antipodal points.

\subsection{The "complex" path of integration: The proper localising term} \label{cpath}

In this subsection we write down the localising term for the vector multiplet that must be used when the path of integration is the one proposed in \citep{BZ}. As for the matter multiplet one can use any of the localising terms written down in previous sections. The reality conditions for the vector multiplet in this case are
 \begin{eqnarray}
A_\theta= A_\theta^*, ~ A_\phi= A_\phi^*, ~  \hat{A}_t=\hat{A}_t^*,%Im[A_t]=  -\sigma  ,
~ \sigma=\sigma^*, ~ D={\red-}D^*.% ~ \lambda=\lambda^*, ~\bar{\lambda}=\bar{\lambda}^*.
\label{ComplexCondBZ}
\end{eqnarray}
The right localising term to use is
\begin{eqnarray}
Q_\epsilon V^{B-Z}_{vector}&:=&Q_{\epsilon} \left(~~\left(\overset{\bullet}{Q_{\epsilon} \lambda}\right)_{B-Z} \lambda\right), \label{locVBZ} \\
\left(\overset{\bullet}{Q_{\epsilon} \lambda}\right)_{B-Z}&:=&\nonumber \left(Q_\epsilon \lambda\right)^* \bigg|_{\hat{A}^*\rightarrow \hat{A}, ~ \sigma^*\rightarrow \sigma, ~ D^* \rightarrow {\red-}\,D}.\\\nonumber
\end{eqnarray}
We say is the right one, because it has positive definite bosonic part along \eqref{ComplexCondBZ}. The difference between \eqref{locVBZ} and \eqref{locV} is the following $Q_\epsilon$ exact term
\begin{eqnarray}
-2\, Q_\epsilon \left( D \,\epsilon\lambda\right).
\end{eqnarray}
The expansion of \eqref{locVBZ} with the form of the algebra \eqref{algebra1} gives the following bosonic
\begin{eqnarray}
 Q_\epsilon V^{B-Z}_B:=\left(F_{1 2}+ \hat{\mathcal{D}}_3 \sigma +i D \right)^2+\left(\hat{F}_{13}\right)^2+ \left(\hat{F}_{2 3}\right)^2, \label{QVvec2BZ}
\end{eqnarray}
and fermionic parts
\begin{eqnarray}
 Q_\epsilon V^{B-Z}_F:=i \,\bar{\lambda}^\dagger_2\, \overleftarrow{\hat{\mathcal{D}}}_t \, \lambda_2. \label{QVvecFBZ}
\end{eqnarray}
Notice that only a pair of fermionic DOF's remains dynamical in this localising action. This is reflection of the cohomological cancellations, in fact these are the DOF's orthogonal to $\epsilon$. When computing the one loop determinants, we will see how this fact is cardinal to obtain the correct result given in \citep{BZ}. The $\lambda_1$ and $\bar{\lambda}^\dagger_1$ are fermionic zero modes of \eqref{QVvecFBZ}. In order no to get a non vanishing result we must not integrate over $\lambda_1$ and $\bar{\lambda}^\dagger_1$.

We can also use the localising term analog to the one used in Appendix \ref{loft} \eqref{BosAc}, which as explained there, must not be confused with \eqref{locVBZ}, namely
\begin{eqnarray}
Q_\epsilon V^{B-Z}_{vector}&:=&Q_{\epsilon} \left(~~\left(\overset{\bullet}{Q_{\epsilon} \lambda}\right)^{II}_{B-Z} \lambda\right) \label{QVectorBZNew}
\\
\left(\overset{\bullet}{Q_{\epsilon} \lambda}\right)^{II}_{B-Z}&:=&-\frac{1}{2} F_{\mu \nu} \epsilon^\dagger\gamma^{\mu \nu} \,{\red-} \,D \epsilon^\dagger+i (\mathcal{D}_\mu\sigma) \epsilon^\dagger \gamma^\mu.
\end{eqnarray}
After expanding this term with the help of the covariant form of the algebra \eqref{algebra} we obtain similar bosonic and fermionic parts as in \eqref{QVvec2BZ} and \eqref{QVvecFBZ} respectively
\begin{eqnarray}
 Q_\epsilon V^{B-Z}_B:=\left(F_{1 2}+ \mathcal{D}_3 \sigma +i D \right)^2+\left(F_{13}-\mathcal{D}_2\sigma\right)^2+ \left(F_{2 3}+\mathcal{D}_1\sigma\right)^2, \label{QVvec2BZNew}
\end{eqnarray}
\begin{eqnarray} &&\nonumber\\
Q_\epsilon V^{Vector}_F&=&-( \bar{\lambda}_1^\dagger   ~~~ \bar{\lambda}_2^\dagger   )  \left(\begin{array}{c c}
 %-
 0 &   i ( \overleftarrow{\mathcal{D}_{1}}-i  \overleftarrow{\mathcal{D}_2} )\\
 %-
0 & -i \overleftarrow{\hat{\mathcal{D}}_3}
 \end{array}\right) \left(\begin{array}{c} \lambda_1 \\  \lambda_2 \end{array}\right).  \label{QVectorStuckelbergBZNew}\\
 \nonumber
\end{eqnarray}
The determinant of the Dirac matrix in \eqref{QVectorStuckelbergBZNew} is formally zero, because the mode $\bar{\lambda}_1, \lambda_1$ are zero modes of the corresponding localising operator. In order no to get a non vanishing result we will not integrate over $\lambda_1$ and $\bar{\lambda}^\dagger_1$.

The bosonic term \eqref{QVvec2BZNew} is not positive definite under the conditions \eqref{ComplexCondBZ} but instead under
 \begin{eqnarray}
A_\theta= A_\theta^*, ~ A_\phi= A_\phi^*, ~ {\red A_t=A_t^*},%Im[A_t]=  -\sigma  ,
~ \sigma=\sigma^*, ~ D={\red-}D^*.% ~ \lambda=\lambda^*, ~\bar{\lambda}=\bar{\lambda}^*.
\label{ComplexCondBZNew}
\end{eqnarray}
%In principle, one can use any of these two choices, \eqref{locVBZ} or \eqref{QVectorBZNew}.

 We prefer to use the path of integration \eqref{ComplexCondBZ}, thenceforth \eqref{locVBZ} is the proper choice and from now on when we refer to the complex path of integration we intend the use of that localising term.

 The zero locus configurations in the path \eqref{ComplexCondBZ} are given by the fluxes on $\mathbb{S}_2$
 \begin{eqnarray}
 F_{12}= \frac{\mathfrak{m}}{2 R^2},
 \end{eqnarray}
 where
 \begin{eqnarray}
 \mathfrak{m}:=\mathfrak{m}^{\mathbb{R}}+\mathfrak{m}^{\mathbb{C}},
 \end{eqnarray}
is the combination of a couple of GNO quantised elements, $\mathfrak{m}^{\mathbb{R}}$ coming from the real configuration for $\sigma$ \eqref{locusN} and $\mathfrak{m}^{\mathbb{C}}$ coming from the complex configuration for $D$
\begin{eqnarray}
D:=i \, \frac{\mathfrak{m}^{\mathbb{C}}}{2 R^2}. \label{ComplexFlux}
\end{eqnarray}
From now on we will focus on the real path of integration introduced in previous sections \eqref{ComplexCond}. However, the computations of one loop determinants for the matter sector are equivalent in both paths of integration  \eqref{ComplexCond} and \eqref{ComplexCondBZ}. The subtle difference comes out in the computation of the vector multiplet one loop determinants. For completeness of analysis, in Appendix \ref{VecBZ} we compute the one loop determinants of the vector multiplet in the complex path \eqref{ComplexCondBZ}, namely with the terms \eqref{QVvec2BZ} and \eqref{QVvecFBZ}.

\section{The 3D TT Index on $\mathbb{S}_2\times \mathbb{S}_1$: "real" vs "complex" path}

In this section we write down the two path integral representations of the index that we promised. The first representation will be a "product" of two blocks. The two blocks being the partition functions of the TT theory on $\mathbb{S}_2\times I$ with $I$ being a semicircle \footnote{The $Q_\epsilon$ and $\tilde{Q}_\epsilon$ TT theories localise to the same result when placed on $\mathbb{S}_2$ times an open segment with the boundary conditions we are choosing.}. There will be two possibilities of matching the fluxes from one side to the other. If the match from $I_N$ to $I_S$ is continuous, then we are quantising the variational problem {\bf VP-II}. The latter, is a functional integration over the real path of fields of the  theory \eqref{refeq0} on $\mathbb{S}_2 \times \mathbb{S}_1$. In this way we reproduce the result of \citep{BZ} but integrating along a section of hermitian fields.

 Thereafter, we write down a second representation: The path integral of, either $Q_\epsilon$ or $\tilde{Q}_\epsilon$-TT theory on $\SS /\{t=0\}$ which is nothing but the quantisation of the variational problem {\bf VP-I} on $\SS /\{t=0\}$. In this way we conclude that in order to have coincidence between the integration along the "real" and "complex" paths of integration when we integrate the theory {\bf VP-I} along the former path \footnote{Along the complex path of integration in order to have fluxes, we just need to consider the regular BPS solutions \eqref{ComplexFlux} as done in \citep{BZ}.}, we must consider contributions coming from real BPS configurations that do not belong to $\mathbb{S}_1$ but to $\mathbb{S}_1/\{t=0\}$  .

\subsection{Functional space of integration:  boundary state on $\mathbb{S}_2\times I_N(S)$}  \label{subsec}
The main scope of this subsection is to define the functional space of integration we will use to compute the gaussian path integrals defining the one loop determinants needed in the "semiclassical" limit $\tau\rightarrow \infty$ of localisation. On passing by, we highlight one of the positive outcomes of considering localising density lagrangians with vanishing supersymmetric variations (even taking into consideration total derivatives): The wave function of a boundary state.

 Let us start by thinking about what we called the "3D $Q_\epsilon$ TT theory"
\begin{eqnarray}
\mathcal{L}_N:=Q_\epsilon V_N=\bigg(\eqref{locV},\eqref{MatterAc}\bigg),\,  ~~~ \mathcal{L}^{CS}_N:=\eqref{ModifiedCS}.
\end{eqnarray}
defined on $\mathbb{S}_2\times I_N$. The analysis for the $\tilde{Q}_\epsilon$ TT theory is completely analog and hence can be reproduced following the same line of reasoning. In this subsection we consider the Chern Simons coupling to vanish, namely  $k=0$.

 The action defined by the Lagrangian density above (excluding the Chern-Simons term  $\mathcal{L}^{CS}_N:=\eqref{ModifiedCS}$ whose $Q_\epsilon$ variation will be discussed below), is $Q_\epsilon$ invariant and the domain space $S_2\times I_N$ is compact; consequently we are not forced to impose boundary conditions for the vector and matter multiplets. Namely supersymmetry and finiteness of the action $\mathcal{L}_N$ are already guarantied. The integral over $\mathbb{S}_2\times I_N$ will be a functional of the set of boundary conditions we choose.

  Let us denote the set of asymptotic behaviours we would like to impose about $t=0$ or $t_0$ by
\begin{eqnarray}
\Psi_\tau^I:=BPS^I+\frac{1}{\sqrt{\tau}}\,\psi^I \label{branch}
 \end{eqnarray}
 with $\tau$ being the localisation "Planck scale" parameter and $I$ an index running over the set of $BPS$ saddles. Equation \eqref{branch} can be interpreted as a branching of the boundary asymptotic behaviour $\Psi^I$ at $t=0$ or $t_0$  in terms of the redefinition of fields used in order to take the semiclassical limit $\tau\rightarrow \infty$ of localisation.
 \def\Nst{\<Z\, \|\psi^I \>}
\def\Sst{\> \psi^\tilde{I} \|\tilde{Z}\>}

 Let us denote by ${Z_N}_\tau$ the path integral of $e^{-\tau \int_{\SIN} Q_\epsilon V_N}$ and by $Z_\tau[\Psi_\tau]$ the same functional integral, but after imposing the boundary conditions $\Psi_\tau$ at $\partial I _{N}$ for the integrated fields. When the localisation limit is performed upon ${Z_N}_\tau[\Psi_\tau]$, as summarised in this equation below
 \begin{eqnarray}
{Z_N}_\tau[\Psi_\tau]=Z_\infty[\Psi_\infty]=\sum_{I} \partial Z_N[BPS^I,\psi^I]:=\sum_{I} \<Z_N\, \|\psi^I \>,
\end{eqnarray}
one remains with a functional $``\partial Z_N"$ of the $I-th$ $Q_\epsilon$-BPS charges and the boundary data $\psi^I$. Notice that we refer to $\psi^I$ as an asymptotic behaviour. This is because the boundary data encoded in $\psi^I$ could be given by values of the fields but also by the values of $\partial_t$- derivatives of the fields at $0$ and $t_0$. The function $\partial Z$ is called the wave function of the boundary state: the state defined by the boundary conditions $\psi^I$. In bra and kets terminology, $\Nst$  is the saddle approximation to $Z_N$ around $I$-th BPS state with boundary data $\psi^I$.

From now on, we will use the following terminology

\begin{itemize}
\item  ~~~~~The bras of $Z$,  $\< Z \|  \Circle\>$~~:= ~$Q_\epsilon V_N$ - wave functions of the $\partial I_N$ boundary state $\Circle$.
\item  ~~~~~The kets of $Z$, $\<\Circle\| Z\>$~~:= ~  $\tilde{Q}_\epsilon V_S $ - wave functions of the $\partial I_S$ boundary state $\Circle$.
\end{itemize}

Notice that we have omitted the suffices $N$ and $S$ in the bra and ket respectively. The bra and ket denote the partition function with domain $\SIN$ and $\SIS$ respectively; however, at some points we will use the suffices in order to avoid confusion.

The partition function on $\SIN$ is recoverable out of the space of wave functions of boundary states. In fact,  when we take the Chern Simons coupling $k$ to vanish ($k=0$), the partition function on $\SIN$ is
\begin{eqnarray}
Z_N:=\sum_{I} \int[D \psi^I] \<  Z_N \| \psi^I\>. \nonumber
\end{eqnarray}
\paragraph{ Wave function of boundary state}
 Let us illustrate with a rather simple example of the matter multiplet sector  with r-charge $q=0$ and without flavour charges around a BPS solution \eqref{locusN} with flux $\mathfrak{m}$
\begin{eqnarray}
{\<Z_{matter}\, \|\psi^\mathfrak{m} \>}&:=& \int_{\psi^\mathfrak{m}} [D\bar{\phi} D\phi D\bar{\psi} D\psi D \bar{F} D F]\,e^{- S^{(2)}_{\mathfrak{m}}-S^{bdry}_{\mathfrak{m}}}  \\
S^{(2)}_{\mathfrak{m}}&:=& \int_{\SIN}-\bar{\phi}^\dagger \left(\Box_{\mathfrak{m}}-b\right)\phi-\bar{\psi}^\dagger \hat{\slashed{D}}_\mathfrak{m} \psi+\bar{F}^\dagger F \label{MatterQuadFlucAc}
 \\
 S^{bdry}_{\mathfrak{m}}&:=& \,\int_{\mathbb{S}_2} \left(\bar{\phi}^\dagger \partial_3 \phi-i \, \bar{\psi}^\dagger \gamma_3 P^- \psi\right) \bigg|^{t_0}_{0},\label{BTMTTR}
 %\\
%\psi^\mathfrak{m}_{t=0}=\psi^\mathfrak{m}_{t=\pi}&:=&
\end{eqnarray}
with
\begin{eqnarray}
\Box_\mathfrak{m}&:=&\mathcal{D}^2_t+\frac{1}{2}\left(\mathcal{D}_+^{(b)}\mathcal{D}^{(b-1)}_- +\mathcal{D}_-^{(b)}\mathcal{D}^{(b+1)}_+\right),%\left(\partial_t- i \rho(u)\right)^2+\left(\partial^2_\theta +\cot{ \theta}  \partial_\theta+\csc^2{\theta}\left(\partial_\phi -i\,s \cos{\theta}\right)^2 \right)
  \\
\hat{\slashed{D}}&:=& \left(\begin{array}{c c}
 %-
 i \hat{\mathcal{D}}_t &~~~   \frac{i}{R} \mathcal{D}^{(b)}_+\\
 %-
 \frac{i}{R} \mathcal{D}^{(b-1)}_- & -i \hat{\mathcal{D}}_t
 \end{array}\right), \label{diracMatter}
\end{eqnarray}
with $b:=-\frac{\rho(\mathfrak{m})}{2}$ \footnote{Note that our $b$ is $\frac{1}{2}$ the one used in \citep{BZ}.} and
\begin{eqnarray}
\hat{\mathcal{D}}_t&:=& \partial_t -i \rho(u)
\\
\mathcal{D}^{(s\pm1)}_\pm&:=& \partial_\theta \mp \frac{i}{\sin \theta} \partial_\phi \mp \frac{s}{\sin \theta}. \label{raising}
\end{eqnarray}
\def\stwo{\mathbb{S}_2}
Notice that in our convention, $\mathcal{D}^{(s+1)}_{+}: Y^{s}_{j j3}\rightarrow Y^{s+1}_{j j3}$, whereas for the convention used in \citep{BZ}, $\mathcal{D}^{(s+1)}_{+}: Y^{s+1}_{j j3}\rightarrow Y^{s+2}_{j j3}$ \footnote{The $\mathcal{D}^{(s+1)}_+$ lowers the eigenvalues of the $\mathbb{S}_2$ part of $\Box_{\mathfrak{m}}$, namely the magnetic Laplacian on $\mathbb{S}_2$, from $j(j+1)-s^2$ by $2s+1$, namely to $j(j+1)-(s+1)^2$.  In contradistinction, it raises the magnetic level $s$ of the spin spherical harmonics $Y^{s}_{j j_3}$ by a unit, namely from $s$ to $s+1$.  The opposite statement can be said for $\mathcal{D}^{(s)}_-$. In summary, the action of the operators $\mathcal{D}^{(s\pm1)}_{\pm}$ on magnetic spherical harmonics can be represented as $\mathcal{D}^{(s\pm1)}_{\pm}: Y^{s}_{j,j_3}\rightarrow Y^{s\pm1}_{j j_3} $ with $Y^{s}_{j j_3}$ being the spin $s$ spherical harmonic in $\stwo$. }. The boundary term in \eqref{BTMTTR} has two components, the first one, comes from the partial integration that was performed in order to arrive to the more appealing form of the action \eqref{MatterQuadFlucAc} and the second comes from the specific $Q_\epsilon V$ term used to localise, which in this case was \eqref{LBOSF0}+\eqref{LFERF0}. Notice that the total derivative coming from the bosonic part of \eqref{LBOSF0} integrates trivially on $\mathbb{S}_2\times I$. Our boundary conditions guaranty the vanishing of the aforementioned boundary term.

We define the space of functional integration on $0<t<t_0$ to be
\begin{eqnarray}
\mathcal{K}^{t_0}_{\tiny\begin{array}{c}{\blue even} \\ or \\ {\red odd} \end{array}}:=\left( \underbrace{ \bigg(Y^{b}_{j,j_3}\bigg)}_{\phi}\otimes \underbrace{\bigg(Y^{b}_{j,j_3}\bigg)}_{\bar{\phi}} \otimes \underbrace{\bigg(\begin{array}{c}_1Y^{b}_{j,j_3}\\ \otimes \\_2Y^{b-1}_{j,j_3} \end{array}\bigg)}_{ _\alpha\psi} \otimes \underbrace{\bigg(\begin{array}{c} _1Y^{b}_{j,j_3}\\  \otimes \\_2Y^{b-1}_{j,j_3} \end{array}\bigg)}_{_\alpha\bar{\psi}}\otimes \underbrace{ \bigg(Y^{b}_{j,j_3}\bigg)}_{F}\otimes \underbrace{\bigg(Y^{b}_{j,j_3}\bigg)}_{\bar{F}}
\right)\otimes\left(\underset{\small \begin{array}{c}{\blue k\in 2\mathbb{Z}} \\ or \\ {\red k\in 2\mathbb{Z}+1} \end{array}}{\oplus}{\frac{e^{ i \pi k \frac{t}{t_0} }}{\sqrt{t_0}}}\right). \nonumber\\ \label{functionalSpace}
\end{eqnarray}
A very important point is that for the {\color{blue} even} and {\color{red} odd} quantisations \eqref{functionalSpace}, {\color{blue} $k \in 2 \mathbb{Z}$} and {\color{red} $k \in 2 \mathbb{Z}+1$}, the boundary term \eqref{BTMTTR} vanishes identically.

We have omitted the direct sums $\underset{j\geq j_{min}}{\oplus} \left(\left(\cdot\right)_{ -j \leq j_3\leq j} \right)$ that act upon each of the internal factors in the direct products $\otimes$ inside the first parenthesis, in order not to make too clumsy the notation. The $j_{min}$ is defined as follows
\begin{eqnarray}
j_{min}:= \left\{ \begin{array}{c c} |b| & \text{Bosons}\\    |b| &~~ \alpha=1~ \text{"Chiral" Fermions}\\ |b-1| &~~ ~~ \alpha=2 ~ \text{ "Anti-Chiral" Fermions}\end{array}\right\}
\end{eqnarray}

The relative differences between the magnetic level $s$ of the spin spherical harmonics $Y^{s}_{j j_3}$ in \eqref{functionalSpace} is determined from consistency with the supersymmetry transformations \eqref{algebraChiral}. For instance from the $O(\frac{1}{\sqrt{\tau}})$ part of the $Q_\epsilon$ algebra \eqref{algebraChiral} and after the redefinition \eqref{FBPS} one gets

\begin{eqnarray*}
Q_\epsilon {\bar{\phi}^\dagger}= -\bar{\psi}^\dagger_2, ~~~ Q_\epsilon \bar{\psi}^\dagger_2=-\bar{F}^\dagger \\
Q_\epsilon \psi_1=- i \hat{\mathcal{D}}_t \phi \\
Q_\epsilon \psi_2= -\frac{i}{R}\mathcal{D}^{(b-1)}_-  \phi.
\end{eqnarray*}
From the third line we conclude that when the magnetic level associated to the fluctuation $\phi$ is assumed to be $b$, the one associated to $\psi_2$ is $b-1$ and the one associated to $\psi_1$ is $b$. Then by analysing the $O(\frac{1}{\sqrt{\tau}})$ of $\tilde{Q}_\epsilon$ algebra \eqref{algebraChiralSouth} and after the redefinition \eqref{FBPS} one closes the cycle to magnetic levels written in \eqref{functionalSpace} after supposing the level of $\bar{\phi}^\dagger$ to be $b$.

The operators $\Box_{\mathfrak{m}}$ and $\hat{\slashed{D}}_{\mathfrak{m}}$ have the following non vanishing expectation values on the relevant spin spherical harmonics $Y^{(s)}_{j,j_3}$ in \eqref{functionalSpace}
\def\dirac{\hat{\slashed{D}}_{\mathfrak{m}}}
\begin{eqnarray}
\int_{\SIN}  ({{}_\alpha}{\bar{\psi}^{b,\,k}_{j j_3}})^\dagger  \,  \hat{\slashed{D}}_{\mathfrak{m}} \, \, {{}_\beta}\psi^{b,\,k}_{ j j_3}&=&\left\{\begin{array}{cc}\left(\begin{array}{cc} - \,\left( \frac{\pi}{t_0} k-\rho(u) \right)&i\left(1+ j-b\right)\\-i\left(j+b\right)  & \, \left(\frac{\pi}{t_0}k-\rho(u)\right) \end{array}\right)& j\geq \max\left(|b|,|b-1|\right), \\ \\ \left(\begin{array}{cc} \CIRCLE & \CIRCLE \\ \CIRCLE  &  \, \left(\frac{\pi}{t_0}k-\rho(u)\right) \end{array}\right) &~|b-1|=j<|b|,\\ \\ \left(\begin{array}{cc} -\,\left( \frac{\pi}{t_0} k-\rho(u) \right)& \CIRCLE\\\CIRCLE  &\, \CIRCLE \end{array}\right)& |b|=j<|b-1|,\end{array} \right\} \nonumber\\ \label{Point}\\
\int_{\SIN} ({\bar{\phi}^{b,\,k}_{j j_3}})^\dagger\, \Box_{\mathfrak{m}} \, \phi^{b, \,k}_{j j_3}&=& -\left(\left(\frac{\pi}{t_0}k-\rho(u)\right)^2+j\left(j+1\right)-b^2\right) \label{ExpValue}.
\end{eqnarray}
The basis vectors being normalised as follows
\begin{eqnarray}
\int_{\SIN} {_\alpha \bar{\psi}^{b,\,k}_{j j_3}}^\dagger \,   _\alpha \psi^{b,\,k^\prime}_{ j^\prime j^\prime_3}&=&\delta^{k k^\prime} \delta_{j j\prime} \delta_{j_3 j_3^\prime}, ~~ \alpha=1,2. \\
\int_{\SIN} ({\bar{\phi}^{b,\,k}_{j j_3}})^\dagger\,  \phi^{b^\prime, \,k^\prime}_{j^\prime j^\prime_3}&=& \delta^{k k^\prime} \delta_{j j\prime} \delta_{j_3 j_3^\prime}.
\end{eqnarray}
Under these orthonormalisation conditions the  sandwich of $\dirac$  and $\Box_{\mathfrak{m}}$ between states with different labels $(j,\,j_3, k)$, vanishes.

Let us explain the meaning of the $\CIRCLE$'s in \eqref{Point} in the case $|b-1|=j<|b|$. The meaning of the $\CIRCLE$ in the alternative case  $|b|=j<|b-1|$ follows straightforwardly. For instance, when $|b-1|=j<|b|$, there is not "chiral" mode $_1 \psi$ but there is "antichiral" mode $_2 \psi$ \citep{BZ}. In this case the $\CIRCLE$'s at the positions $(1,1)$, $(1,2)$ and $(2,1)$ of the corresponding $2\times2$ matrix in \eqref{Point} denote the absence of these components due to the lack of "chiral"(resp. "antichiral") mode $_1 \psi$. Notice that the determinant of the fermionic Dirac matrix in the first line of \eqref{Point}, coincides with the eigenvalue of the bosonic operator \eqref{ExpValue} $(\Box_{\mathfrak{m}}-b)$. As $\Box_{\mathfrak{m}}-b$ is the bosonic operator (complex scalar) action \eqref{MatterQuadFlucAc} and the Dirac operator \eqref{Point} is the fermionic counterpart in \eqref{MatterQuadFlucAc} one could naively conclude that bosonic and fermionic determinants of \eqref{MatterQuadFlucAc} should cancel each other out due to the fact mentioned in the previous sentence. However, as will be shown in a while, cancellation between bosonic/fermionic determinants does not follow, due to the lack of "chiral/antichiral ground state" at level $j=min(|b|,|b-1|)$. The dots $\CIRCLE$ in the second and third line of \eqref{Point} represent the absence of such "chiral/antichiral" ground state at the respective spin level $j$ \citep{BZ}.

Let us continue with our example of boundary state. Should we select the following boundary conditions at both $t=0$ and $t=t_0$
\begin{eqnarray}
\mathcal{X}(t \sim t_0)=\mathcal{X}(t \sim 0):=c_\mathcal{X}\,\mathcal{X}^{b, \, {\blue k=0}}_{j {j_3}}(\theta,\phi,t), ~~~ j\geq\max\left(|s|,\,|s-1|\right)\nonumber\\
\mathcal{X}:=\{\phi, \, \bar{\phi},\, _1\psi,\,  _2\psi ,\, _1\bar{\psi},\, _2\bar{\psi},\, F,\, \bar{F}\}, \label{BCSel}
\end{eqnarray}
the functional space \eqref{functionalSpace} collapses to that single state with the following semiclassical distribution density
\begin{eqnarray}
\<Z_{matter} \| \mathcal{X}\>=e^{ \left(\<\Box_{\mathfrak{m}}\>_{j,b,u}-b\right)\left(\bar{\phi}^\dagger\phi\right)^{b\, k=0}_{j,j3}+\left(\bar{\psi}^{b, k=0}_{j, j3}\right)^\dagger \<\hat{\slashed{D}}_\mathfrak{m} \> \psi^{b\, k=0}_{j,j3}-\left(\bar{F}^{b\, k=0}_{j,j3}\right)^\dagger F^{b\, k=0}_{j,j3}},
\end{eqnarray}
where by $\<\>$ we mean the expectation values posted in \eqref{ExpValue}. By $t\sim 0$ and $t\sim t_0$ in \eqref{BCSel}, we mean the asymptotic behaviour of $\mathcal{X}$ in a vicinity of $0$ and $t_0$. Notice that the selection of the mode ${\color{blue} k=0}$ in \eqref{BCSel}, determines not only the value of $\mathcal{X}$ but also their derivatives $\partial_t \mathcal{X}$ at $0$ and $t_0$.

\paragraph{On the $\mathcal{K}_{\blue even}$ and $\mathcal{K}_{\red odd}$ KK mode space} \label{OddEven}

The  KK basis states $e^{i\frac{\pi}{t_0} k t}$ in spaces \eqref{functionalSpace} are not physical when one works with hermitian fluctuations (because they are complex). This is the case of the vector multiplet fluctuations. However these complex KK states can be used in any case, to compute the determinant of $i \,\partial_t$. It is convenient to work with these modes, because they are eigenstates and thence further diagonalisation procedures are not needed. The complex space spanned by these complex basis states is
\begin{eqnarray}
\sum_{k} c_k e^{i \frac{\pi}{t_0} k t}.
\end{eqnarray}
In the case of hermitian fluctuations one needs to restrict to the subspace defined by
\begin{eqnarray}
c_{-k}=c_k^*,
\end{eqnarray}
which is the real vector space
\begin{eqnarray}
\left\{\sin\left( \frac{\pi}{t_0} k \, t\right)\oplus \cos\left( \frac{\pi}{t_0} k\,  t\right)  \right\}_{\oplus k} \text{ with } ~~ k\in \left\{\begin{array}{c} { \blue 2 \mathbb{Z}} \geq 0 \\ \, { \red 2 \mathbb{Z}+1}>0   \end{array}\right\}. \label{ecuacNO}
\end{eqnarray}
Notice that when $t_0 \rightarrow 2 \pi^- $ \eqref{ecuacNO} contracts to
\begin{eqnarray}
\left\{\sin\left( \frac{1}{2} k \, t\right)\oplus \cos\left( \frac{1}{2} k\,  t\right)  \right\}_{\oplus k} \text{ with } ~~ k\in 2 \mathbb{Z} \geq 0.  \label{ecuacNO2}
\end{eqnarray}
In \eqref{ecuacNO} and \eqref{ecuacNO2} one can further restrict to $\sin$ or $\cos$ or a combination of them by imposing Dirichlet or Neumann boundary conditions upon the fluctuations at $t=0$ and $t_0$. Specifically
\begin{eqnarray}
\begin{array}{ccc}
(D,D)&\rightarrow& \sin \\
(N,N)&\rightarrow&\cos.
\end{array}
\end{eqnarray}
As already explained at the end of the previous section, we will use the first of these restrictions for the vector multiplet case, in order to annihilate the boundary variations of the modified Chern-Simons term in a gauge invariant way. We stress that, for the matter multiplet, the selection of {\color{blue} $even$} or {\color{red} $odd$} KK modes does not have to do with preservation of supersymmetry, even though both choices preserve supersymmetry. In previous sections we argued that the selection of one of the quantisations {\blue even}/ {\red odd}  has to do with the correct definition of variational problem. In this section the evident consequence of the choices {\blue even}/ {\red odd}  is the cancelation of the boundary term \eqref{BTMTTR}. Moreover, let us stress, that in order to have an orthonormal basis of KK modes on the interval $I=(0,t_0)$ we must choose either {\blue even} or {\red odd} quantisation but we can't have both at the time. In the latter line of thought, the choice of {\blue even} or {\red odd} is a freedom we have, in order to define the functional space of integration $\mathcal{K}^{t_0}$.  %Notice also, that we can use (D,D) conditions to cancel \eqref{BTMTTR} but this is not necessary.

\subsection{3D TT index on $\mathbb{S}_2$ times an interval}\label{QuantiEven}

In this subsection we write down the relevant one loop determinants on $\SIN$ and $\SIS$.
 The functional integration of \eqref{MatterQuadFlucAc} (together with integration over the boundary conditions $\psi^\mathfrak{m}$) over the functional space \eqref{functionalSpace}, is easy to perform with the use of \eqref{ExpValue}. The final result for the one loop determinant around the BPS configuration $\left(\mathfrak{m}_N,u_N\right)$ is
\begin{eqnarray}
\int [D\psi^\mathfrak{m}]_{\mathcal{K}^{t_0}}\<Z_{matter} \| \psi^{\mathfrak{m}_N}\>&=&\frac{Det_{\mathcal{K}^{t_0}}\dirac}{ Det_{\mathcal{K}^{t_0}}\left(\Box_{\mathfrak{m}_N}-b_N\right)}:= \prod_{\small \begin{array}{c}{\blue k\in 2\mathbb{Z}} \\ or \\ {\red k\in 2\mathbb{Z}+1} \end{array}}  \left(\frac{\pi}{t_0}k-\rho(u_N)\right)^{2b_N-1}\nonumber\\
&\equiv&Z^N_{matter}(\mathfrak{m}_N,\mathcal{K}^{t_0}). \label{OLM}
\end{eqnarray}

To arrive to this formula above one must write down the quotient between the fermionic
\begin{eqnarray}
\det_{(k,\rho)} \dirac=\left\{
\begin{tabular}
[c]{ccc}
$\prod_{j\geq\max(|b|,|b-1|)}\left(\left(\frac{\pi}{t_0}k-\rho(u)\right)^2+j\left(j+1\right)-b\left(b-1\right)\right)^{2j+1}$&  \     & \\
$\times \left(\frac{\pi}{t_0}k-\rho(u)\right)^{2 b-1}$ &  \  if & $|b-1|=j<|b|$\\
$\times \left(\frac{\pi}{t_0}k-\rho(u)\right)^{-2 b +1}$&   \ if & $|b|=j<|b-1|$%
\end{tabular}
\right., \nonumber\\ \label{DetFer}
\end{eqnarray}
and bosonic determinant (inverse of the partition function of the complex scalar)
\begin{eqnarray}
\det_{(k,\rho)}{\left(\Box_{\mathfrak{m}}-b\right)}=\prod_{j\geq |b|}\left(\left(\frac{\pi}{t_0}k-\rho(u)\right)^2+j\left(j+1\right)-b\left(b-1\right)\right)^{2j+1}.\label{DetBos}
\end{eqnarray}
The determinants \eqref{DetFer} and \eqref{DetBos} are computed out of the relevant matrix elements, which are given in equations \eqref{Point} and \eqref{ExpValue}. In this case $b:=-\frac{\rho(\mathfrak{m})-q_R}{2}$.
Notice that we have dropped some minus signs. Those signs can be absorbed by the transformations $(k,\rho ) \rightarrow (-k,-\rho)$ which are involutions of the set of KK modes and weights we are going to multiply over. At this point, is straightforward to check that $\forall b$
\begin{eqnarray}
\frac{\det_{(k,\rho)}\dirac}{\det_{(k,\rho)}\left(\Box-b\right)}=\left(\frac{\pi}{t_0}k-\rho(u)\right)^{2b-1}.
\end{eqnarray}
To obtain \eqref{OLM} we multiply over the modes $k$ belonging to the preferred KK spectrum: {\blue even} or {\red odd}.

 After an analog computation, but for the vector multiplet, as described in details in Appendix \ref{1loopdets}, one gets

\begin{eqnarray}
Z^N_{vector}(\mathfrak{m}_N,\mathcal{K}^{t_0})%=\frac{Det_{\mathcal{K}^{t_0}}\dirac}{\pm Det^{\frac{1}{2}}_{\mathcal{K}^{t_0}}\left(\Box-b\right)}
:= \prod_{\small \begin{array}{c}{\blue k\in 2\mathbb{Z}} \\ or \\ {\red k\in 2\mathbb{Z}+1} \end{array}}  \left(\frac{\pi}{t_0}k-\alpha(u_N)\right)^{-\alpha(\mathfrak{m}_N)+1}.
\end{eqnarray}
In analog way as previously described in the case of $\SIN$, on $\SIS$ one obtains for the matter multiplet the following contribution coming from the one loop super-determinant

\begin{eqnarray}
\int [D\psi^\mathfrak{m}_S]_{\mathcal{K}^{2 \pi-t_0}}\<\psi^{\mathfrak{m}_S} \|  Z_{matter} \>&=&\frac{Det_{\mathcal{K}^{2\pi-t_0}}\dirac}{ Det_{\mathcal{K}^{2\pi- t_0}}\left(\Box-b_S\right)}:= \prod_{\small \begin{array}{c}{\blue k\in 2\mathbb{Z}} \\ or \\ {\red k\in 2\mathbb{Z}+1} \end{array}}  \left(\frac{\pi}{2\pi -t_0}k-\rho(u_S)\right)^{2b_S-1}\nonumber \\&\equiv&Z^S_{matter}(\mathfrak{m}_S,\mathcal{K}^{2\pi-t_0}),
\end{eqnarray}
and for the vector multiplet

\begin{eqnarray}
Z^S_{vector}(\mathfrak{m}_S,\mathcal{K}^{2\pi-t_0})%\equiv\int [D\psi_V^{\mathfrak{m}_S}]_{\mathcal{K}^{2 \pi-t_0}}\<  \psi_V^{\mathfrak{m}_S}\| Z_{vector}\>%=\frac{Det_{\mathcal{K}^{t_0}}\dirac}{\pm Det^{\frac{1}{2}}_{\mathcal{K}^{t_0}}\left(\Box-b\right)}
:= \prod_{\small \begin{array}{c}{\blue k\in 2\mathbb{Z}} \\ or \\ {\red k\in 2\mathbb{Z}+1} \end{array}}  \left(\frac{\pi}{2\pi- t_0}k-\alpha(u_S)\right)^{-\alpha(\mathfrak{m}_S)+1}.\label{OLV}
\end{eqnarray}

Notice, that the functional form of the $I_N$ and $I_S$ one loop determinants are the same. In fact, one obtains the same answer by using either $Q_\epsilon$ or $\tilde{Q}_\epsilon$ at each of the segments. The same holds for Chern-Simons contributions. Henceforth, one can use any of the supercharges to perform localisation on one side or the other, the result is the same. The one point that must not be forgotten is that we are considering different theories at $I_N$ and $I_S$ and hence there is not relation between BPS solutions on one side or the other, yet.

Notice that these expressions for the one loop contributions are not regularised yet. Also in order not to make the presentation too clumsy we have omitted the product over weights $\rho$ and $\alpha$. We will include these products in the final expressions.

As a final thought, we try to elucidate whether  the {\blue even} quantisation could or could not be obtained out of the {\red odd} one by a shift $\Delta u$ of the holonomies. Notice that {\blue even} and {\red odd} even one loop determinants are equivalent if there exists a shift $\Delta u$ of the holonomies such that $\forall \alpha$ and $\forall \rho$
\begin{eqnarray}
\alpha^*(\Delta u) = k_\alpha^\prime \frac{\pi}{t_0}\label{AbelianFactor0}\\
\rho^*(\Delta u) = k_\rho^\prime \frac{\pi}{t_0}\label{AbelianFactor}
\end{eqnarray}
with $k_{\alpha,\rho}^\prime\in 2 \mathbb{Z}+1$.
Generically, such $\Delta u_i$ with $i=1,\ldots,rank(\mathcal{G})$ do not exist. In this section we restrict ourselves to the analysis of simple Lie algebra's $\mathcal{G}$.  Let us focus on the particular case $t_0=\pi$ and $\mathcal{G}=su(3)$ which reduces \eqref{AbelianFactor0} to 
\begin{eqnarray}
\left\{{\alpha^*}\right\}\cdot\Delta u&=&\left\{ \begin{tabular}{ccc}
$\left(~1,~1\right)\cdot $&$(-1,~2) \cdot$ &$(~2,-1) \cdot $\\
$\left(-1,-1\right) \cdot $ &$(~1,-2) \cdot $ & $(-2,~1) \cdot$
\end{tabular} \right\}\cdot \left(\begin{tabular}{c}$\Delta u_1 $\\ $\Delta u_2 $ \end{tabular}\right)\\&\in& \left\{ \begin{tabular}{ccc}
$2 \mathbb{Z}+1$ & $2 \mathbb{Z}+1$& $2 \mathbb{Z}+1$ \\
$2 \mathbb{Z}+1 $ & $2 \mathbb{Z}+1 $ & $2 \mathbb{Z}+1$
\end{tabular} \right\}
\end{eqnarray}
which is solved by 
\begin{eqnarray}
\left(\begin{tabular}{c}$\Delta u_1$ \\$ \Delta u_2 $ \end{tabular}\right)\in\left(\begin{tabular}{c}$\frac{1}{2}+\mathbb{Z}$ \\ $ \frac{1}{2}+\mathbb{Z}$  \end{tabular}\right). \label{SolDiscrete}
\end{eqnarray}
If we choose the matter representation $\{\rho^*\}$ to be the ones with HW $\left(1,0 \right)$ (fundamental) or $\left(0, 1\right)$ (antifundamental) then \eqref{SolDiscrete} solves \eqref{AbelianFactor} too. However, \eqref{SolDiscrete} does not solve \eqref{AbelianFactor} generically. For instance, if we choose  other irreps like for instance $\left(2,0\right)$ or $(0,2)$ or $(2,2)$ whose set of non zero weights are
\begin{eqnarray}
\{\rho^*\}=\left\{ \begin{tabular}{ccc}$
\left(~2,~0\right)$&$(0,~1) $ &$(-2,2)$  \\
$\left(1,-1\right)$  &$(~-1, ~0) $  &$(0,~-2)$ 
\end{tabular} \right\} \text{  or  } \left\{ \begin{tabular}{ccc}
$\left(~0,~2\right)$&$(1,~0)$  &$(2,-2)$  \\
$\left(-1,1\right)$  &$(~0, ~-1)$   &$(-2,~-0)$ 
\end{tabular} \right\}\\
or  \left\{ \begin{tabular}{cccccc}
$\left(~2,~2\right)$&$(0,~3)$  &$(-2,4)$ & $(3,0)$ &$(4,-2)$& $(1,1)$  \\
$\left(2,-1\right)$  &$(~3, ~-3)$   &$(-1,~ 2)$ & $(1,-2)$& $(2,-4)$&$(-3,3)$\\
$(-2,1)$&$(-4,2)$&$(-1,-1)$&$(-3,0)$&$(0,-3)$&$(-2,-2)$ 
\end{tabular} \right\},
\end{eqnarray}
is easy to check, that \eqref{SolDiscrete} does not solve \eqref{AbelianFactor}. In conclusion, {\blue even} and {\red odd} one loop determinants are not (generically) related to each other by a shift in the holonomy $\Delta u$. 
% in the case of simple algebra's because in this case the number of roots $\alpha$ is $rank(\mathcal{G})$ and they are all linearly independent, thence \eqref{AbelianFactor0} is invertible, and we can solve for the $\Delta u_i$, $i=1,\ldots,rank(\mathcal{G})$. Once the set of $\Delta u_i$ that solves \eqref{AbelianFactor0} is found, no freedom remains to solve for the matter sector equations \eqref{AbelianFactor} unless matter is in the adjoint of $\mathcal{G}$. In the latter case, the $\rho$'s are roots $\alpha$'s and \eqref{AbelianFactor} follows from \eqref{AbelianFactor0}. For generic irreps $\{\rho\}$, and in specific for the ones used in this paper (see next section), \eqref{AbelianFactor} is not consistent with \eqref{AbelianFactor0} and in consequence one can not map {\color{blue} even}  to {\color{red} odd} quantisation by mean of a shift $\Delta u$ of the holonomies $u$.

We are now in conditions to write down the partition function along the "real path" of the $\mathcal{N}=2$ 3D TT theory on $\mathbb{S}_2\times(0,t_0)$ with both {\blue even} and {\red odd} quantisation. After collecting Chern Simons part and one loop determinants for vector and matter multiplets the final result can be written as follows
\begin{eqnarray}
B_{{ \blue even}({\red odd})}(t_0):=\frac{1}{|W|}\sum_{\mathfrak{m}} \int \frac{d^{r}u}{(2\pi)^r} ~  B_{{ \blue even}({\red odd})}(\mathfrak{m},u,t_0),\label{Block}
\end{eqnarray}
where 
\begin{eqnarray}
B_{{ \blue even}({\red odd})}(\mathfrak{m},u;t_0):=e^{i k \frac{t_0}{2 \pi} u\cdot \mathfrak{m}} \prod_{\alpha^*,~k\in {\blue  2\mathbb{Z}}({\red 2\mathbb{Z}+1})}\left(\frac{\pi}{t_0}k-\frac{\alpha(u)}{2\pi}\right)^{-\alpha(\mathfrak{m})+1}
\nonumber\\ \times \prod_{\rho^*,~k\in {\blue  2\mathbb{Z}}({\red 2\mathbb{Z}+1})}\left(\frac{\pi}{t_0}k-\frac{\rho(u)}{2\pi}\right)^{2b-1}. \label{Block2}
\end{eqnarray}
In obtaining the final expressions \eqref{Block}-\eqref{Block2} we have used the substitution $u\rightarrow \frac{u}{2\pi}$. We have used the letter $B$ to denote the partition function on segments \eqref{Block}, because \eqref{Block} can be thought of as a block in a sense that will be explained in the next paragraph.

\subsection{Factorisation of the 3D TT Index:  Variational problem {\bf VP-II}} \label{FaCnes}

Let us summarise what we have done and what we want to do next to end up this section. So far, we have computed the partition function along "real path" on halves of $\mathbb{S}_1$.  A natural question that comes to mind, is whether one can express the formula for the partition function on $\mathbb{S}_1$ given in \citep{BZ} (with the sum over fluxes included) as a product of partition functions along the "real path" of some theories on halves of $\mathbb{S}_1$: blocks. As reviewed in \ref{cpath}, the formula of \citep{BZ} can be interpreted as the partition function of the corresponding Lagrangian \eqref{QVectorBZNew} along a "complex path"  \eqref{ComplexCondBZNew}. The previously mentioned factorisation, can be obtained out of localisation of the partition function associated to the variational problem {\bf VP-II}. In summary, what we will find is that the sum over fluxes that we have called the $3D$ TT index \citep{BZ}, can be seen as the localisation of the path integral along the real path of the variational problem {\bf VP-II}.

Let us start by shallowly reviewing some analog results in the literature. Some cases have been studied, where a partition function $Z_{\mathcal{M}}$ of $3D$ $\mathcal{N}=2$ theories on certain compact spaces $\mathcal{M}={\mathbb{S}^b_3},~\mathbb{S}_2\times\mathbb{S}_1$, factorises to a "product" of two blocks $B_{D_2\times \mathbb{S}_1}$(holomorphic) and $\bar{B}_{D_2\times \mathbb{S}_1}$(antiholomorphic). Without entering in details, these blocks are expressed as
\begin{eqnarray}
B_{D_2\times \mathbb{S}_1}=\sum_{V} B_{D_2\times \mathbb{S}_1}(V),
\end{eqnarray}
a sum over the smooth BPS vacua $V$ of the localisation formula for the partition function $B_{D_2\times \mathbb{S}_1}(V)$ of the original theory placed on $D_2 \times \mathbb{S}_1$ about the vacuum $V$.
The "product" of blocks is not defined through a simple multiplication but as a convolution with kernel $\mathcal{G}(V_1,V_2)$ that can be schematically represented as
\begin{eqnarray}
Z_{\mathcal{M}}=\sum_{V_1}\sum_{V_2} \bar{B}_{D_2 \times \mathbb{S}_1}(V_1) \mathcal{G}(V_1,V_2) B_{D_2 \times \mathbb{S}_1}(V_2).
\end{eqnarray}
That proposal was introduced and thoroughly studied in \citep{Pasquetti,Pasquetti2} and in several latter contributions, and states that the path integration of generic $3D$ $\mathcal{N}=2$ theories on compact manifolds $\mathcal{M}$ "factorise" in terms of the so called holomorphic and antiholomorphic blocks with specific kernel $\mathcal{G}(V_1,V_2)$.
 
In analog but different fashion and from the results presented in the previous subsection, we are able to show that the formula given in \citep{BZ} for the 3D TT index that we will denote from now on as $Z^{\mathbb{C}}_{\SS}$, can be factorised as follows
\begin{eqnarray}
=\sum_{V_1}\sum_{V_2} B_{ \blue even }_{\mathbb{S}_2 \times (0,\pi)}(V_1) \mathcal{G}(V_1,V_2) B_{\red odd}_{\mathbb{S}_2 \times (0,\pi)}(V_2), \label{factorise}
\end{eqnarray}
where $B_{\blue even}_{\mathbb{S}_2 \times (0,\pi)}(V_1)$ and $B_{\red odd}_{\mathbb{S}_2 \times (0,\pi)}(V_2)$ are the two variants of partition functions of the 3D $\mathcal{N}=2$ TT theory on $\mathbb{S}_2\times(0,\pi)$ that we defined in the previous subsection. 

 Let us work out the RHS of the precise proposition

$ Proposition$
 \begin{eqnarray}
Z^{\mathbb{C}}_{\SS}&=&\frac{1}{|W|^2}\sum_{\mathfrak{m}_1}\sum_{\mathfrak{m}_2}\int \frac{d^{r}u_1}{(2\pi)^r}\int \frac{d^{r}u_2}{(2\pi)^r} B_{\blue even}(\mathfrak{m}_1,u_1;\pi) \mathcal{G}(\mathfrak{m}_1,u_1;\mathfrak{m}_2,u_2) B_{\red odd}(\mathfrak{m}_2,u_2;\pi).\nonumber \\&&\mathcal{G}(\mathfrak{m}_1,u_1;\mathfrak{m}_2,u_2):=|W|\delta^{(r)}_{\mathfrak{m}_1\pm\mathfrak{m}_2,0}
\delta^{(r)}_{\left(u_1\pm u_2\right)},\label{facNew}
\end{eqnarray}
where the blocks are defined in \eqref{Block}. Let us work out piece by piece, the integrand of the RHS in the first line of \eqref{facNew}, which reduces to
 \begin{eqnarray}
 \frac{1}{|W|}\sum_{\mathfrak{m}_1} \int \frac{d^{r}u_1}{(2\pi)^r} B_{\blue even}(\mathfrak{m}_1,u_1;\pi)B_{\red odd}(\pm\mathfrak{m}_1,\pm u_1;\pi). \label{u}
 \end{eqnarray}
Is easy to see that the partition function associated to the variational problem {\bf VP-II} is the $+$ sign choice in the second block in \eqref{u} with the further selection of {\blue even} and {\red odd } KK modes in $I_N$ and $I_S$, respectively. At this point we can discard the $-$ sign option, however we will keep it till the end to make an interesting observation.
 
 The product of the Chern-Simons classical part in $B_{\blue even}$ and $B_{\red odd}$ in \eqref{u} is
 \begin{eqnarray}
 e^{i\frac{k}{2 } {\red  u\cdot \mathfrak{m}}}\times e^{ i\frac{k}{2}{ \blue u\cdot \mathfrak{m}}}=e^{ i k{  u\cdot \mathfrak{m}}},
 \end{eqnarray}
 The product of the matter sector determinants in $B_{\blue even}$ and $B_{\red odd}$ is
 \begin{eqnarray}
 &=&\prod_{{\blue k\in 2\mathbb{Z}},~\rho^*}  \left(k-\frac{\rho(u)}{2 \pi}\right)^{-\rho(\mathfrak{m})+q_R-1} \times \prod_{{\red k\in 2\mathbb{Z}+1},~\rho^*}  \left(k\mp\frac{\rho(u)}{2\pi}\right)^{\mp \rho(\mathfrak{m})+q_R-1} \\&=&\prod_{{\blue k\in 2\mathbb{Z}},~\rho^*}  \left(k-\frac{\rho(u)}{2 \pi}\right)^{-\rho(\mathfrak{m})+q_R-1} \times \prod_{{\red k\in 2\mathbb{Z}+1},~\pm\rho^*}  \left(k-\frac{\rho(u)}{2\pi}\right)^{-\rho(\mathfrak{m})+q_R-1}\\
 &=& \prod_{ k\in \mathbb{Z},~\rho^*}\left(k-\frac{\rho(u)}{2\pi}\right)^{-\rho(\mathfrak{m})+q_R-1}\\
 &=&\prod_{~\rho^*}\left(\frac{C_{reg}(\rho)}{\sin{\frac{\rho(u)}{2}}}\right)^{\rho(\mathfrak{m})-q_R+1}, \label{matfac1loop}
 \end{eqnarray}
 where $C_{reg}(\rho)$ is an arbitrary constant that comes from the regularisation of the product over the KK modes $k$.
From the second line to the third we have assumed that the set of weights $\{\rho\}$ remains invariant under the transformation $\rho \rightarrow - \rho$. By $\rho^*$ we intend the exclusion of the vanishing weight. The result for the vector multiplet 1-loop determinant is obtained in the same way and it coincides with \eqref{matfac1loop} after particularising to $q_R=2$ and the weights $\{\rho^*\}$ to the non zero roots $\{\alpha^*\}$.
 
 \paragraph{Final formula}

Having the tree level Chern-Simons phases and one loop contributions, we are in conditions to write down the final result of the RHS of \eqref{facNew}  %Thence by summing over the "glued" saddle points $(\mathfrak{m}_i,u_i)$ $i=1,\ldots,r$  the final formula is
 \begin{eqnarray}
 %\sum_{Saddle points=m }\int \left(\text{Irreducible $\delta_\epsilon$ -Multiplet of Zero Modes around fluxes}\right) % \int_{\mathbb{R}+i \eta} d^r D_0 \text{ with } \eta \in \mathbb{R}
\frac{1}{|W|}\sum_{ \mathfrak{m}^r }\int_{ \cup_{i=1}^{r} A_{i}} \frac{d^r u}{(2\pi)^r}  ~e^{i k u \cdot \mathfrak{m} } Z^{1-loop}_{Vector}Z^{1-loop}_{Matter}.  \label{PartitionFunction}
\end{eqnarray}
After particularising to $C^{Matter}_{reg}=\frac{i}{2}$ and $C^{Vector}_{reg}=\frac{i}{2}~sign\left(\rho\right)$ we finally obtain
\begin{eqnarray}
Z^{1-loop}_{Vector} &:=&   \prod_{\alpha \in \mathcal{G}>0} \bigg( 1-x^\alpha \bigg) ,  \nonumber\\
Z^{1-loop}_{Matter}&:=& \prod_{\rho \in \mathcal{R}} \bigg( \frac{x^{\frac{\rho}{2}}}{1-x^\rho} \bigg)^{\rho(\mathfrak{m}) - q_R+1},\label{Vector1LoopBulk}
\end{eqnarray}
with $x:=e^{i u}$. The reality condition on the fields \eqref{ComplexCond}, specifically the one on $\hat{A}_t$, implies the moduli components $u_i$ must be integrated along real segments. Equation \eqref{PartitionFunction} % telling that the $Z^{TT}_{\mathbb{S}_2\times \mathbb{S}_1}$
is a sum of Fourier transforms of 1-loop determinants.  The experienced reader could be puzzled about the absence of the VEV of $\sigma$. In our computation, dependence  with no derivative acting on $\sigma$ was completely absorbed in the redefined potential $\hat{A}$. Thenceforth the VEV of $\sigma$ is the $Im[u]$ and in consequence it is annihilated by our reality conditions \eqref{ComplexCond}. In Appendix \eqref{Inte} we show how this path of integration coincides with the Jeffrey-Kirwan (JK) up to the so called "boundary contributions".
 
Equation \eqref{PartitionFunction} together with \eqref{Vector1LoopBulk} and the JK prescription is precisely the result for the 3D TT index obtained  in \citep{BZ} or equivalently the result of 3D TT index integrated along the path \eqref{ComplexCondBZ}. In appendix \ref{VecBZ} we explain how to obtain the vector multiplet one loop determinant along the complex path \eqref{ComplexCondBZ}, which is the only result that could possibly differ with the computation along the real path \eqref{ComplexCond}, because in both cases the localising term (expanded about the relevant BPS saddles) and reality conditions for the matter multiplet are the same and consequently the one loop determinants for matter are the same. Finally, after regularisation, one arrives to \eqref{PartitionFunction} after integration along \eqref{ComplexCond}. We do not repeat the intermediate steps (results) described in the previous sentences because they are the same steps (results) that we have already performed (obtained) in the analysis of integration along the real path \eqref{ComplexCond} in this section to arrive to \eqref{PartitionFunction}-\eqref{Vector1LoopBulk}. In equation \eqref{RealLine} we will summarise how to obtain the integration along the "real path" \eqref{ComplexCond} on $\mathbb{S}_1$ with a point excluded, out of the variational problem {\bf VP-I} and previous results in this section.

Notice that we have selected one block with {\blue even} quantisation and the other one with {\red odd} in such a way the product of 1 loop determinants becomes the 1 loop determinant on $\mathbb{S}_1$. As argued in the vicinity of equation \eqref{AbelianFactor0}, for generic matter gauge representations $\{\rho\}$ the combination ({\blue even} ,{\red odd}) is the only way to obtain the 1 loop determinant on $\mathbb{S}_1$ out of the three possible products ({\blue even}, {\blue even}), ({\red odd}, {\red odd}) and ({\blue even} ,{\red odd}).

\subsection{Localisation of the variational problem \text{ \bf VP-I} }
As motivated in the Introduction, one of our initial goals was to see from scratch whether the result of integration along the complex path \eqref{ComplexCondBZ} coincides with integration along the real path \eqref{ComplexCond} of \eqref{lQ}. In this subsection we focus on the quantisation of the variational problem {\bf VP-I} and as we shall see below the results coincide, but only if when integrating along the real path we include real BPS configurations that are singular on $\mathbb{S}_1$. 

Let us show the aforementioned coincidence by analysing the results we have obtained in this section. The integration along the real path of fields %with $Q_\epsilon$ or $\tilde{Q}_\epsilon$ localising terms
after localising upon non trivial fluxes \eqref{locusN}( resp. \eqref{locusS}) that are real $Q_\epsilon$( resp. $\tilde{Q}_\epsilon$)-BPS configurations and singular on $\mathbb{S}_1$ (regular on $\mathbb{S}_1/\{t=0\}$) is the partition function on a segment of length $t_0=2\pi$, namely
\begin{eqnarray}
Z^{\mathbb{R}}_{TT}&:=&B_{{ \blue even}({\red odd})}(t_0)
,~~~~~~t_0\rightarrow 2 \pi^-\nonumber \\
&:=&\frac{1}{|W|}\sum_{\mathfrak{m}} \int \frac{d^{r}u}{(2\pi)^r} ~  B_{{ \blue even}({\red odd})}(\mathfrak{m},u,2\pi). \label{RealLine}
\end{eqnarray}
The explicit expression of $B_{{ \blue even}({\red odd})}(\mathfrak{m},u,t_0)$ was written down in \eqref{Block2}. Then by using the same regularisation recipe used in \eqref{matfac1loop} we obtain the same result \eqref{Vector1LoopBulk}, as if we had integrated along the complex path \eqref{ComplexCondBZ} or \eqref{ComplexCondBZNew}.

In this way, we have shown that in order to have coincidence between integration along the real path \eqref{ComplexCond} and integration along the complex path \eqref{ComplexCondBZ} we must include in the former (when working with the variational problem {\bf VP-I}), real BPS solutions that do not live on $\mathbb{S}_1$: \eqref{locusN} or \eqref{locusS} with $t_0\rightarrow 2\pi$; but on $\mathbb{S}_1/\{t=0\}$. Namely we are forced to exclude a point out of $\mathbb{S}_1$.

\paragraph{Brief observation}
First, let us summarise what we have deviced in order to obtain sum over fluxes when integrating along the real path:
\begin{itemize}
\item  1) We have localised with different supercharges on both half-circles say $Q_\epsilon$ and $\tilde{Q}_\epsilon$: $I_N=(0,\pi)$ and $I_S=(\pi,2\pi)$ and match fields continuously from on side to the other. In summary, to quantise the variational problem {\bf VP-II} with {\blue even} and {\red odd} KK modes on $I_N$ and $I_S$ respectively.
\item 2) To use {\bf VP-I} and exclude a point out of $\mathbb{S}_1$.
\end{itemize}
To finish up this section we comment about what could be the meaning of the minus sign choice on \eqref{facNew}. In this paper we restrain to make a brief comment, we hope however to come back to these issues in future work. 
Let us end up this section by briefly elaborating upon a singular real BPS configurations we dealt with.  
Let us focus on the factorised representation \eqref{facNew} of $Z^{\mathbb{C}}_{TT}$ with minus sign choice
\begin{eqnarray}
Z^{\mathbb{C}}_{TT}= \frac{1}{|W|}\sum_{\mathfrak{m}_1} \int \frac{d^{r}u_1}{(2\pi)^r} B_{\blue even}(\mathfrak{m}_1,u_1;\pi)B_{\red odd}(-\mathfrak{m}_1, - u_1;\pi). \label{u2}
\end{eqnarray}
We interpret the {\blue even} and {\red odd} blocks \eqref{u2} as coming from the localisation formula on the two halves of $\mathbb{S}_1$, $(0,\pi)$ and $(\pi,2\pi)$, indistinctly, about vacua with flux $\mathfrak{m}_1$ and $-\mathfrak{m}_1$ respectively.  As briefly stated in the paragraph below equation \eqref{OLV}, we can indistinctly use either $Q_\epsilon$ or $\tilde{Q}_\epsilon$ at both halves of $\mathbb{S}_1$ to obtain the {\blue even}/{\red odd} blocks. If we use $Q_\epsilon$ on both half-circles, then following what we have learnt in section \ref{recast},  specifically equation \eqref{locusN}, we can write down the $Q_\epsilon$-BPS vacua we localise about on each half as
  \begin{eqnarray}
\left\{ \begin{array}{cccccc} F_{12 }&=&\frac{\mathfrak{m}_1}{2 R^2}. & ~~~ \sigma&=&- \,\frac{\mathfrak{m}_1}{2 R^2
 }\, t , ~~~~~~ ~~~  0<  t < \pi.\\F_{12}&=&-\frac{\mathfrak{m}_1}{2 R^2} &  ~~~\sigma&= &+ \,\frac{\mathfrak{m}_1}{2 R^2
 } \, (t-2 \pi) , ~~~  \pi<  t < 2\pi. \end{array}\right. \label{'t Hooft}
 \end{eqnarray}
We draw attention to the fact that \eqref{'t Hooft} is a singular solution of the Bogomolnyi equation \eqref{Bogomolnyi}. In fact, \eqref{'t Hooft} is a 't Hooft modified solution to the Bogomolnyi equations on $\SS$ with 't Hooft insertions at $0$ and $\pi$. A 't Hooft insertion/operator, as defined in Section 9 of \citep{KapustinWitten}, induces a discontinuity in the $\mathbb{S}_2$ -magnetic flux at a specific point in $\mathbb{S}_1$. These operators generate singular solutions to the Bogomolnyi equations on $\mathbb{S}_2\times\mathbb{R}$ which are called 't Hooft modified. In \eqref{'t Hooft} the insertion at $t=0$ changes the flux from $\mathfrak{m}_1$ to $-\mathfrak{m}_1$ and the one at $t=\pi$ changes it back from $-\mathfrak{m}_1$ to $\mathfrak{m}_1$. In the conclusions we will mention a potential extension originated out of the latter observation. However, as it is not the scope of the present work, we will not elaborate more on this issue, we leave that to future work.

\section{A toy example}

\subsection{ Mass deformation of $U(N)$ with $N_f=N$ fundamental and $N_a=N$ antifundamental matter multiplets }

As promised, we finalise by exploring the large $N$ behaviour of an example of unitary matrix model in question. We focus on a massive deformation of $U(N)_{k=0}$ with $N_f=N$ chiral-antichiral pairs in the fundamental and $N_a=N$ matter multiplets in the anti fundamental. By massive deformation we mean the presence of a $U(1)$ flavour Wilson line along $\mathbb{S}_1$, $v$, with $Re[v]=0$. This is equivalent to consider massive matter content with mass $Im[v]>0$. As we shall commute the summation $\sum_{\mathfrak{m}}$ with the integration $\int_{JK} d u_i$, we will see that the "positive" chiral-antichiral poles will get relocated after summation is performed upon the integrand.

In this case we have $N$ integrations (for a while we will replace $N_a$ by $N_f$)

\begin{eqnarray}
Z_{U(N)}&=&\sum_{\mathfrak{m}} \int^{2\pi}_{0}\frac{d^N u}{(2 \pi)^N N!}~% \xi^{- \sum_i \mathfrak{m}_i}
\frac{   \prod_{i<j} 4\sin^2{\left(\frac{u_i-u_j}{2}\right)} %e^{i k u\cdot \mathfrak{m}}
}{\prod_{i}\left(-2 i \sin \frac{u_i-v}{2}\right)^{N_f Q^+_i} \left(2 i\sin \frac{u_i+v}{2} \right)^{N_f Q^-_i} }, \nonumber\\
%&=&\sum_{\mathfrak{m}}\frac{1}{N!} \int^{\pi}_{-\pi}\frac{d^N u}{(2 \pi)^N}~\frac{   \prod_{i<j} 4\sin^2{\left(\frac{u_i-u_j}{2}\right)} e^{i k (u+\pi)\cdot \mathfrak{m}}}{\prod_{i}\left(-2 i \cos \frac{u_i-v}{2}\right)^{N_F Q^+_i} \left(2 i\cos \frac{u_i+v}{2} \right)^{N_F Q^-_i} } \\
%&=&\sum_{\mathfrak{m}}\frac{1}{N!} \int^{2\pi}_{0}\frac{d^N u}{(2 \pi)^N}~\frac{   \prod_{i<j} 4\sin^2{\left(\frac{u_i-u_j}{2}\right)} e^{i k u\cdot \mathfrak{m}}}{\prod_{i}\left(4  \sin \frac{u_i-v}{2} \sin \frac{u_i+v}{2} \right)^{N_F Q_\mathfrak{n}} } \prod_i\left(\frac{\sin \frac{u_i+v}{2}}{\sin \frac{u_i-v}{2}}\right)^{N_F \mathfrak{m_i}}
%\\
\end{eqnarray}
where $Q^\pm_i:=\pm \mathfrak{m}_i+\mathfrak{n}-q_R+1$. % The topological $U(1)$ Wilson line $\xi$ has been redefined as $\xi\rightarrow (-1)^{N_F} \xi$. In the limit $\xi \rightarrow \infty$ the geometric series converges uniformly inside the contour of integration $S_1$ as its radius of convergence goes with $\xi$, \citep{}. We shall assume $k=0$.
 The indices $\mathfrak{m}_i$ are summed over the region
$\mathfrak{m}_i \geq -\mathfrak{n} + q_R$. This is the region where the positive charges have poles. After solving the geometric series one obtains

\begin{eqnarray}
&=& \int^{2\pi}_{0}\frac{d^N u}{(2 \pi)^N N!}~\frac{   \prod_{i<j} 4\sin^2{\left(\frac{u_i-u_j}{2}\right)}} %e^{i k u\cdot \mathfrak{m}}
{\prod_{i}\left(4 \sin \frac{u_i+v}{2} \right)^{N_f \left(2 \mathfrak{n}-2q_R+1\right)} \left(% \xi
 \left(\sin{ \frac{u_i-v}{2}}\right)^{N_f} - \left(\sin \frac{u_i+v}{2}\right)^{N_f}\right)}  .  \nonumber\\
\end{eqnarray}

It is convenient to change variables to $z_i := e^{i u_i}$. Thence

\begin{eqnarray}
Z_{U(N)}%&=&\sum_{\mathfrak{m}}\frac{e^{-i N_F N v(\mathfrak{n}-q_R+1)}}{N!}\oint_{S_1} \left(\frac{d z_i}{2 \pi i}\right)^N (-1)^{N_F \sum_i \mathfrak{m}_i}\prod_{i<j} (z_i-z_j)^2  \prod_i  \frac{z_i^{\bigg(N_f (\mathfrak{n}-q_R+1)-N%+k \mathfrak{m}_i
%\bigg)}}{  \left(1-z_i e^{i v} \right)^{N_f Q^-_i } \left(z_i-e^{i v}\right)^{N_f Q^+_i}} \nonumber \\
&=& \frac{e^{-i N_f N v(\mathfrak{n}-q_R+1)}}{N!}\times\nonumber\\ &&\oint_{S_1}  \left(\frac{d z_i}{2 \pi i}\right)^N \frac{\prod_{i<j} (z_i-z_j)^2}{\prod_i \left(1-z_i e^{i v} \right)^{N_f \left(2 \mathfrak{n}-2q_R+1\right) }}  \prod_i  \frac{z_i^{\bigg(N_f (\mathfrak{n}-q_R+1)-N%+k \mathfrak{m}_i
\bigg)}}{%\left(%\xi
% \left(z_i-e^{i v}\right)^{N_f }-\left(z_i e^{i v}-1 \right)^{N_f }\right)
P_{N_f}(z_i)
}, \nonumber\\ \label{PFunc2}
\end{eqnarray}
with
\begin{eqnarray}
P_{N_f}(z):=\left( \left(z-e^{i v}\right)^{N_f}-\left(z e^{i v}-1 \right)^{N_f }\right)= (1- e^{i N_f v}) \prod^{N_f}_{a=1} (z -z_{(a)}), ~ \label{Poli}
\end{eqnarray}
and $N_f=N$. Ignoring the potential poles at $z_i=0$ in the integrand of \eqref{PFunc2} (as in \citep{BZ} for $k=0$) and from the Cauchy theorem, it comes out

\begin{eqnarray}
Z_{U(N)}=\frac{e^{-i N^2 v(\mathfrak{n}-q_R+1)}}{(1-e^{i N v})^N} \frac{%\prod^N_{a \neq b=1} | z_{(a)} -z_{(b)}|
  \prod^N_{a=1} z_{(a)}^{N (\mathfrak{n}-q_R)}}{\prod^N_{a=1} \left(1-z_{(a)} e^{i v} \right)^{N \left(2 \mathfrak{n}-2q_R+1\right) }}. \label{PFunc3}
\end{eqnarray}

Where $z_(a)$ with $a=1\, \ldots, N$ are the $N$ roots of \eqref{Poli} \footnote{In the cancellation of the Vandermonde determinant there comes out a phase that we have implicitly absorbed with the topological phase. }.
Is easy to show that the partition function \eqref{PFunc3} can be written as
\begin{eqnarray}
Z_{U(N)}:=e^{ N^2 F[\mathfrak{n},v]}, \label{FEn0}
\end{eqnarray}
where
\begin{eqnarray}
F[\mathfrak{n},v]&:=&F_1(N)%-F_1(N)
+(2 \mathfrak{n}-2 q_R+1) F_2(N)- (\mathfrak{n}-q_R) F_3(N),\label{FEn1} \\\nonumber
\end{eqnarray}
\begin{eqnarray}
F_1(N)&:=& - i v (\mathfrak{n}-q_R+1)- \log{ (1-e^{i N v})^{\frac{1}{N}}},
\\
%F_1(N) &:=&  -\int_{\mathcal{C}} d z^\prime_1 dz_\prime_2 \rho(z^\prime_1) \rho(z^\prime_2) \log |z^\prime_1-z^\prime_2|\\
F_2(N) &:=&- \int_{\mathcal{C}} \rho(z^\prime) \log (1-z^\prime e^{i v}), \\
 F_3(N)&:=&- \int_{\mathcal{C}}dz^\prime \rho(z^\prime)\log z^\prime,
  \label{FreeE}
\end{eqnarray}
and
\begin{eqnarray}
\rho(z):= \frac{1}{N} \sum^{N}_{a=1} \delta(z-z_{a}), ~~ \int_\mathcal{C} \rho=1.
\end{eqnarray}
 $\rho$ is the density of zeroes of the polynomial $P(z)$. The path $\mathcal{C}$ is the support of $\rho$ in the complex plane $z$. We comment next on the support $\mathcal{C}$ for large values of $N$.

For  complex flavour lines with $Re[v]=0$ ($\eta:=e^{i v} \in \mathbb{R}$), the polynomial $P(z)$ is self-reciprocal, namely
\begin{eqnarray}
P(z)=(-1)^{N+1} z^N P(\frac{1}{z}).
\end{eqnarray}
%It is not odd to find this kind of polynomials in physics. For instance they appear in the study of the ferromagnetic Ising model \citep{}.
It is theorem that such polynomials have either all their zeroes at $S_1: |z|=1$ or they distribute symmetrically with respect to $S_1$. In the former case, the necessary and sufficient condition is that $P^\prime(z)$ must have its zeroes in the interior of $S_1$ or onto $S_1$. Here will not attempt to use these analytical facts, we will just try a numerical exploration of this toy model. For a nice read on a recent perspective on the topic of self-reciprocal polynomials and their zeroes, please refer to \citep{Suzuki}.

\begin{figure}[ht!]
\centering
\includegraphics[width=70mm]{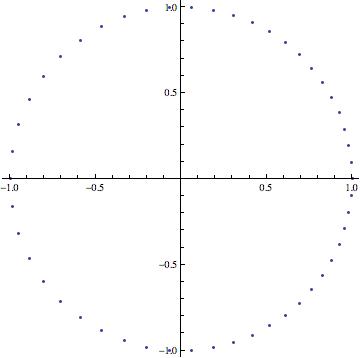}
\includegraphics[width=70mm]{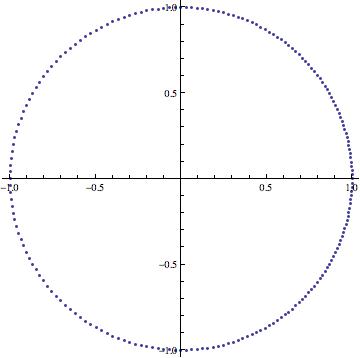}
\caption{The $N$ roots of $P_N(z)$ for $\eta=e^{i v}=\frac{1}{8}$. For $N$ equal $50$ (Left) and 200 (Right) they are on $S_1$. \label{Fig1}}
\end{figure}

 In figure \ref{Fig1} we show the $N$ zeroes of $P_N(z)$ in the complex plane $z$, for $\eta=\frac{1}{8}$ and $N=50$ and $N=200$ respectively. We have stopped at $N=200$ because beyond this value the numerical result provided by Mathematica was not trustable enough. In the case of small $\eta<1$ we have always found that the support of zeroes of $P_N(z)$ is $\mathcal{C}=S_1$.

\begin{figure}[ht!]
\centering
\includegraphics[width=90mm]{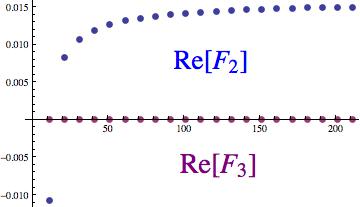}
\caption{The real part of the components, $F_2(N)$ and $F_3(N)$ of the free energy \eqref{FEn1} for  $N$ ranging between $11$ and 211 at step 10.  \label{Fig2}}
\end{figure}

We have been able to check numerically that generically for $\eta<1$, as shown in figure \ref{Fig2} for the case $\eta=\frac{1}{8}$, the real parts of $F_1$, $F_2$ and $F_3$ saturate for large enough $N$. The smaller the $\eta$, the larger the mass, the larger the maximal $N$ we were able to explore in each case. %Thenceforth our semi-analytic study has led us to observe that in our toy example, a massive deformation with  $\eta=e^{i v}=\frac{1}{8}$,
In summary
\begin{eqnarray}
Re[\log Z]|_{\eta =\frac{1}{8}}  \sim  N^2 \left(\mathfrak{n}+ \dots \right),\label{ReLogZ}
\end{eqnarray}
in contradistinction to the twisted ABJM case studied in \citep{BZ2, BZ3} but a priori expected from comparison with similar hermitian models \citep{Marino, Review}. The dots in \eqref{ReLogZ} represent terms without dependence on the magnetic fluxes $\mathfrak{n}$.  Specifically for $\mathfrak{n}=0$ and generic choice for $q_R$ there remains a universal contribution proportional to $N^2$, as straightforward to infer out of \eqref{FEn0} and \eqref{FEn1}.
\section{Summary}
The main scope of this work was to find an alternative path integral representation to justify the presence of the sum over fluxes $\mathfrak{m}$ in the topologically twisted index of \citep{BZ} upon integration over real fields and without sacrificing positive definiteness of the bosonic part of the localising action. %In this way one works with a positive definite bosonic localising part. %The magnetic fluxes \eqref{flux} were not considered to be part of the real locus in \citep{BZ}. To have such fluxes into a locus the authors relaxed the reality conditions of the auxiliary field $D$. In this way the configurations \eqref{flux} + \eqref{D} become part of the locus of both supercharges $Q$ and $\tilde{Q}$.
% In \citep{BZ} it was shown that, even though \eqref{flux} + \eqref{D} is not a saddle point when real conditions are imposed upon the integrated fields ("real path"), it does contribute to the integration along the "real path", %. To argue the contribution, the authors noticed the consequences
%  due to the integration over the super manifold of zero modes $(u_0, \bar{u}_0, \lambda_0, \bar{\lambda}_0, D_0)$.

  An important point we stress upon,  is that on the real path, the space of zero locus solutions of the localising term used in \citep{BZ} is defined by the Bogomolnyi equations. On $\mathbb{S}_2 \times \mathbb{S}_1$ the set of smooth solutions to these equations is given by the set of flat connections on $\mathbb{S}_2$ --which is trivial -- and constant values for $\sigma$ and $A_t$. In approaching the problem with the same supercharge all along $\mathbb{S}_1$  --namely following the approach that we have called {\bf VP-I}--,  one needs to exclude a point out of the $\mathbb{S}_1$ in order to consider solutions with fluxes on $\mathbb{S}_2$, or equivalently, to consider real BPS vacua which are singular on $\mathbb{S}_1$. In the latter case, we have checked that the expression for the 3D TT Index of \citep{BZ} is recovered after integration over the real path of fields by following the usual localisation method and without the need to sacrifice positive definiteness of the bosonic part of the action. We have shown the mechanism behind making these singular solutions part of the vanishing locus of the corresponding positive definite $Q_\epsilon (\tilde{Q}_\epsilon)V$ localising terms -- on the "real" path (as explained in the introduction)--. 

  Following the same line of reasoning, we constructed and integrated out the localising term with positive definite bosonic part along the complex path of integration, reproducing in alternative manner the result of \citep{BZ}.

  We have found the $3D$ TT Index on $\mathbb{S}_2$ times an open segment by carefully managing total derivatives. We have checked that the 3D TT Index on $\SS$ of \citep{BZ} can be factorised into two blocks:  the partition functions of $3D$ $\mathcal{N}=2$ $Q_\epsilon$ TT theory on $\mathbb{S}_2\times (0,\pi)$ and $3D$ $\mathcal{N}=2$ $\tilde{Q}_\epsilon$ TT theory on $\mathbb{S}_2\times (\pi,2\pi)$, respectively, upon localisation of the variational problem {\bf VP-II}.  At the level of one loop determinants, the matching condition \eqref{MatchingCond} must be enlarged with the selection of complementary sets of $\mathbb{S}_1$ KK modes on $(0,\pi)$ and $(\pi,2 \pi)$. We have dubbed such modes as {\blue even} and {\red odd} for obvious reasons that were explained in subsection \ref{subsec}. 
  
  The selection of a  glueing prescription among BPS vacua from one side $(0,\pi)$ to the other $(\pi,2\pi)$, is specified by the matching condition \eqref{MatchingCond} and implies the identification of fluxes from one side to the other. The selection of $+$ sign in \eqref{facNew} corresponds to localising the variational problem {\bf VP-II} -- mentioned in the previous paragraph --, specifically to the matching conditions \eqref{MatchingCond}. %, if we select the minus sign in the kernel \eqref{facNew} the "effective" zero locus BPS solution on $\SS$, can  be identified with a solution to the $Q_\epsilon$(or $\tilde{Q}_\epsilon$) BPS Bogomolny equations on $\SS$ with 't Hooft operators inserted at the junction points $t=0$ and $t=\pi$ (For instance, as written in \eqref{'t Hooft}). By "effective" zero locus BPS solution on $\SS$ we meant the union of the BPS solution used at $(0,\pi)$ and $(\pi,2\pi)$. There are infinitely many more singular solutions to the BPS equations on $\SS$ than the one linked to the 3D TT index. A natural question that comes to mind is what could be the relevance of such solutions from the holographic dual viewpoint. 
 Instead, the choice $-$ on \eqref{facNew} seems to be related to allowing 't Hooft modified solutions of the Bogomolnyi equations. An interesting point to pursue in the future, could be to study the dual description of theories that allow more general 't Hooft modified BPS vacua. The aforementioned generalisations, are deformations of the 3D TT index and it would be interesting to interpret the presence of such domain walls or 't Hooft deformations %\footnote{Perhaps an analysis inspired by the work done by the authors in \citep{Atish} could be pursued in the long run.}
%in relation to domain walls in a different context,
in the holographic dual set up. It would be interesting to provide an interpretation of the latter generalisations, from the point of view of microstate counting or deformations of $AdS_4$ supersymmetric black holes \citep{BZ2,BZ3,BZ4,Gnecchi}. Those interesting lines of analysis will be pursued elsewhere.

\section*{Acknowledgments}

The author thanks Francesco Benini for his comments and suggestions, Victor I. Giraldo Rivera and Lorenzo di Pietro for their careful reading and critics on a previous version of the manuscript and to Leopoldo A. Pando Zayas,  Junya Yagi, M. Nouman Muteeb and Antonio Sciarappa for useful conversations at different stages of this work.   Thanks a lot to an anonymous referee for critics and suggestions and to Massimo Porrati and Sameer Murthy for drawing my attention to a subtle and important point. A grateful acknowledgement to The Abdus Salam ICTP, scientific and administrative staffs, for their kind hospitality during a couple of months of work. The work of the author is supported by CONICET.

 \appendix

 \section{Geometry and Conventions} \label{conventions}

 In \citep{BZ} it was proposed a formula for the partition function %of an $N=2$
of a gauge theory with two supercharges $(Q, \tilde{Q})$, with the same R-charge, on $\mathbb{S}_2\times \mathbb{S}_1$
\begin{eqnarray}
ds^2= R^2(d\theta^2+\sin{\theta}^2 d\phi^2)+dt^2 \nonumber \\
\eta_{a b}=Diag(1,1,1), ~ e^1_\theta= R, ~~ e^2_\phi= R \sin \theta, ~~ e^3_t = 1.
\label{metric}\end{eqnarray}
$0 \leq \theta < \pi$, $0 \leq \phi, ~t <2\pi$ and $t \sim t+2\pi$, and type A semi-topological twisting along $\mathbb{S}_2$. Namely with a background $U(1)_R$ gauge potential
\begin{eqnarray}
V_\mu&:=& \mp \frac{1}{2} \omega^{21}_\phi \delta^\phi_\mu, \label{vmu0}
%\\&=&\mp \frac{u}{\sqrt{1-u^2}} \delta^x_\mu&\\&=&\pm \frac{1}{2 y} \delta^x_\mu.
\end{eqnarray}
with $\omega$ being the torsionless spin connection of \eqref{metric}.

 The supercharges $Q$ and  $\tilde{Q}$ are parameterised by two killing spinors (KS) $\epsilon$ and $\tilde{\epsilon}$ with R-charge $-1$. The KS satisfy the following equations
\begin{eqnarray}\nonumber\\
\mathcal{D}_\mu \epsilon &:=&\left(\partial_\mu +\frac{1}{4}\gamma_{M N} \omega^{M N}_{~~\mu}- i V_\mu\right)\epsilon=0, ~ \mathcal{D}_\mu \tilde{\epsilon}=0.
\label{KSpinor}\\\nonumber
\end{eqnarray}
where $\gamma_{M N}:=\frac{[\sigma_M, \sigma_N]}{2}$ and $\sigma_M$ are the Pauli matrices.
Under \eqref{vmu0} the solution to the KS equations are  $\epsilon=\tilde{\epsilon}=\epsilon_{\pm}$ with $\epsilon_{\pm}$ a constant eigenspinor of $\omega_\phi^{a b}\gamma_{ab}=2 \omega_\phi^{21}\gamma_{21}=- 2 \,i \,\cos{ \theta } \sigma_3$, with eigenvalue $\pm1$ respectively
\begin{eqnarray}
\sigma_3 \epsilon_\pm = \pm  \epsilon_\pm. \label{ksform}
\end{eqnarray}

 We will represent flat and curved indices by numbers $(1, 2, 3)$
and greek letters $(\theta, ~\phi,~ t)$ respectively. Our convention for spinor bilinear are summarised in the following set of definitions
\begin{eqnarray}
\Xi^\dagger \Psi := {\Xi^\dagger}^{\beta}C_{\beta \alpha } \Psi^\alpha% =%\left\{
	%\begin{array}{clcl}
		%\Psi^\dagger \Xi & \text{ commuting spinors} \\
		%-\Psi^\dagger \Xi & \text{ anticommuting spinors }\end{array}
%\right.
 \nonumber\\
~~ \text{          where       }  ~~ {\Xi^\dagger}^\beta := \left(C_{\beta \gamma} \Xi^\gamma\right)^* \text{  and  } C_{\beta \gamma}:= i \gamma_{2}= \left( \begin{array}{c c} 0 & 1 \\-1 & 0  \end{array}\right).\\\nonumber
\end{eqnarray}

 %We stress $\Xi^{\dagger}$-Ours $ \equiv {\Xi^c}^{\dagger}$-Benini-Zaffaroni. Notational tip: Our SUSY parameter spinors will be commuting. The spinor physical fields are anti commuting.

 This is the same definition written down in equation A.2 of  \citep{Closset}. % http://arxiv.org/pdf/1212.3388v3.pdf.
  By complex conjugation $*$, we mean transconjugation on the gauge algebra representation.
As already mentioned in the bulk of the manuscript, the transformation parameter spinors $\epsilon, \bar{\epsilon}$ are commuting. Physical spinor fields are anti commuting, hence transformation parameter spinors and physical spinor fields commute. For self completeness we write down the following set of identities
\begin{eqnarray}
~\chi \cdot \epsilon :=\chi_\beta \epsilon^\beta, ~ \epsilon_\beta= \epsilon^\alpha C_{\alpha \beta},\nonumber \\~ \epsilon ^\alpha=C^{\alpha \beta} \epsilon_\beta,~ C^{\alpha \beta} C_{\alpha \beta}=2, ~ C^{-1}=C^T=-C . \nonumber\\ C \gamma_\mu C^{-1}=-\gamma_\mu^T, ~ \gamma^{\mu \nu}:= i \epsilon^{\mu \nu}_{~~\beta} \gamma^\beta.\\\nonumber
\end{eqnarray}
As for the covariant derivatives
\begin{eqnarray} \nonumber\\
\mathcal{D}_\mu:=\left\{
	\begin{array}{clcl}
		\left(\,\partial_\mu+\frac{1}{4}\gamma_{M N} \omega^{M N}_{~~\mu}-i A_\mu \,- \, i q_R V_\mu \right) & \text{ on spinors} \\
		\left(\,\partial_\mu-i A_\mu \, - \, i q_R V_\mu \right) & \text{ on scalars }\end{array}.
\right. \label{CovDer} \\ \nonumber\end{eqnarray}
In due time, the spin connection shall be absorbed in the $U(1)_R$ magnetic flux term by a redefinition of $R$-charge.  The difference between the effective $R$-charge felt by a "spin up" and "down" chiral spinor, like for instance the one between the components of the gaugino $\lambda_1$ and $\lambda_2$,  will be $2$. 

The gauge field strength components
\begin{eqnarray}\nonumber\\
F_{\mu \nu}:=\partial_\mu A_\nu-\partial_\nu A_\mu -i [A_{\mu}, A_{\nu}].  \\\nonumber
\end{eqnarray}

 \subsection{Gauge Multiplet}\label{loft}

% In this subsection we consider $(A_t, A_\theta, A_\phi,\sigma, D)$ to be hermitian.
In this appendix in essence we will follow the early literature on the topic, specifically \citep{Kapustin, Alday, Cadabra}. %$A_t$ will not be considered hermitian but $A_t+i \sigma$ do. The latter of the conditions on $A_t$ implies that $[Im[A_\mu], \sigma]=0$. As shown in the main body of the paper, we will gauge away $A_t$. Thence, a posteriori, $A_t$ will not be a physical degree of freedom.
%To perform or check some of these computations both Mathematica and  Cadabra \citep{Cadabra} were used at some point. 
Let the covariant derivative on the gaugini be defined as
 \begin{eqnarray} \nonumber\\
 \mathcal{D}_\mu \lambda&=&\left(\partial_\mu+\frac{1}{4}\gamma_{M N} \omega^{M N}_{~~\mu}-i [A_\mu,\circ] -i  V_\mu\right) \lambda \nonumber \\
\mathcal{D}_\mu \bar{\lambda}&=&\left(\partial_\mu+\frac{1}{4}\gamma_{M N} \omega^{M N}_{~~\mu}-i [A^*_\mu, \circ] -i V^*_\mu\right)\bar{\lambda}. \\\nonumber
 \end{eqnarray}
As mentioned before, the spin connection shall be absorbed in the $U(1)_R$ magnetic flux term by a redefinition of $R$-charge.  The difference between the effective $R$-charge felt by a "spin up" and "down" chiral spinor, like for instance the components of the gaugino $\lambda_1$ and $\lambda_2$,  will be $2$.

In our conventions the $\gamma$'s are the Pauli matrices are hermitian. Let us define the object

\begin{eqnarray}
\overset{\bullet}{Q_{\epsilon} \lambda}%&=&-\frac{1}{2} F_{\mu \nu}^* \epsilon^\dagger\gamma^{\mu \nu} - D^* \epsilon^\dagger-i (\mathcal{D}_\mu\sigma)^* \epsilon^\dagger \gamma^\mu\\
&:=&-\frac{1}{2} F_{\mu \nu} \epsilon^\dagger\gamma^{\mu \nu} +D \epsilon^\dagger+i (\mathcal{D}_\mu\sigma) \epsilon^\dagger \gamma^\mu, \label{eqbullet} \\ \nonumber
\end{eqnarray}
that will be part of the definition of localising term
\begin{eqnarray}\nonumber\\
Q_\epsilon \left( \left(\overset{\bullet}{Q_\epsilon \lambda}\right) \lambda \right).\label{locTApp}\\\nonumber
\end{eqnarray}
We stress, that the $\bullet$ must not be confused with the $\dagger$ operation. In fact $(Q_\epsilon \lambda)^\dagger$ is very similar to $\overset{\bullet}{(Q_\epsilon \lambda)}$ (the difference being for complexified fields). The former is obtained from the latter by substituting $A_\mu$, $D$ and $\sigma$ in \eqref{eqbullet} by their complex conjugated $A_\mu^*$, $D^*$ and $\sigma^*$.

We stress that the object $\overset{\bullet}{(Q_\epsilon \lambda)}$ defined in Subsection \ref{recast} and the one defined in this subsection are no the same; their difference is
\begin{eqnarray}
\eqref{eqbullet}-\eqref{eqbullet0}=2 i (\mathcal{D}_\mu\sigma) \epsilon^\dagger \gamma^\mu. \label{mismatch}
\end{eqnarray}
This difference arises because in \eqref{eqbullet0} we substitute $\hat{A}_3^*:=A_3^*-i \sigma^*$ and $\sigma^*$ by
$\hat{A}_3=A_3+i \sigma$ and $\sigma$ respectively. Meanwhile in \eqref{eqbullet} we substitute $A_3^*$ and $\sigma^*$ by $A_3$ and $\sigma$ respectively%. The the former transformation contradicts the latter. This incongruence is the cause of the mismatch. 
We apologise for the abuse of notation.

From \eqref{mismatch} we conclude that the difference between the $Q_\epsilon$ exact terms
\begin{eqnarray}
\eqref{locTApp}- \eqref{locV}=Q_\epsilon\bigg(2 i (\mathcal{D}_\mu\sigma) \epsilon^\dagger \gamma^\mu\lambda\bigg),
\end{eqnarray}
is again $Q_\epsilon$ exact and not only $Q_\epsilon$ closed. After continuing the computation of the bosonic and fermionic parts with \eqref{locV} instead of \eqref{locTApp} and by using the covariant form of the algebra \eqref{algebra}, we arrive to the same result obtained in subsection \ref{recast}, where after using \eqref{effPot} we obtained \eqref{QVvec2} and \eqref{QVectorStuckelberg}. Instead of reporting the previously mentioned computation, in this subsection we report the computation that follows from the use of \eqref{locTApp}. As a nice difference we will obtain a covariant localising term in contradistinction with the analog results \eqref{QVvec2} and \eqref{QVectorStuckelberg}. However, as shall be seen in a while, this new term is not positive definite under the reality conditions \eqref{ComplexCond}. Instead, it will be positive definite under the more orthodox reality conditions

\begin{eqnarray}
A_\theta= A_\theta^*, ~ A_\phi= A_\phi^*, ~ {\red A_t=A_t^*},%Im[A_t]=  -\sigma  ,
~ \sigma=\sigma^*, ~ D=D^*.% ~ \lambda=\lambda^*, ~\bar{\lambda}=\bar{\lambda}^*.
\label{ComplexCondApp}
\end{eqnarray}

In the results to come, the reader will find the killing vector $v$, which is defined as

\begin{eqnarray}
v := \epsilon^\dagger \gamma^\mu \epsilon \partial_\mu= \partial_{t}.
\end{eqnarray}

Upon expanding \eqref{locTApp} with the covariant form of the algebra \eqref{algebra}, we obtain

\begin{eqnarray}
Tr\left( (\overset{\bullet}{Q_{\epsilon} \lambda}) Q_{\epsilon} \lambda\right)&=& \epsilon^\dagger \cdot \epsilon ~Tr \bigg[ \left ( \star F _\mu+ %i
 \mathcal{D}_\mu \sigma\right )^2 +D^2\bigg], \label{BosAc} \\
Tr\left( Q_{\epsilon} (\overset{\bullet}{Q_{\epsilon} \lambda})   \lambda \right) &=& \epsilon^\dagger \cdot \epsilon~ Tr \bigg(-i (\mathcal{D}_\mu\bar{\lambda})^\dagger \gamma^\mu \lambda +i [\bar{\lambda}^\dagger,\sigma] \lambda \bigg), \\\nonumber
\end{eqnarray}
with $ \star F _\mu := \frac{1}{2} \epsilon_\mu^{~\nu \beta} F_{\nu \beta}$.
\eqref{BosAc} is positive definite under the reality conditions \eqref{ComplexCondApp}.
%Notice that in this case it comes out $|D|^2$ not $D^2$. We have supposed $A_\mu$ and $\sigma$ to be real.
%After long but straightforward computation, the total SUSY variation of $Q_{\epsilon}((Q_{\epsilon}\lambda)^\dagger \lambda)$ \footnote{Notice this is not the same as $Q_{\epsilon}Q_{\epsilon}(i \lambda^\dagger \lambda)\equiv 0$.} is shown to vanish identically. Namely
%\begin{eqnarray}
%Q_{\epsilon}\left(Q_{\epsilon}((Q_{\epsilon}\lambda)^\dagger \lambda)\right) \equiv 0.
%\end{eqnarray}
%\begin{eqnarray}\nonumber
%\\
%\epsilon^\dagger \epsilon {D}^{\mu}{\left( - i \bar{\lambda} {\gamma}^{\nu} \epsilon {F}_{\mu \nu} + \frac{1}{2}\, i {\epsilon}_{\mu}^{ ~~\nu \beta} \bar{\lambda} \epsilon {F}_{\nu \beta} - i \bar{\lambda} {\gamma}_{\mu} \epsilon D - \bar{\lambda} \epsilon {D}_{\mu}\,  \sigma - {\epsilon}_{\mu}^{~~ \nu \beta} \bar{\lambda} {\gamma}_{\beta} \epsilon {D}_{\nu}\,  \sigma\right)}\, \nonumber\\
%\end{eqnarray}
Finally

\begin{eqnarray}
\mathcal{L}_{YM}&=&Tr\left[\frac{1}{2} F^2+\mathcal{D}_\mu \sigma^2+D^2+i \bar{\lambda}^\dagger \gamma^\mu \mathcal{D}_\mu \lambda +i \bar{\lambda}^\dagger [\sigma,\lambda]\right]\label{YMaction}\\
&=& Tr\left[  Q_{\epsilon}\left( (\overset{\bullet}{Q_{\epsilon} \lambda}) \lambda \right)- \epsilon^\dagger\cdot \epsilon ~ \mathcal{D}_\mu\left(2 ~(*F)^\mu \sigma+ i \bar{\lambda}^\dagger \gamma^\mu \lambda\right) \right],
\end{eqnarray}
The boundary term is congruent with the equivalent C.6 in \citep{Drukker}. In this case the localising action is
\begin{eqnarray}
Q_\epsilon V_{Vector}\equiv \mathcal{L}_{SYM}+Tr[ \mathcal{D}_\mu\left(2 (*F)^\mu \sigma+ i \bar{\lambda}^\dagger \gamma^\mu \lambda\right)] =Q_{\epsilon}\left( (\overset{\bullet}{Q_{\epsilon} \lambda}) \lambda \right). \label{QVector} \\\nonumber
\end{eqnarray}
From \eqref{BosAc} is straightforward to see that this localisation action vanishes evaluated at the configurations $\sigma^-_{(0)}$ \eqref{sigma}.

%is supersymmetric in a manifold with boundaries.

%\begin{eqnarray}
%Q_{\epsilon}\left(Q_{\epsilon}\left( ({Q_{\epsilon}\lambda})^\dagger \lambda \right)\right)=0%v^\mu \mathcal{D}_\mu \left(\right)
%\end{eqnarray}

\subsection{Matter multipets}
\label{Matter}
%Let the complexified gauge transformations be
%\begin{eqnarray}
%\delta_{gauge} \phi &=& i \Lambda^{a} T_a \phi \nonumber
%\\
%\delta_{gauge}\bar{\phi} &=& i \Lambda^{a}%^*
%  T_{a}
  %^*
%   \bar{\phi},\end{eqnarray}
%where $T_a$ is not hermitian for the moment.
Let the covariant derivatives be
\begin{eqnarray}
\mathcal{D}_\mu \phi&=& \left(\partial_\mu-i A_\mu -i q V_\mu -i W_\mu \right) \phi, \nonumber \\
\mathcal{D}_\mu \bar{\phi}^\dagger&=& \left(\partial_\mu+i A%^*
_\mu +i q V%^*
_\mu+i W%^*
_\mu\right) \bar{\phi}^\dagger, \nonumber\\
\mathcal{D}_\mu \psi&=&\left(\partial_\mu+\frac{1}{4}\gamma_{M N} \omega^{M N}_{~~\mu}-i A_\mu -i (q-1) V_\mu- i W_\mu \right)\psi, \nonumber \\
\mathcal{D}_\mu \bar{\psi}^\dagger&=&\left(\partial_\mu+\frac{1}{4}\gamma_{M N} \omega^{M N}_{~~\mu}+i A%^*
_\mu +i (q-1) V%^*
_\mu+i W
%^*
_{\mu}\right)\bar{\psi}^\dagger, 
\end{eqnarray}
where $W_\mu$ is a background flavour connection.

Let us split the orthodox localising Lagrangian for matter \eqref{LOCMATORT} into parts $\mathcal{I}$ and $\mathcal{II}$,   $\mathcal{L}:=\mathcal{L}^{\mathcal{I}}+\mathcal{L}^{\mathcal{II}}$ where
\begin{eqnarray}
\mathcal{L}^{\mathcal{I}}:= Q_\epsilon V^{\mathcal{I}},  ~~  \mathcal{L}^{\mathcal{II}}:= Q_\epsilon V^{\mathcal{II}},
\end{eqnarray}
with
\begin{eqnarray}
V^{\mathcal{I}}&:=& i \,\epsilon \gamma^\mu \psi \mathcal{D}_\mu \bar{\phi}^\dagger+F \bar{\psi}^\dagger \epsilon^c+ i \, \bar{\phi}^\dagger \sigma\, \epsilon\psi, \\
V^{\mathcal{II}}&:=&i \bar{\phi}^\dagger \,\epsilon\lambda\, \phi, \\ V^{\mathcal{I}}+V^{\mathcal{II}}&=&V_{\text{ Eq. } \eqref{MatterAc}}+ 2\, i \, \bar{\phi}^\dagger \sigma\, \epsilon\psi. \label{EqRefNew}
\end{eqnarray}
 Expanding this localising terms, we get the following bosonic and fermionic terms

\begin{eqnarray}
\mathcal{L}_{bosonic}^{\mathcal{I}}&=&   {D}^{\mu}{\bar{\phi} }^\dagger  \, {D}_{\mu}{\phi} + i \, \epsilon^{\mu \nu}_{~~\beta}v^{\beta} {D}_{\mu}\bar{\phi}^\dagger {D}_{\nu} \phi
\nonumber\\&& +v^\mu ({D}_{\mu}{\bar{\phi})^\dagger} (\sigma \phi) \,  + v^\mu  (\bar{\phi}^\dagger \sigma) ({D}_{\mu}{\phi})  + (\bar{\phi}^\dagger \sigma)(\sigma \phi) +\bar{F}^\dagger F,  \label{LBOS1}
\end{eqnarray}

\begin{eqnarray}
\mathcal{L}_{fermionic}^{\mathcal{I}}&=&i\, \bar{\psi}^\dagger {\gamma}^{\mu} {D}_{\mu}{\psi}-i \bar{\psi}^\dagger \sigma \psi\,-i \, %\frac{1}{2}
\bar{\phi}^\dagger\, \bar{\lambda}^\dagger \psi %+ \frac{i}{2} \bar{\phi}^\dagger\,  \bar{\lambda}^\dagger \slashed{v} \psi
-\,  \bar{\psi}^\dagger P^- \lambda \,\phi - i\,\mathcal{D}_\mu \bigg( \bar{\psi}^\dagger P^+\gamma^\mu\psi    \bigg) . \label{LFER1}
\end{eqnarray}

\begin{eqnarray}
\mathcal{L}^{\mathcal{II}}_{bosonic}%=- 2 i \, \bar{\phi}^\dagger\, Q_{\epsilon} \left( \overset{\bullet}{Q_{\epsilon}\sigma} \right)\phi
&=& \frac{1}{2} \epsilon^{\mu \nu}_{~~\beta} v^\beta\, \bar{\phi}^\dagger \, F_{\mu \nu}\, \phi
+i \bar{\phi}^\dagger D \phi+ v^\mu \bar{\phi}^\dagger(\mathcal{D}_\mu \sigma)\phi \label{LBOS2}\\&=& \bar{\phi}^\dagger \bigg( (*F_\mu+\mathcal{D}_\mu \sigma)v^\mu+i D \bigg) \phi \nonumber
\\ &&\\
\mathcal{L}^{\mathcal{II}}_{fermionic}%= -2 i\, Q_{\epsilon} \bar{\phi}^\dagger \, \overset{\bullet}{Q_{\epsilon}\sigma} \, \phi
&=& -i \bar{\psi}^\dagger \epsilon \, \epsilon^\dagger \lambda \phi = -i\,  \bar{\psi}^\dagger P^+ \lambda \,\phi \\ \nonumber\label{LFER2}
\end{eqnarray}
The second term in \eqref{LBOS1} and the first in \eqref{LBOS2} combine to give
\begin{eqnarray}
\mathcal{D}_{\mu}\left( i \, \epsilon^{\mu \nu}_{~~\beta}v^{\beta} \bar{\phi}^\dagger {D}_{\nu} \phi\right)-  \, \epsilon^{\mu \nu}_{~~\beta}v^{\beta} \bar{\phi}^\dagger \left(q V_{\mu \nu}+W_{\mu \nu}\right)\phi,
\end{eqnarray}
after using
\begin{eqnarray}
\left(F_{\mu \nu}+q V_{\mu \nu}+ W_{\mu \nu}\right) \phi \equiv i [\mathcal{D}_\mu, \mathcal{D}_\nu ] \phi.
\end{eqnarray}
The $V_{\mu \nu}$ and $W_{\mu \nu}$ above, are the field strengths of the $U(1)_R$ and flavour background connections respectively.

The third and fourth terms in \eqref{LBOS1} combine with the third term in \eqref{LBOS2} to give a total derivative
\begin{eqnarray}
\mathcal{D}_\mu\left( v^\mu \bar{\phi}^\dagger\sigma \phi \right).
\end{eqnarray}

After adding up \eqref{LBOS1} and \eqref{LBOS2}
\begin{eqnarray}
\mathcal{L}_{bosonic}&=& ({D}^{\mu}{\bar{\phi}})^\dagger  \, {D}_{\mu}{\phi} +  \, \bar{\phi}^\dagger\left(i D- \epsilon^{\mu \nu}_{~~\beta}v^{\beta} \left(q V_{\mu \nu}+W_{\mu \nu}\right) \right)\phi
\nonumber\\&&   + (\bar{\phi}^\dagger \sigma)(\sigma \phi) +\bar{F} F+\,\mathcal{D}_{\mu}\left( i \, \epsilon^{\mu \nu}_{~~\beta}v^{\beta} \bar{\phi}^\dagger {D}_{\nu} \phi +  v^\mu \bar{\phi}^\dagger\sigma \phi \right). \label{LBOS}
\end{eqnarray}
After adding up \eqref{LFER1} and \eqref{LFER2}
\begin{eqnarray}
\mathcal{L}_{fermionic}&=&i \, \bar{\psi}^\dagger {\gamma}^{\mu} {D}_{\mu}{\psi}- i \bar{\psi}^\dagger \sigma \psi - i \,
\bar{\phi}^\dagger\, \bar{\lambda}^\dagger \psi- i\,  \bar{\psi}^\dagger \lambda \,\phi- i\,\mathcal{D}_\mu \bigg( \bar{\psi}^\dagger P^+\gamma^\mu\psi    \bigg). \label{LFER} \\\nonumber
\end{eqnarray}
$\mathcal{L}_{bosonic}+\mathcal{L}_{fermionic}$ is the same as equation (2.17) of \citep{BZ} up to total derivatives that can be immediately read out of \eqref{LBOS} and \eqref{LFER}.

\begin{eqnarray}
\mathcal{L}_{here}-\mathcal{L}_{there}=\,\mathcal{D}_{\mu}\left( i \, \epsilon^{\mu \nu}_{~~\beta}v^{\beta} \bar{\phi}^\dagger {D}_{\nu} \phi +  v^\mu \bar{\phi}^\dagger\sigma \phi-i\, \bar{\psi}\,^\dagger P^+ \gamma^{\mu} \psi \right). \label{OursBZ} \\\nonumber
\end{eqnarray}

\subsection{The orthodox $\tilde{Q}_\epsilon$ localising terms} \label{OrtTermAnti}
For completeness of presentation in this section we post the orthodox $\tilde{Q}_\epsilon$ localising terms for vector and matter multiplets. For the vector multiplet
\begin{eqnarray}
-\tilde{Q}_\epsilon Q_\epsilon \bigg( \bar{\lambda}^\dagger\lambda -4 \sigma D \bigg), \label{locVCov2}
\end{eqnarray}
After expanding this term one gets
\begin{eqnarray}
\mathcal{L}^{Vector}_B&=& \frac{1}{2} F_{\mu \nu}^2+\left(\mathcal{D}_\mu\sigma\right)^2+D^2- i\, \sigma \{\bar{\lambda}^\dagger,\lambda\} -i\sigma \{\bar{\lambda}^\dagger, P^- \lambda\} \\&{\red -}& 2\, i\, \mathcal{D}_\mu \bigg({\, \red- \,}\sigma v^\nu F^\mu_{~\nu}-v^\mu \sigma D\, {\red-} \, i \sigma D^\mu \sigma - \epsilon^{\mu \nu \beta} v_\beta \, \sigma \mathcal{D}_\nu \sigma  \bigg), \nonumber\\
\mathcal{L}^{Vector}_F&:=&i \bar{\lambda}^\dagger \slashed{D} \lambda +i\, \sigma \{\bar{\lambda}^\dagger, P^- \lambda\}-i \mathcal{D}_\mu \left(\bar{\lambda}^\dagger \gamma^\mu P^{{\red+}} \lambda\right).
\end{eqnarray}
From \eqref{LocVCov} and \eqref{locVCov2} and after using the following identities
\begin{eqnarray}
\bar{\lambda}^\dagger \gamma^\mu P^{{\red+}} \lambda\, - \, \bar{\lambda}^\dagger P^-\gamma^\mu  \lambda&=& v^\mu \left( \bar{\lambda}^\dagger \lambda\right), \\
i\, tr \mathcal{D}_\mu \left( \epsilon^{\mu \nu \beta} v_\beta \, \sigma \mathcal{D}_\nu \sigma\right) &=& \, tr [\sigma,[F_{12},\sigma]] =0, \label{sigmaFsigma}
\end{eqnarray}
one arrives to the expected result
\begin{eqnarray}
\bigg \{ Q_\epsilon, \tilde{Q}_\epsilon  \bigg \}  \bigg( \bar{\lambda}^\dagger\lambda -4 \sigma D \bigg)&=&\,i\, v^\mu \mathcal{D}_\mu\left( \bar{\lambda}^\dagger \lambda -4 \sigma D\right).
\end{eqnarray}
In \eqref{sigmaFsigma} it is used the fact $\sigma$ is only charged under the dynamical gauge group. Namely neutral under $R-$  and flavour symmetries.
The $\tilde{Q}_\epsilon$ orthodox localising term for matter is
\begin{eqnarray}
\tilde{Q}_\epsilon Q_\epsilon \bigg( \bar{\psi}^\dagger\psi +2\, i \, \bar{\phi}^\dagger \sigma \phi \bigg). \label{MatterAc2}
\end{eqnarray}
After expanding this term one gets
\begin{eqnarray}
\mathcal{L}^{Matter}_B&:=&D^{\mu}{\bar{\phi}}^\dagger  \, {{D}}_{\mu}{\phi} +\bar{\phi}^\dagger\,\left(%-\hat{\mathcal{D}}_3\sigma+
 i D\, +    \sigma^2 \,- \epsilon^{\mu \nu}_{~~\beta}v^{\beta} \left(q V_{\mu \nu}+W_{\mu \nu}\right) \right) \phi \nonumber\\&+& \bar{F}^\dagger F%+\left(\hat{\mathcal{D}}_3 \bar{\phi}\right)^\dagger\sigma \phi
\,{ \red - }\,\mathcal{D}_{\mu}\left(% v^\mu \bar{\phi}^\dagger \sigma \phi +
v^\mu\, \bar{\phi}^\dagger \sigma \phi \,{ \red -}\,i \, \epsilon^{\mu \nu}_{~~\beta}v^{\beta} \bar{\phi}^\dagger {D}_{\nu} \phi \right),\nonumber \\  && \nonumber\\ \mathcal{L}^{Matter}_F&:=& i \, \bar{\psi}^\dagger {\gamma}^{\mu} {{D}}_{\mu}{\psi}-i \bar{\psi}^\dagger\, \sigma \,\psi- i\,  \bar{\psi}^\dagger \lambda \,\phi- i \,
\bar{\phi}^\dagger\, \bar{\lambda}^\dagger \psi- i {D}_\mu\left( \bar{\psi}^\dagger \gamma^\mu P^{ \red-} \psi\right).  \\&&\nonumber
\end{eqnarray}
From \eqref{MatterAc} and \eqref{MatterAc2} and after use of the following identity
\begin{eqnarray}
\bar{\psi}^\dagger P^+ \gamma^\mu \psi-\bar{\psi}^\dagger \gamma^\mu P^{ \red-} \psi= v^\mu\,  \left(\bar{\psi}^\dagger \psi\right),
\end{eqnarray}
one gets the expected result
\begin{eqnarray}
\bigg \{ Q_\epsilon, \tilde{Q}_\epsilon  \bigg \}   \bigg( \bar{\psi}^\dagger\psi +2\, i \, \bar{\phi}^\dagger \sigma \phi \bigg)&=&\,i\, v^\mu \mathcal{D}_\mu \bigg( \bar{\psi}^\dagger\psi +2\, i \, \bar{\phi}^\dagger \sigma \phi \bigg). \\\nonumber
\end{eqnarray}
%\section{Integration of vector fluctuations} \label{appReal}
%In this appendix we relax the condition $Im[\hat{A}_t]=0$. Let us see the consequences.
 \section{1-loop determinants and JK Contour}

 \subsection{Zero locus analysis} \label{AppLoop}

  To solve the zero locus equations \eqref{Bogomolnyi} we have first to fix the gauge. We choose
\begin{eqnarray}
\hat{A}_3=u, \label{gf}
\end{eqnarray}
with $\partial_{\theta,~ \phi, ~t}u=0$.
From the Bianchi identity of $\hat{F}_{\mu \nu}$ it follows that $\mathcal{D}_t F_{12}=0$, next we suppose the fields $A_\theta$ and $A_\phi$ are single valued on $\mathbb{S}_1$ and thenceforth it follows that $F_{\theta \phi}$ and $F_{1 2}$ are also single valued. From the condition in the previous equation it follows that
 \begin{eqnarray}
~ [\hat{A}_t, A_\phi]=[\hat{A}_t, A_\theta]=[\hat{A}_t, F_{12}]=0. \label{commEq}
 \end{eqnarray}\footnote{Provided $e^{2 \pi i u}\neq \mathbb{Id}$. This is because the solution to $\hat{F}_{t \theta}=0$, for instance, is $A_\phi(t)=e^{i t u} A_{\phi}(0)e^{-i t u}$ and in consequence if $2 \pi u$ exponentiate to the unit then $A_\phi(t+2\pi)=A_\phi(t)$ and condition \eqref{commEq} is not necessary anymore. This condition will be needed again when dealing with 1 loop determinants. In fact, it will be assumed all along this work. } Equation \eqref{commEq} together with the second line in \eqref{Bogomolnyi}  and \eqref{gf} gives
\begin{eqnarray}
\partial_t A_\phi= \partial_t A_\theta=\partial_t F_{\theta \phi}=0.
\end{eqnarray}
What we have not mentioned so far is the dependence on $\theta$ and $\phi$ of $\sigma$, $A_\theta$ and $A_\phi$. Here we will assume the GNO condition
\begin{eqnarray} 
F_{1 2 }=\frac{ \mathfrak{m} }{2 R^2
}  \implies  \int_{\mathbb{S_2}} F_{\theta \phi}\, d\theta\wedge d\phi=2 \pi \mathfrak{m} ~~\text{  with  }~~~ \forall_{\rho} ~e^{2 \pi\, i\, \rho(\mathfrak{m}) }=1, \label{GNO2} 
\end{eqnarray}
for the constant magnetic flux $\mathfrak{m}$ -- for instance, the latter assumption guaranties invariance under large gauge transformations of the classical Chern-Simons contribution--. The $\rho$ in \eqref{GNO2} stands for an arbitrary weight of the corresponding Lie algebra. %However in principle there could be more general solutions to the Bogomolnyi equations \eqref{Bogomolnyi} that one could also analyse.
The equation for $F_{12}$ in \eqref{GNO2} is uniquely solved -- under GNO conditions -- for specific values of $A_\theta$ and $A_\phi$ (up to trivial gauge transformations) that will be defined below.

 The only missing equation $\mathcal{D}_3 \sigma=\frac{ \mathfrak{m}}{2 R^2
 }$ can be solved uniquely for $\sigma$ (by demanding independence of $u$ on $\mathfrak{m}$ and $\sigma(t=0)=\sigma_0$),%. Requiring the solution to be independent on the holonomy $u$ or equivalently, to be regular in the limit to $u=0$ one gets
\begin{eqnarray}
 \sigma^N = \sigma_0- \frac{\mathfrak{m}}{2 R^2
 } \,t, ~~~  0< t < t_0, \label{GNO3}
\end{eqnarray}
with $\sigma$, $\sigma_0$ and $\mathfrak{m}$ in the Cartan sub algebra of $u$, $A_\theta$ and $A_\phi$
%On $S_2\times S_1$ a space of topologically trivial solutions to \eqref{Bogomolnyi} is
%\begin{eqnarray}
%\{F_{\mu \nu}=0, \,
%\sigma=\text{ Constant } \} /\text{Gauge transformations}. \label{locus}
%\end{eqnarray}

 The gauge invariant classes of saddle points \eqref{GNO2}-\eqref{GNO3} can be parameterised by the following gauge potentials
 \begin{eqnarray}
 \hat{A}_{(0)}= -\frac{
\mathfrak{m}}{2% R^2
}
\cos{\theta}  d\phi +u%{A_{(0)}}_3(t)
 dt  ,%\nonumber\\
%\tilde{F}_{1 2 }_{(0)}=\frac{ \mathfrak{m} }{2 R^2}, ~~~ \sigma_{(0)}= \sigma_0- \frac{\mathfrak{m}}{2 R^2} t,
\label{monopoles0}  \label{'tHooftLine}
\end{eqnarray}
were $\mathfrak{m}$, $u$ and $\sigma_{(0)}$ represent mutually commuting elements of the gauge algebra. As already said, we stress that the weight vector components of the generators $\mathfrak{m}$ are supposed to be GNO quantised -- the weights of $\mathfrak{m}$ belong to the Co-root lattice--.

As for the matter multiplets, let the contour of integration for the fields $\phi$, $\bar{\phi}$, $F$ and $\bar{F}$ be \eqref{ComplexCondChiral}. The $Q_\epsilon$-BPS configurations along the path \eqref{ComplexCondChiral} are given by \eqref{QeBPS}.
Notice that the condition $\hat{\mathcal{D}}_{3} \phi=0$ holds independently of the reality condition on $\hat{A}_3$. We remark this fact, because later on it will be needed to turn on a complex flavour line $A^F_3$ along $\mathbb{S}_1$. The result of the zero locus analysis presented in this section still holds in that case.

We must also notice, that to make an arbitrary linear variation of the matter action \eqref{LBOSF0}+\eqref{LFERF0} at the zero locus \eqref{locusGauge}+\eqref{QeBPS}
\begin{eqnarray}
\frac{\delta QV^{matter}}{\delta \Phi} \bigg|_{\eqref{locusGauge}+\eqref{QeBPS}} \cdot \delta \Phi,~~~ ~
 \Phi:=\left(\phi, \bar{\phi}, F, \bar{F}\right),
 \end{eqnarray}
vanish, it is needed the scalar fields $\bar{\phi} \text{ and }  F$ to be solutions of %the EOM
\begin{eqnarray} 
\frac{\delta Q V_{matter}}{\delta \phi} \bigg|_{\eqref{locusGauge}+\eqref{QeBPS}} &=&\frac{\delta Q V_{matter}}{\delta \bar{F} } \bigg|_{\eqref{locusGauge}+\eqref{QeBPS}} =0. \label{EOMChiral} 
\end{eqnarray}
Otherwise the zero locus \eqref{QeBPS} will not be stable.

 However as we are going to integrate along the path \eqref{ComplexCondChiral} it follows that at the zero locus \eqref{QeBPS} the following conditions hold,
\begin{eqnarray}
\hat{\mathcal{D}}_3 \bar{\phi}%=\sigma \bar{\phi}
= \mathcal{D}_1\bar{\phi}+i  \mathcal{D}_2 \bar{\phi}= F=0. \label{QeBPS2}
\end{eqnarray}
In fact \eqref{QeBPS2} coincides with the $\tilde{Q}_\epsilon$-BPS conditions for $\bar{\phi}$ and consequently, it is a solution of the EOM coming from $QV_{matter}$ \eqref{EOMChiral}. This is guarantied from the fact that for the reality conditions \eqref{ComplexCondChiral} $\left(QV_{matter}\right)_{Bos}\geq0$ and saturates for the $Q_\epsilon$ BPS real configurations \eqref{QeBPS}. In words, the unique solution (saddle point) to
\begin{eqnarray}
Q_\epsilon V_{matter}=0,
\end{eqnarray}
along \eqref{ComplexCondChiral} is \eqref{QeBPS}. In a generic path of integration this will not be case.

As in \citep{BZ}, we will avoid the zones in the Coulomb branch that allow for non trivial solutions to \eqref{QeBPS}. From the integration on $\mathbb{S}_2\times \mathbb{S}_1$ of the zero locus equations \eqref{QeBPS} and \eqref{QeBPS2} follows
\begin{eqnarray}
(\hat{A}_{(0)})_{2,3} \centerdot \phi=(\hat{A}_{(0)})_{2,3}\centerdot \bar{\phi}=0, \label{LocChi}\end{eqnarray}
where by $"\centerdot"$ we mean the product associated to the gauge representation carried by $\phi$ and $\bar{\phi}$.
 If the non trivial components $(\hat{A}_{(0)})_2$ and $(\hat{A}_{(0)})_3$ do not share a common zero eigenvalue under the action $"\centerdot"$ the matter multiplet zero locus is
\begin{eqnarray}
\bar{\phi}=\phi=0. \label{ChiralBPS}
\end{eqnarray}
Condition \eqref{ChiralBPS} eliminates the quadratic mixing of matter and gauge fluctuations around saddle points. We consider gauge representations where \eqref{ChiralBPS} is the only solution to \eqref{LocChi} at least for some region of values of the moduli $(\hat{A}_{(0)})_2$ and $(\hat{A}_{(0)})_3$. For example, with $\phi$ and $\bar{\phi}$ in the adjoint where $"\centerdot"$ stands for the commutator $"[~ , ~ ]"$, clearly \eqref{ChiralBPS} does not hold as the only solution of \eqref{LocChi}. For other irreps \eqref{ChiralBPS} generically holds in some region of the moduli space $\left((\hat{A}_{(0)})_2, (\hat{A}_{(0)})_3\right)$.%, except for specific points in the moduli space, as we shall explicitly see in subsection \ref{1loopdets}, while manipulating 1-loop determinants \footnote{For instance, condition \eqref{Invertibility} excludes points of the moduli space in question, where non trivial zero modes exist.}.

%, like for instance, when every component in $\hat{A}_{(0)}$ exponentiates trivially. At such kind of points new massless modes arise for $\phi$ and $\bar{\phi}$.

%\subsubsection*{The classical on shell actions}

 \subsection{Integrality condition on $\rho(h_i)$}\label{ChevalleyAp}

We remind here the definition of Chevalley basis of a semi-simple Lie algebra $g$ of rank $r$ which is generated by  the $3 r$ set of generators $\{h_i, e_i, f_i\}_{i=1,\ldots, r}$ with commutation relations
\begin{eqnarray}
[h_i, h_j]&=&0 ,\nonumber \\ \newline [h_i, e_j] &=&C_{j i} e_j, \nonumber \\ \newline
[h_i, f_j]&=&-C_{j i} f_j, \nonumber \\\newline ~ [e_i, f_j]&=& \delta_{i j} h_j, \label{Chevalley}
\end{eqnarray}
where
\begin{eqnarray}
C_{ij}\in \mathbb{Z}, \label{StructureConstant}
\end{eqnarray}
are the components of the Cartan matrix of $g$. The roots are $\alpha_i= \{C_{i,1}, \ldots, C_{i,r}\}$. The remaining $dim(g)-3r$ generators are generated by the adjoint actions
\begin{eqnarray}
ad(e_i)^n e_j&,& \\
ad(f_i)^n f_j&,& \text{ with }    1\leq n \leq 1- C_{i j} \text{ and } i\neq j.
\end{eqnarray}
The remaining structure constants can be derived out of \eqref{Chevalley}. They are integers too.

 The $e_i$ acting onto a state $|h_1, \ldots, h_r\>_{\{\rho\}}$ in a representation $\{\rho\}$ with eigenvalues $\rho(h_j)$ under the action of $h_j$, with $j=1, \ldots, r$ will define a new state, with eigenvalues $\rho(h_j)-C_{j i}$ with respect to $h_j$. In similar fashion the $f_i $'s acting onto the state $|h_1, \ldots, h_r \>_{\{\rho\}}$ will define a new state with eigenvalues of $h_j$ shifted from $\rho(h_j)$ to $\rho(h_j)+C_{j i} $. From this analysis and \eqref{StructureConstant}, one can conclude that highest weight representations (and their sums) will necessarily obey
\begin{eqnarray}
\rho(h_i) \in \mathbb{Z}, \label{rhoZ}
\end{eqnarray}
for every weight $\rho$, in the basis in which the algebra takes the form \eqref{Chevalley}. This is, a highest weight $\rho_{HW}$ will necessarily obey $\rho_{HW}(h_i)\in \mathbb{Z}$ because their Dynkin labels are positive integers ($\rho_{HW}:= n_i h_i^{\star} $ with $n_i:=\rho_{HW}(h_i) \in \mathbb{N}$ \footnote{This is because we have defined $h^{\star}_i(h_j)%= \frac{2 \<\alpha_i, h_j \>}{\<\alpha_i, \alpha_i \>}
 :=\delta_{i j}$ with $h^\star_i$ identified, in this way, as the dual to the Cartan generator $h_i$. $h^*_i$ can also be thought as a state with eigenvalues $h^{\star}_i(h_j)=\delta_{i j}$ under $h_j$.  Each one of the states $h^\star_i$ is a HW state of an irreducible representation, better known as fundamental. Specifically, they carry a "spin $\frac{1}{2}$" representation of the $SU(2)$ whose Cartan generator is $h_i$. In our conventions the weight vector (fundamental) of the state $h^\star_i$ is simply $\Lambda_i:=||\delta_{ij}||_{j=1,\ldots, r}$. From this point on, we will abuse of $h^*_i$ to denote the weights (fundamental) of its dual state.} by definition.  Additionally, every weight in the HW representation is obtained out of $\rho_{HW}$ by performing a multiple subtraction of the roots $\alpha_i:= C_{i j} h^\star_j$ ($C_{i j}:=\alpha_i(h_j)$).  Thence from \eqref{StructureConstant} it follows \eqref{rhoZ}.

Finally we show a property that was used in the main body of the manuscript. Let $u:= u_i h_i$ be a generic element in the Cartan. We remind that for $h$ and $l$ mutually commuting elements of the algebra then
\begin{eqnarray}
\rho(\#h):=\# \rho(h), ~~ \rho(h+l):=\rho(h)+\rho(l).\end{eqnarray}\footnote{This second equation means that the sum of eigenvalues of the operators $h$ on $l$ is the eigenvalue of the operator $h+l$. This statement has sense only if $h$ and $l$ are mutually commuting. This is because the full representation must be composed by eigenstates of both $h$ and $l$ simultaneously.}
Thenceforth
\begin{eqnarray}
e^{ i \rho(u+2 \pi h_i)}=e^{ i \rho(u)},
\end{eqnarray}
and we conclude that any function $f[e^{i\rho(u)}]$ is invariant under $u_i\rightarrow u_i+2\pi$.

 \subsection{1-loop determinants on $\SS /{p}$ } \label{1loopdets}

In this subsection we first show that the quadratic expansion of the chiral multiplets unorthodox $Q_\epsilon$ ($\tilde{Q}_\epsilon$) localising term around the $Q_\epsilon$($\tilde{Q}_\epsilon$)-locus that we have found in previous subsection, coincides with the quadratic expansion obtained in \citep{BZ} (In their case, when the zero mode $D_0$ vanishes and supersymmetry gets restored). We consider $t_0\rightarrow 2\pi^-$ in the case of $I_N$ and $t_0\rightarrow 0^+$ in the case of $I_S$. Namely we will study both versions, the 3D $Q_\epsilon$ and $\tilde{Q}_\epsilon$ TT theories on $\SS /\{t=0\}$. The result of using either $Q_\epsilon$ or $\tilde{Q}_\epsilon$ is the same in the end.

As shall be noticed below, the 1 loop determinant of a vector multiplet is periodic along the real direction of the moduli $u_i$ with periodicity $2\pi$ (Our conventions can be found in appendix \ref{ChevalleyAp}). As for matter multiplets, the period $2\pi$ is related to both gauge invariance and absence of parity anomaly. In fact, a matter multiplet in a complex representation will have, generically, a 1 loop contribution with a period of $4\pi$. Indistinctly, under a shift of $2\pi$ in $u$, or under a parity inversion $u\rightarrow-u$, the one loop contribution of a mode with gauge weight $\rho$, that is an integer power of $\sin{\frac{\rho(u)}{2}}$, changes by a factor of $-1$ to that same power. That minus sign could survive in specific representations $\left\{\rho\right\}$ and in that case, the consequence of its presence is two-folded: 1-loop gauge and parity anomaly.

There are various ways to ensure the absence of the aforementioned gauge/parity anomaly out \footnote{A sufficient condition without the use of Chern-Simons terms can be found in \citep{Romo}. We can also solve the gauge/parity anomaly by the use of Chern-Simons terms, however we don't want to impose any constraint on the Chern-Simons content of the theory and consequently we must constraint the gauge representation carried by the matter content.}. One possible solution is to consider matter multiplets in real representations. Another possibility would be to consider more than one matter multiplet in such a way their anomalies cancel each other out. One such example comes from considering two matter multiplets in complex conjugated representations of the gauge group in question. As shall be explicitly seen in the case of gauge group $U(N)$ the product of the fundamental and antifundamental matter multiplets' 1 loop factors becomes $2\pi$ instead of $4\pi$ periodic.  Finally, we also elaborate about an issue pointed out in \citep{BZ},  the possible presence of "boundary contributions" and their geometrical meaning.

 Let us start by the matter multiplet. The bosonic and fermionic parts of the action for quadratic fluctuations coming from  \eqref{LBOSF0} (after trivial integration by parts) and from \eqref{LFERF0}, are
\begin{eqnarray}
S^{quad}_{B}&=& \int_{\mathbb{S}_2\times \mathbb{S}_1} \delta \bar{\Phi} \cdot \mathcal{O}_{B} \cdot\delta \Phi \\
S^{quad}_{F}&=&\int_{\mathbb{S}_2\times \mathbb{S}_1} \delta \bar{\Psi}^\dagger\cdot \mathcal{O}_{F} \cdot \delta \Psi
\end{eqnarray}
with
\begin{eqnarray}
\mathcal{O}_{B}&=& \left\{
	\begin{array}{clcl}
		-\hat{\mathcal{D}}^\mu \hat{\mathcal{D}}_\mu+ \hat{\mathcal{D}}_3\sigma^+ - \epsilon^{\mu \nu}_{~~\beta}v^{\beta} \left(q V_{\mu \nu}+W_{\mu \nu}\right)&~~~~~Q_\epsilon: ~ 0< t < 2\pi \\ &&\\
		-\hat{\mathcal{D}}^\mu \hat{\mathcal{D}}_\mu{\red-} \hat{\mathcal{D}}_3\sigma^- - \epsilon^{\mu \nu}_{~~\beta}v^{\beta} \left(q V_{\mu \nu}+W_{\mu \nu}\right) &~~~~~\tilde{Q}_\epsilon: ~0<t<2\pi  \end{array}\right\}, \nonumber  \\
\mathcal{O}_{F}&=&  i \hat{\slashed{\mathcal{D}}}. \label{OpBZ}
 \end{eqnarray}
 In contradistinction with the operators coming out of \eqref{LBOS} and \eqref{LFER}\footnote{This is, by discarding in those equations total derivatives. that for the zero locus we are considering provide non trivial contributions. }, found after working with the form of the algebra \eqref{algebraChiral0},  these operators do not look like the analog ones in \citep{BZ}. However as we shall see next, when evaluated on their corresponding zero loci, they will be the same.
%First, we note that by defining
 %\begin{eqnarray}
%A_{(0)}_3(t) := u- i( \sigma_{(0)} -\frac{\mathfrak{m}}{2 R^2}
% t). \label{TransTothooft}
%\end{eqnarray}
%we can cancel part of the contribution of the zero locus background value for $\sigma
%$, $\sigma_{(0)}:= \sigma_0 -\frac{\mathfrak{m}}{2 R^2}
% t$ as the effective background gauge potential becomes
% \begin{eqnarray}
%\hat{A}_{(0)}= -\frac{\mathfrak{m}}{2 R^2}\cos{\theta} %-1)
%%d\phi +u~ dt . \label{'tHooftLine}
%\end{eqnarray}
%We notice that the choice \eqref{TransTothooft} does not cancel every possible effect due to the presence of the background profile \eqref{sigma}. In fact,
The term $\mathcal{D}_3 \sigma^\pm%_{(0)}
=\mp\frac{\mathfrak{m}}{2 R^2}$ in \eqref{OpBZ} provides the contribution that in the approach followed by \citep{BZ}, comes from the non trivial (complex) saddle value of the auxiliary field $D$ that we have written in \eqref{D}. Specifically, from the quadratic term proportional to $D$ in \eqref{LBOS}. When evaluated in their corresponding supersymmetric zero loci,  the operators \eqref{OpBZ} are the same as the analog ones in \citep{BZ}  and thenceforth is straightforward to check that the final result for the 1-loop determinant is the same after changing $u_i$ by $\frac{u_i}{2\pi}$
 \begin{eqnarray} 
Z^{1-loop}_{Matter}\equiv \prod_{\rho \in \mathcal{R}} \bigg( \frac{x^{\frac{\rho}{2}}}{1-x^\rho} \bigg)^{\rho(\mathfrak{m}) - q_R+1},  \label{matter1Loop}
\end{eqnarray}
with $x=e^{i u}$. Where $\mathcal{R}=\{\rho\}$ is a representation of the gauge algebra. In subsection \ref{subsec} we have computed this one loop contribution for $\SIN(I_S)$. The computation is completely analog in this case. See appendix  A.3 in \citep{BZ} for other approaches to derive \eqref{matter1Loop}.

Now let us show that in the Chevalley basis (See appendix \ref{Chevalley}) and for non gauge/parity anomalous gauge representations, or as supposed in this work, representations that remain invariant under the inversion $\rho \rightarrow -\rho$, the 1 loop determinant for the chiral multiplet is invariant under the transformation $u_i\rightarrow u_i+2\pi \mathbb{Z}$. An important property to keep in mind is that

\begin{eqnarray}
Z^{1-loop}_{Matter}[u+2 \pi h_i]=\left(\prod_{\rho}(-1)^{ \rho(h_i)(\rho(\mathfrak{m})-q_R+1)   }\right) Z^{1-loop}_{Matter}[u], \label{periodicity}
\end{eqnarray}
were $h_i$ is any of the $r$ Cartan generators. In arriving to \eqref{periodicity} we have used the property $e^{i 2 \pi \rho( h_i) }=1$, that comes from the fact that the scalar product between the weights in a given representation $\{\rho\}$ and the fundamental weights of the algebra $h_{i}$ obeys the integrality condition $\rho(h_i) \in \mathbb{Z}$, that we have elaborated upon in appendix \ref{ChevalleyAp}% equation \eqref{rhoZ}
. From \eqref{periodicity} and $2\rho(h_i) \rho(\mathfrak{m})  \in 2\mathbb{Z}$ it follows that if representation $\mathcal{R}$ is invariant under $\rho \rightarrow -\rho$ then
\begin{eqnarray}
\prod_{\rho}(-1)^{ \rho(h_i)(\rho(\mathfrak{m})-q_R+1)}=\prod_{\rho>0}(-1)^{ 2\rho(h_i)\rho(\mathfrak{m})}=1
\end{eqnarray}
and consequently
\begin{eqnarray}
Z^{1-loop}_{R}[u_i+2 \pi]=Z^{1-loop}_{R}[u_i].
\end{eqnarray}
 Similarly from \eqref{periodicity} and $2\rho(h_i) \rho(\mathfrak{m})  \in 2\mathbb{Z}$ is immediate to prove that $\left(Z_R Z_{\bar{R}}\right)[u_i+2\pi]=\left(Z_R Z_{\bar{R}}\right)[u_i]$ where by $Z_{R}$ and $Z_{\bar{R}}$ we mean the one loop contributions of a couple of chiral multiplets with coincident quantum numbers but with complex conjugated gauge representations $R$ and $\bar{R}$.

%As for the asymptotic behaviour in the limit $u_i\rightarrow \pm i \infty$ \begin{eqnarray}
%Z_{Chiral} \rightarrow \prod_{\rho} (\text{Sign}{Tr \rho})^{|\rho| \cdot m}e^{\pm i \rho \cdot u ~(n-q_R+1)}  \left(1+ ~ subleading~ terms\right).
%\end{eqnarray}

While analysing the JK procedure and the potential presence of the so called "boundary contributions", we will find useful to know about  the asymptotic behaviour in the limit $u_i\rightarrow \pm i \infty$. Also, as we shall see next, this asymptotic behaviour for $Z_\mathfrak{m}$ involves $\mathfrak{m}$ only through the constant phase $(-1)^{\sum_i \mathfrak{m}_i}$. As already argued when we use a complex representation for the matter sector we should also add up a complementary matter multiplet in the complex conjugated representation with the same additional charges (R-charge, etc).
%This is because we do not want an anomalous gauge symmetry. %Consequently every weight $\rho$ will be accompanied by its complex conjugated $-\rho$ and thence
%\begin{eqnarray}
%Z_{Chiral} Z^{c.c.}_{Chiral} \rightarrow \prod_{\rho^*} (-1)^{\rho \cdot m}e^{\pm i |\rho| \cdot u ~(n-q_R+1)} \left(1+ ~ subleading~ tems\right). \label{AsymptoticCC}
%\end{eqnarray}
%Where by $|\rho|$ we mean the vector whose components are the absolute value of the components of $\rho$.
So, to illustrate clearly, we start by writing down a particular example of \eqref{matter1Loop} for matter in the fundamental and anti-fundamental of $U(N)$
\begin{eqnarray}
Z^{Fund-U(N)}_{Matter}&=&\prod_{i=1}^{N}\left(\frac{i}{2}\csc{\frac{u_i-v}{2}}%\frac{e^{ \frac{i}{2} u_i }}{1-e^{i u_i }}
\right)^{m_i+n -q_R+1}\nonumber\\
Z^{AntiFund-U(N)}_{Matter}&=&\prod_{i=1}^{N}\left(-\frac{i}{2}\csc{\frac{u_i+ v}{2}}
%\frac{e^{ -\frac{i}{2} u_i }}{1-e^{-i u_i }}
\right)^{-m_i+n-q_R+1}
\end{eqnarray}
where $ v$ and $\mathfrak{n}$, %with $Re[v]$ and $Im[v]$\in \mathbb{R}_+$
are a $U(1)$ flavour complex Wilson line along $\mathbb{S}_1$ and flux onto $\mathbb{S}_2$, respectively. Notice that for $u_j \rightarrow \pm i \infty$ with $j=1 \ldots N$ the asymptotic expansion of the product
\begin{eqnarray}
Z^{Fund-U(N)}_{Matter}Z^{AntiFund-U(N)}_{Matter}&=& (-1)^{\sum_i m_i}e^{\mp i \sum_{i} u_i|\mathfrak{n}-q_R+1|} \left(1+  ~sub.~terms \right) \label{AsymptoticFundAntiFund} 
\end{eqnarray}
will only depend on $\mathfrak{m}$ through the phase $(-1)^{\sum_i
\mathfrak{m}_i}$ that can be absorbed with the $U(1)$ topological holonomy \eqref{TopSym}. Thence, as the integrand will be independent on $\mathfrak{m}$ and blindly following the argument given in section 2.3.4 of \citep{BZ} if $k=0$ we will simply ignore the presence of "boundary contributions". To the reader which is not familiar with this terminology, it shall be seen in a while what is meant by "boundary contributions". Notice that it does not matter whether one takes the plus or minus sign in the limit, the asymptotic behaviour \eqref{AsymptoticFundAntiFund} will not approach zero. %This is because we have chosen a matter content with a spectrum of weights invariant under the inversion $\rho\rightarrow -\rho$. Should we have chosen only a fundamental matter multiplet the one loop determinant would have vanished asymptotically as $u\rightarrow \pm i \infty$ provided .

%\subsubsection*{Vector multiplet}

Next, we move on to the study the 1 loop contribution of the vector multiplet.
We expand the localising $Q_\epsilon V$ term \eqref{QVvec2}, around the zero locus \eqref{monopoles0} to get

\begin{eqnarray}
{\mathcal{L}_B^{Vector}}_{quadratic}%&=& |\delta D-i ( \delta F_{1 2}+ \delta \hat{\mathcal{D}}_3 \sigma )|^2+|\delta \hat{F}_{13}- i \delta \hat{F}_{2 3}|^2 \nonumber \\
 &=& \left\{
	\begin{array}{clcl}
		 \delta D^2+( \delta F_{1 2}+ \delta \hat{\mathcal{D}}_3 \sigma )^2+(\delta \hat{F}_{13})^2+ (\delta \hat{F}_{2 3})^2&~~~~~Q_\epsilon: ~ 0 <  t < 2\pi \\&&\\
		 \delta D^2+( \delta F_{1 2}- \delta \hat{\mathcal{D}}_3 \sigma )^2+(\delta \hat{F}_{13})^2+ (\delta \hat{F}_{2 3})^2 &~~~~~ \tilde{Q}_\epsilon: ~0<t<2\pi  \end{array}\right\}  \nonumber \\ \label{DiffQVectorB} \\ \nonumber\\
{\mathcal{L}_F^{Vector}}_{quadratic}&=& \left\{
	\begin{array}{clcl}
		(\delta \bar{\lambda}_1^\dagger   ~~~ \delta\bar{\lambda}_2^\dagger   )  \left(\begin{array}{c c}
 i\hat{\mathcal{D}}_t &  0\\
 \frac{i}{R} \mathcal{D}^{(b-1)}_-  & -i\hat{\mathcal{D}}_t
 \end{array}\right) \left(\begin{array}{c}\delta \lambda_1 \\  \delta \lambda_2 \end{array}\right) &~~~~~Q_\epsilon: ~ 0 < t  < 2\pi \\ &&\\
		(\delta \bar{\lambda}_1^\dagger   ~~~ \delta\bar{\lambda}_2^\dagger   )  \left(\begin{array}{c c}
 i\hat{\mathcal{D}}_t &   \frac{i}{R} \mathcal{D}^{(b)}_+\\
  0& -i\hat{\mathcal{D}}_t
 \end{array}\right) \left(\begin{array}{c}\delta \lambda_1 \\  \delta \lambda_2 \end{array}\right) &~~~~~\tilde{Q}_\epsilon: ~0<t<2\pi  \end{array}\right\} \label{DiffQVectorF}\\ \nonumber
\end{eqnarray}
where in this section $b=-\frac{\rho(\mathfrak{m})}{2}$.

% and
%\begin{eqnarray}
%\mathcal{D}_t&:=& \partial_t -i \rho(u)
%\\
%\mathcal{D}^{(s)}_\pm&:=& \partial_\theta \mp \frac{i}{\sin \theta} \partial_\phi \mp \frac{s}{\sin \theta}.
%\end{eqnarray}

%In passing from the first to the second line in \eqref{DiffQVectorB}, we have used the reality conditions \eqref{ComplexCond}.
In \eqref{DiffQVectorF} we have integrated by parts $\hat{\mathcal{D}}_t$, $\mathcal{D}^{(b)}_-$ and $\mathcal{D}^{(b-1)}_+$. This discarded total derivative vanishes because we are considering (D,D) boundary conditions for the vector multiplet in order to annihilate the supersymmetric variation of the modified Chern-Simons terms.  %Notice that \eqref{DiffQVectorB} and \eqref{DiffQVectorF} depend on $\sigma$ through the effective potential $\hat{A}$ and thence the explicit time dependence in $\sigma$, $h_m t$, is cancelled after using the definition \eqref{TransTothooft}.

The BRST term is
\begin{eqnarray}
Q\left(\bar{\delta c}\delta \hat{A}_t\right)= \delta \bar{ c}\hat{\mathcal{D}}_t \delta c+ \delta \bar{b}\delta\hat{A}_t \label{BRSTAc}
\end{eqnarray}
and the non ``covariant" BRST transformations are
\begin{eqnarray}
Q_{B} \delta \hat{A}_\mu=\hat{\mathcal{D}}_\mu\delta c, ~~ Q_{B} \delta \bar{c}=\delta \bar{b}.
\end{eqnarray}
After imposing the gauge fixing condition
\begin{eqnarray}
\delta \hat{A}_t=0, \label{gfixingcond}
\end{eqnarray}
the  quadratic expansion for bosonic density of Lagrangian is
\begin{eqnarray}
\frac{1%Csc(\theta)^2
}{R^2}\left( \tilde{\mathcal{D}}_- \delta A_{+} -\tilde{\mathcal{D}}_+ \delta A_{-}\pm R ~ \hat{\mathcal{D}}_t\delta \sigma\right)^2 \nonumber\\+ \frac{2 %Csc(\theta)^2
}{R^2}\left((\hat{\mathcal{D}}_t \delta A_{+})^2 +(\hat{\mathcal{D}}_t \delta A_-)^2\right), \label{EQreferencia}
\end{eqnarray}
where
\begin{eqnarray}
\delta A_+&:=& \frac{1}{2}(\delta A_\theta+ \csc{ \theta}\, \delta A_\phi)\nonumber \\
\delta A_-&:=&\frac{1}{2}(%-
\delta A_\theta %+
- \csc{ \theta} \, \delta A_\phi). \\
 \tilde{\mathcal{D}}_\pm&:=&\left( \mathcal{D}_\theta \pm \csc{\theta}  \, \mathcal{D}_\phi\right).
 \end{eqnarray}
%Notice that the $\tilde{\mathcal{D}}_{\pm}$ are not the raising operators defined in \eqref{raising}, $\mathcal{D}^{(s)}_{\pm}$, so they must not be confused.
%To save space, by the $\pm$ in the \eqref{EQreferencia} we mean the $I_N$ and $I_S$ quadratic expansions respectively.

The basis of functions on $\mathbb{S}_2$ to use in this case will be
\begin{eqnarray}
(\delta A_{+}, ~\delta A_{-}, ~ \delta \sigma, ~\delta \bar{c}, ~ \delta c) \rightarrow (Y^{(b-1)}_{j,j_3},~ Y^{(b-1)}_{j,j_3}, ~ Y^{(b)}_{j,j_3},~ Y^{(b)}_{j,j_3},~ Y^{(b)}_{j,j_3}), \label{b37} 
\end{eqnarray}
with
\begin{eqnarray}
j \geq (|b-1|,~ |b-1|, ~ |b|, ~ |b|, ~  |b|)  ~\text{   and   }   -j \geq j_3 \geq j.
\end{eqnarray}
The $j$ and $j_3$ grow at step 1.  We consider the roots $\rho$ to be outside of the Cartan sub algebra values, in order not to get a vanishing determinant. The $Y^{(b)}_{j,~ j_3}$ above are the spin $b$ spherical harmonics (See \citep{WuYang, Cremonesi,Gomis}). The magnetic levels in \eqref{b37} are selected from demanding consistency with supersymmetry, as we shall explain in a while.

After the change of variables on each patch
\begin{eqnarray} 
\delta \sigma \rightarrow \delta \sigma_{\pm}:=  \delta \sigma\pm \frac{1}{R}  \hat{\mathcal{D}}_t^{-1}\left(\tilde{\mathcal{D}}_+\delta A_{-} -\tilde{\mathcal{D}}_- \delta A_{+} \right),
\end{eqnarray}
the functional integration over $\delta \sigma_\pm$ (notice this change of variable has unit Jacobian) gives
\begin{eqnarray}
\frac{1}{\sqrt{\det_{\SS/\{t=0\}}{\hat{\mathcal{D}}_t^2}}}=\prod_{\rho} \prod_{j\geq |b|} \prod_{n \in \mathbb{Z}}(2 \pi n-\rho(u) )^{-2 j-1}. \label{IntegrationDeltaSigma}
\end{eqnarray}
%The $\rho^*$, $n^*$ means exclusion of the Cartan root $\{0,\ldots,0\}$. Namely exclusion of the zero mode fluctuation, otherwise \eqref{IntegrationDeltaSigma} diverges.
The operator $\hat{\mathcal{D}}_t$ is invertible provided
\begin{eqnarray}
\rho(u)\neq 2\pi \times \mathbb{Z}. \label{Invertibility}
\end{eqnarray}
If \eqref{Invertibility} holds then \eqref{IntegrationDeltaSigma} is finite. Condition \eqref{Invertibility} kills the zero modes of $\hat{\mathcal{D}}_t$ and implies the elimination of the Cartan roots (weights) $\rho=\{0,\ldots,0\}$.

 The contribution \eqref{IntegrationDeltaSigma} cancels out the functional integration of $(\delta \bar{c}, \delta c)$, the Fadeev-Popov determinant, which equals $\det_{\SS/\{t=0\}} \hat{\mathcal{D}}_t$, over the same functional space. Finally, we integrate $(\delta A_+, \delta A_-)$ over the functional space in $\SS/\{t=0\}$ to get
 \begin{eqnarray}
\bigg(\prod_{\rho} \prod_{j\geq |b-1|} \prod_{n}(2 \pi n- \rho(u) )^{-2 j-1}\bigg) \times \bigg(\prod_{\rho} \prod_{j\geq |b-1|} \prod_{n}(2 \pi n- \rho(u) )^{-2 j-1}\bigg). \label{ApAm}
 \end{eqnarray}

 The quadratic density Lagrangian for the gaugino fluctuations was written in \eqref{DiffQVectorF}. Notice that the off diagonal operator does not affect the value of the determinant to compute. In this sense, the use of the non covariant localising term \eqref{locV} and \eqref{locV2}  simplifies the job of computing one loop determinants. One must compute the determinants on $\SS/\{t=0\}$ by expanding the space of functions on $\mathbb{S}_2$ spanned by the following basis of functions
\begin{eqnarray}
(\delta \lambda_1 , ~ \delta \bar{\lambda}_1^\dagger,~\delta \lambda_2,~ \delta\bar{\lambda}_2^\dagger )
\rightarrow (Y^{(b)}_{j,j_3}, ~ Y^{(b)}_{j,j_3}, ~ Y^{(b-1)}_{j,j_3}, ~ Y^{(b-1)}_{j,j_3})  \label{b43}
\end{eqnarray}
and
\begin{eqnarray}
j \geq (|b|,~|b|,~|b-1|, ~ |b-1|).  
\end{eqnarray}
 The relative magnetic levels are obtained from consistency with the supersymmetry transformations \eqref{algebra1} and \eqref{algebra1South}. Specifically from their $O(\frac{1}{\sqrt{t}})$ term after expanding with \eqref{FBPS} one gets
\begin{eqnarray}
Q_\epsilon \delta A_1=i \, \frac{1}{2} \, \delta \bar{\lambda}^\dagger_2,~~  Q_\epsilon \delta A_2=-\frac{1}{2} \, \delta\bar{\lambda}^\dagger_2, \nonumber\\
Q_\epsilon \delta \lambda_1= \delta D%+\mathcal{D}_+ A_- +   \mathcal{D}_-  A_+
+\hat{\mathcal{D}}_t \delta \sigma_+ - i [\delta \hat{A}_t, \sigma^N(t)], \nonumber\\
Q_\epsilon \delta \lambda_2= \hat{\mathcal{D}}_t\delta A_1 + i \,\hat{\mathcal{D}}_t \delta A_2 -\mathcal{D}^{(b-1)}_- \delta A_3.
\end{eqnarray}
 From the equations in the first and third lines we can infer that the magnetic levels of $\delta A_\theta$ and $ \delta A_\phi$ must coincide with the ones of $\delta \bar{\lambda}^\dagger_2$ and $\delta \lambda_2$. From the second line we conclude that $\delta D$, $\delta \sigma_+$ and $\delta \lambda_1$ have coincident magnetic level. A similar analysis with the BRST transformation shows that $c$ and $\bar{c}$ must have the same magnetic level as $\delta \hat{A}_t$ which must be the same as $\delta \lambda_1$. After analysing the $O(\frac{1}{\sqrt{\tau}})$ terms in the expansion of $\tilde{Q}_\epsilon$ algebras one concludes that the magnetic levels are the ones written in \eqref{b37} and \eqref{b43}.

After functional integration of the quadratic localising action of the gaugino and antigaugino linear fluctuations one gets the factors
 \begin{eqnarray}
\bigg(\prod_{\rho} \prod_{j\geq |b|} \prod_{n}(2 \pi n- \rho(u) )^{2 j+1}\bigg) \times \bigg(\prod_{\rho} \prod_{j\geq |b-1|} \prod_{n}(2 \pi n- \rho(u) )^{2 j+1}\bigg),  
 \end{eqnarray}
whose product with \eqref{ApAm} simplifies to
 \begin{eqnarray}
 \prod_\rho \prod_n \left(2 \pi n - \rho(u)\right)^{-\frac{\rho(\mathfrak{m})}{2}+1}& =& \prod_\rho \left( C^V_{reg}(\rho) \sin{\frac{\rho(u)}{2}}\right)^{-\frac{\rho(m)}{2}+1}\nonumber\\ &=& \prod_\rho \left(-2 i ~ sign\left(\rho\right) \sin{\frac{\rho(u)}{2}}\right)^{-\frac{\rho(m)}{2}+1} \nonumber\\
 &=&(+1)^{2 \delta(\mathfrak{m})} \prod_{\rho>0} \left(1-e^{i \rho(u)}\right),
 \end{eqnarray}
 Where $\{\rho\}$ is the set of roots $\{\alpha\}$ and $\delta:=\frac{1}{2} \sum_{\alpha>0} \alpha$ is the Weyl vector ($2 \delta \in \mathbb{Z} $). The arbitrary regularisation constant $C^{V}_{reg}$ for the vector multiplet is defined to be $-2 i ~sign\left(\rho\right) $ for convenience. Finally, we obtain the known answer
\begin{eqnarray}
Z^{1-loop}_{Vector} \equiv  \prod_{\alpha \in \mathcal{G}>0} \bigg( 1-x^\alpha \bigg) d^r u. \label{Vector1Loop}
\end{eqnarray}

\subsection{Vector multiplet 1- loop determinant in the complex path of integration}
\label{VecBZ}
The quadratic actions coming out of the localising terms \eqref{QVvec2BZ} and \eqref{QVvecFBZ}, along the complex path \eqref{ComplexCondBZ} and after gauge fixing condition \eqref{gfixingcond} is imposed, are
\begin{eqnarray}
\mathcal{L}^{B}_{quadratic}&:=&(i\,\delta \tilde{D} )^2+(\mathcal{D}_t \delta A_1)^2+ (\mathcal{D}_t \delta A_2)^2, \\
\mathcal{L}^{F}_{quadratic}&:=&i \,\delta \bar{\lambda}^\dagger_2\, \overleftarrow{\hat{\mathcal{D}}}_t \, \delta \lambda_2,
\end{eqnarray}
where
\begin{eqnarray}
i\,\delta \tilde{D} &:=&i\,\delta D+\delta F_{1 2}+ \delta\hat{\mathcal{D}}_3 \sigma .
\end{eqnarray}
The $\delta \tilde{D}$ integrates trivially. The integration of $\delta A_1$ and $\delta A_2$ gives after following the same steps outlined in the previous subsection
\begin{eqnarray}
\bigg(\prod_{\rho} \prod_{j\geq |b-1|} \prod_{n}(2 \pi n- \rho(u) )^{-2 j-1}\bigg) \times \bigg(\prod_{\rho} \prod_{j\geq |b-1|} \prod_{n}(2 \pi n- \rho(u) )^{-2 j-1}\bigg). \label{ApAmBZ}
 \end{eqnarray}

The integration of $\delta \lambda_2$ and $\delta \bar{\lambda}^\dagger_2$, multiplied by the integration of $\delta \bar{c}^\dagger$ and $\delta c$ following from the BRST action \eqref{BRSTAc} gives
 \begin{eqnarray}
\bigg(\prod_{\rho} \prod_{j\geq |b-1|} \prod_{n}(2 \pi n- \rho(u) )^{2 j+1}\bigg)\times \bigg(\prod_{\rho} \prod_{j\geq |b|} \prod_{n}(2 \pi n- \rho(u) )^{2 j+1}\bigg). \label{ccAc} 
 \end{eqnarray}
As already mentioned, we will not integrate over the zero modes $\delta \lambda_1$ and $\delta \bar{\lambda}_2$ in order not to obtain vanishing results.
The product of the two contributions \eqref{ccAc} and \eqref{ApAmBZ}, was already shown in the previous subsection to be \eqref{Vector1Loop}, hence concluding our analysis of the complex path of integration used in \citep{BZ}. The analysis and results of the one loop contribution of the matter and Chern-Simons terms in the complex path, are equivalent to the analysis and results that were already presented for the real path of integration, so we do not repeat them.

\subsection{A Jeffrey Kirwan Contour} \label{Inte}

One possible choice of segment of integration for the moduli $u_i$ is
\begin{eqnarray}
A_i := \{u_i \in \mathbb{R}: 0< u_i<2 \pi \}. \label{SegmentA}
\end{eqnarray} \footnote{Notice that the extrema 0 and $2\pi$ are excluded, as consistency with \eqref{Invertibility} demands.}
From the usual Jeffrey Kirwan (JK) perspective, the selection of the chamber \eqref{SegmentA} is analog to the selection of a reference vector $\eta$ with components $\eta_i>0$ $i=1\ldots r$.  Integration over the set of contours \eqref{SegmentA} can be mapped to the computation of residues generated by a set of charges (weights)  $\rho_{i}$ $i=1,\ldots, r$.

\begin{figure}[ht!]
\centering
\includegraphics[width=70mm]{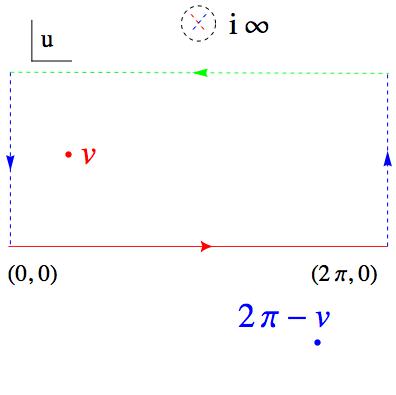}
\includegraphics[width=70mm]{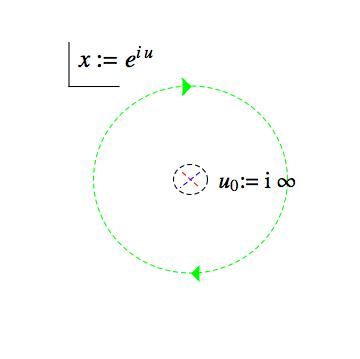}
\caption{The integration path to use in the $u$ complex plane is depicted in red in the figure above. The red (blue) point represents the pole associated with positively (negatively) charged matter multiplets. The position of these poles is determined by the flavour Wilson line along $\mathbb{S}_1$ that we denote as $v$. These poles have images that repeat with period $2 \pi \mathbb{Z}$. The integration along the two disconnected blue lines cancel out each other. The map $x:=e^{i u}$ sends the green line in the left to the $\mathbb{S}_1$ contour in the right, that encloses only the potential pole at the origin $x=0$($u=i \infty$). The integration along the green line equates then to the residue at the point $x=0$ ($u=i \infty$). The red line maps to the circumference of unit radius. The circumference of unit radius will enclose not only the pole at $x=0$ but also the poles associated to the presence of matter multiplets, namely the image of the red point in the figure in the left.  \label{Fig0}}
\end{figure}

We say that the set of charges $\{\rho_i\}_{i=1,\ldots, r}$, with $r=rank(\mathcal{G})$, generate a residue around a point $u_*$ if the integrand of the partition function looks like
\begin{equation}
f[u-u_*]\frac{du_1}{\rho_1 (u-u_{*})}  \wedge  \ldots \wedge \frac{du_r}{\rho_r (u-u_{*})},
\label{Form}
\end{equation}
in a vicinity of $u_*$, with $u_*$ and $\rho_i$ such that
\begin{eqnarray}
\rho_{i}(u_*)=0 ,~~ i=1,\ldots, r, \nonumber\\
Det[\{\rho_{1}, \ldots \rho_{r}\}_{r \times r}] \neq 0, ~~\text{ for   } u \neq u^*. \label{DetPole}
\end{eqnarray}
Condition \eqref{DetPole} is necessary to have a non trivial integration form \eqref{Form}. It is not sufficient because $f[u_*]$ could "accidentally" vanish.
Each one of the $\rho_a$, $a =1,..., r$ is a vector of $r$ elements. By $Det$, in \eqref{DetPole}, we mean the determinant of the $r \times r$ matrix whose columns (or raws) are the $r -$vectors $\rho_a$ with $a=1,\ldots r$.

The Jeffrey Kirwan recipe amounts to select residues generated by $(\rho_1, \ldots \rho_r)$ provided
\begin{eqnarray}
 \eta \in Cone(\rho_{1}, \ldots ,\rho_{r}). \label{JK}
 \end{eqnarray}
 In words, condition \eqref{JK} selects residues coming from poles associated to positively charged $\rho_{i}$ chiral ($\eta_i>0$). This is, iff  $\rho_i>0$ $i=1,\ldots, r$ then $\eta \in Cone(\rho_{1}, \ldots ,\rho_{r})$. The reference vector $\eta$ must not belong to the boundary of the Cone.

Let us see for the case of rank one gauge group how the selection of segment \eqref{SegmentA} can be identified with the JK procedure just reviewed. For that we define the auxiliary closed contour composed by the union of \eqref{SegmentA} and
\begin{eqnarray}
%A_i&=& [0,2 \pi) , \\
B%_i
&=&[i L, 0], \\
C%_i
&=&[2\pi, 2\pi  +i L], \\
{D_L}%_i
&=&[2\pi +i L, i L). %\text{ with } 1 \leq i \leq r.
\end{eqnarray}
%From here on we will focus on the case of rank $r=1$ but the discussion can be easily generalised to higher rank cases.
In the sake of clarity of presentation we have depicted this closed contour in figure \ref{Fig0} to the left. If the integrand of \eqref{PartitionFunction} is $2\pi$- periodic, integration over the segments $B$ and $C$ cancel out \footnote{They run in opposite direction.} and
\begin{eqnarray}
\int_{A \cup B \cup C\cup \mathcal{D}_L} du_1 ~~\ldots=\left(\int_{A} du_1+~ \int_{D_L} du_1\right)~~ \ldots. \label{Contour}
\end{eqnarray}
From \eqref{Contour} and taking the limit $L\rightarrow \infty$ we conclude that
\begin{eqnarray}
\int_{A} du_1~~ \ldots &=& \int_{\Gamma:=A \cup B \cup C \cup D_\infty} d u_1~~ \ldots~+\int_{-D_\infty}d u_1 ~~ \ldots. \\
&=& Res[\cdot,\Gamma]+I_{bdry}%Res[\cdot,+i \infty]
\end{eqnarray}
where by $Res[\cdot, \Gamma]$ we mean the sum of residues on the upper half plane $Im[u]>0$ but discarding the potential  "boundary contribution" (as called in \citep{BZ})
\begin{eqnarray}
I_{bdry}:=\int_{-D_\infty}d u_1.
\end{eqnarray}
 This contribution can be computed by performing the exponential map $u \rightarrow x:=e^{i u}$. In this way the line $D$, the green line in figure \ref{Fig0}, is mapped to a circumference centred at $x=0 ~(u=\infty)$. Consequently $-I_{ bdry}$ equates to the residue of the integrand in the new coordinates $x$, at $x_0=0$.

Before moving on, let us comment about the relation between our procedure of integration and the JK prescription given in \citep{BZ}. Each factor in the matter one loop determinant \eqref{matter1Loop} has a pole at $u_*=0$ provided  $\rho(\mathfrak{m})>-\mathfrak{n}+ q_R-1$. After turning on the complex flavour lines $u\rightarrow u-v$$(\rightarrow u+ v)$ with $Re[v]$ and $Im[v] \in \mathbb{R}_+$ the poles coming from positive (negative) charges which originally are sitting at $u=0$ get shifted to $u= v$ ($u=- v$) which is in the upper (lower) half complex plane $Im[u]>0$, and hence enclosed (excluded) by $\Gamma$. Namely our selection of $\Gamma$ encloses the poles associated to positive charges $\rho$. This is the same result one gets by applying the JK prescription with $\eta>0$. The other choice $\eta<0$ corresponds to close the segment \eqref{SegmentA} over lower half of the complex plane. We will not treat the case of higher rank gauge groups, in any deepness, but the graphical analysis can be generalised to that case.

\bibliographystyle{apsrev4-1}
\bibliography{S2S1}

\begin{thebibliography}{39}%
\makeatletter
\providecommand \@ifxundefined [1]{%
 \@ifx{#1\undefined}
}%
\providecommand \@ifnum [1]{%
 \ifnum #1\expandafter \@firstoftwo
 \else \expandafter \@secondoftwo
 \fi
}%
\providecommand \@ifx [1]{%
 \ifx #1\expandafter \@firstoftwo
 \else \expandafter \@secondoftwo
 \fi
}%
\providecommand \natexlab [1]{#1}%
\providecommand \enquote  [1]{``#1''}%
\providecommand \bibnamefont  [1]{#1}%
\providecommand \bibfnamefont [1]{#1}%
\providecommand \citenamefont [1]{#1}%
\providecommand \href@noop [0]{\@secondoftwo}%
\providecommand \href [0]{\begingroup \@sanitize@url \@href}%
\providecommand \@href[1]{\@@startlink{#1}\@@href}%
\providecommand \@@href[1]{\endgroup#1\@@endlink}%
\providecommand \@sanitize@url [0]{\catcode `\\12\catcode `\$12\catcode
  `\&12\catcode `\#12\catcode `\^12\catcode `\_12\catcode `\%12\relax}%
\providecommand \@@startlink[1]{}%
\providecommand \@@endlink[0]{}%
\providecommand \url  [0]{\begingroup\@sanitize@url \@url }%
\providecommand \@url [1]{\endgroup\@href {#1}{\urlprefix }}%
\providecommand \urlprefix  [0]{URL }%
\providecommand \Eprint [0]{\href }%
\providecommand \doibase [0]{http://dx.doi.org/}%
\providecommand \selectlanguage [0]{\@gobble}%
\providecommand \bibinfo  [0]{\@secondoftwo}%
\providecommand \bibfield  [0]{\@secondoftwo}%
\providecommand \translation [1]{[#1]}%
\providecommand \BibitemOpen [0]{}%
\providecommand \bibitemStop [0]{}%
\providecommand \bibitemNoStop [0]{.\EOS\space}%
\providecommand \EOS [0]{\spacefactor3000\relax}%
\providecommand \BibitemShut  [1]{\csname bibitem#1\endcsname}%
\let\auto@bib@innerbib\@empty
%</preamble>
\bibitem [{\citenamefont {Benini}\ and\ \citenamefont {Zaffaroni}(2015)}]{BZ}%
  \BibitemOpen
  \bibfield  {author} {\bibinfo {author} {\bibfnamefont {F.}~\bibnamefont
  {Benini}}\ and\ \bibinfo {author} {\bibfnamefont {A.}~\bibnamefont
  {Zaffaroni}},\ }\href {\doibase 10.1007/JHEP07(2015)127} {\bibfield
  {journal} {\bibinfo  {journal} {JHEP}\ }\textbf {\bibinfo {volume} {07}},\
  \bibinfo {pages} {127} (\bibinfo {year} {2015})},\ \Eprint
  {http://arxiv.org/abs/1504.03698} {arXiv:1504.03698 [hep-th]} \BibitemShut
  {NoStop}%
%%CITATION = ARXIV:1504.03698;%%
\bibitem [{\citenamefont {Benini}\ \emph
  {et~al.}(2015{\natexlab{a}})\citenamefont {Benini}, \citenamefont {Hristov},\
  and\ \citenamefont {Zaffaroni}}]{BZ2}%
  \BibitemOpen
  \bibfield  {author} {\bibinfo {author} {\bibfnamefont {F.}~\bibnamefont
  {Benini}}, \bibinfo {author} {\bibfnamefont {K.}~\bibnamefont {Hristov}}, \
  and\ \bibinfo {author} {\bibfnamefont {A.}~\bibnamefont {Zaffaroni}},\
  }\href@noop {} {\  (\bibinfo {year} {2015}{\natexlab{a}})},\ \Eprint
  {http://arxiv.org/abs/1511.04085} {arXiv:1511.04085 [hep-th]} \BibitemShut
  {NoStop}%
%%CITATION = ARXIV:1511.04085;%%
\bibitem [{\citenamefont {Hosseini}\ and\ \citenamefont
  {Zaffaroni}(2016)}]{BZ3}%
  \BibitemOpen
  \bibfield  {author} {\bibinfo {author} {\bibfnamefont {S.~M.}\ \bibnamefont
  {Hosseini}}\ and\ \bibinfo {author} {\bibfnamefont {A.}~\bibnamefont
  {Zaffaroni}},\ }\href@noop {} {\  (\bibinfo {year} {2016})},\ \Eprint
  {http://arxiv.org/abs/1604.03122} {arXiv:1604.03122 [hep-th]} \BibitemShut
  {NoStop}%
%%CITATION = ARXIV:1604.03122;%%
\bibitem [{\citenamefont {Chung}\ and\ \citenamefont {Yoshida}(2016)}]{Chung}%
  \BibitemOpen
  \bibfield  {author} {\bibinfo {author} {\bibfnamefont {H.-J.}\ \bibnamefont
  {Chung}}\ and\ \bibinfo {author} {\bibfnamefont {Y.}~\bibnamefont
  {Yoshida}},\ }\href@noop {} {\  (\bibinfo {year} {2016})},\ \Eprint
  {http://arxiv.org/abs/1605.07165} {arXiv:1605.07165 [hep-th]} \BibitemShut
  {NoStop}%
%%CITATION = ARXIV:1605.07165;%%
\bibitem [{\citenamefont {Nedelin}\ \emph {et~al.}(2016)\citenamefont
  {Nedelin}, \citenamefont {Nieri},\ and\ \citenamefont {Zabzine}}]{Nedelin}%
  \BibitemOpen
  \bibfield  {author} {\bibinfo {author} {\bibfnamefont {A.}~\bibnamefont
  {Nedelin}}, \bibinfo {author} {\bibfnamefont {F.}~\bibnamefont {Nieri}}, \
  and\ \bibinfo {author} {\bibfnamefont {M.}~\bibnamefont {Zabzine}},\
  }\href@noop {} {\  (\bibinfo {year} {2016})},\ \Eprint
  {http://arxiv.org/abs/1605.07029} {arXiv:1605.07029 [hep-th]} \BibitemShut
  {NoStop}%
%%CITATION = ARXIV:1605.07029;%%
\bibitem [{\citenamefont {Strominger}\ and\ \citenamefont
  {Vafa}(1996)}]{Strominger}%
  \BibitemOpen
  \bibfield  {author} {\bibinfo {author} {\bibfnamefont {A.}~\bibnamefont
  {Strominger}}\ and\ \bibinfo {author} {\bibfnamefont {C.}~\bibnamefont
  {Vafa}},\ }\href {\doibase 10.1016/0370-2693(96)00345-0} {\bibfield
  {journal} {\bibinfo  {journal} {Phys. Lett.}\ }\textbf {\bibinfo {volume}
  {B379}},\ \bibinfo {pages} {99} (\bibinfo {year} {1996})},\ \Eprint
  {http://arxiv.org/abs/hep-th/9601029} {arXiv:hep-th/9601029 [hep-th]}
  \BibitemShut {NoStop}%
%%CITATION = HEP-TH/9601029;%%
\bibitem [{\citenamefont {Sen}(2008)}]{Sen}%
  \BibitemOpen
  \bibfield  {author} {\bibinfo {author} {\bibfnamefont {A.}~\bibnamefont
  {Sen}},\ }\href {\doibase 10.1088/1126-6708/2008/11/075} {\bibfield
  {journal} {\bibinfo  {journal} {JHEP}\ }\textbf {\bibinfo {volume} {11}},\
  \bibinfo {pages} {075} (\bibinfo {year} {2008})},\ \Eprint
  {http://arxiv.org/abs/0805.0095} {arXiv:0805.0095 [hep-th]} \BibitemShut
  {NoStop}%
%%CITATION = ARXIV:0805.0095;%%
\bibitem [{\citenamefont {Witten}(1991)}]{Witten0}%
  \BibitemOpen
  \bibfield  {author} {\bibinfo {author} {\bibfnamefont {E.}~\bibnamefont
  {Witten}},\ }\href@noop {} {\  (\bibinfo {year} {1991})},\ \Eprint
  {http://arxiv.org/abs/hep-th/9112056} {arXiv:hep-th/9112056 [hep-th]}
  \BibitemShut {NoStop}%
%%CITATION = HEP-TH/9112056;%%
\bibitem [{\citenamefont {{Jeffrey}}\ and\ \citenamefont
  {{Kirwan}}(1993)}]{JK0}%
  \BibitemOpen
  \bibfield  {author} {\bibinfo {author} {\bibfnamefont {L.~C.}\ \bibnamefont
  {{Jeffrey}}}\ and\ \bibinfo {author} {\bibfnamefont {F.~C.}\ \bibnamefont
  {{Kirwan}}},\ }in\ \href@noop {} {\emph {\bibinfo {booktitle} {eprint
  arXiv:alg-geom/9307001}}}\ (\bibinfo {year} {1993})\BibitemShut {NoStop}%
\bibitem [{\citenamefont {Closset}\ \emph {et~al.}(2015)\citenamefont
  {Closset}, \citenamefont {Cremonesi},\ and\ \citenamefont {Park}}]{Closset2}%
  \BibitemOpen
  \bibfield  {author} {\bibinfo {author} {\bibfnamefont {C.}~\bibnamefont
  {Closset}}, \bibinfo {author} {\bibfnamefont {S.}~\bibnamefont {Cremonesi}},
  \ and\ \bibinfo {author} {\bibfnamefont {D.~S.}\ \bibnamefont {Park}},\
  }\href {\doibase 10.1007/JHEP06(2015)076} {\bibfield  {journal} {\bibinfo
  {journal} {JHEP}\ }\textbf {\bibinfo {volume} {06}},\ \bibinfo {pages} {076}
  (\bibinfo {year} {2015})},\ \Eprint {http://arxiv.org/abs/1504.06308}
  {arXiv:1504.06308 [hep-th]} \BibitemShut {NoStop}%
%%CITATION = ARXIV:1504.06308;%%
\bibitem [{\citenamefont {Honda}\ and\ \citenamefont {Yoshida}(2015)}]{Honda}%
  \BibitemOpen
  \bibfield  {author} {\bibinfo {author} {\bibfnamefont {M.}~\bibnamefont
  {Honda}}\ and\ \bibinfo {author} {\bibfnamefont {Y.}~\bibnamefont
  {Yoshida}},\ }\href@noop {} {\  (\bibinfo {year} {2015})},\ \Eprint
  {http://arxiv.org/abs/1504.04355} {arXiv:1504.04355 [hep-th]} \BibitemShut
  {NoStop}%
%%CITATION = ARXIV:1504.04355;%%
\bibitem [{\citenamefont {Closset}\ and\ \citenamefont
  {Kim}(2016)}]{ClossetNew}%
  \BibitemOpen
  \bibfield  {author} {\bibinfo {author} {\bibfnamefont {C.}~\bibnamefont
  {Closset}}\ and\ \bibinfo {author} {\bibfnamefont {H.}~\bibnamefont {Kim}},\
  }\href@noop {} {\  (\bibinfo {year} {2016})},\ \Eprint
  {http://arxiv.org/abs/1605.06531} {arXiv:1605.06531 [hep-th]} \BibitemShut
  {NoStop}%
%%CITATION = ARXIV:1605.06531;%%
\bibitem [{\citenamefont {Pestun}(2012)}]{Pestun}%
  \BibitemOpen
  \bibfield  {author} {\bibinfo {author} {\bibfnamefont {V.}~\bibnamefont
  {Pestun}},\ }\href {\doibase 10.1007/s00220-012-1485-0} {\bibfield  {journal}
  {\bibinfo  {journal} {Commun. Math. Phys.}\ }\textbf {\bibinfo {volume}
  {313}},\ \bibinfo {pages} {71} (\bibinfo {year} {2012})},\ \Eprint
  {http://arxiv.org/abs/0712.2824} {arXiv:0712.2824 [hep-th]} \BibitemShut
  {NoStop}%
%%CITATION = ARXIV:0712.2824;%%
\bibitem [{\citenamefont {Benini}\ \emph {et~al.}(2014)\citenamefont {Benini},
  \citenamefont {Eager}, \citenamefont {Hori},\ and\ \citenamefont
  {Tachikawa}}]{Tachikawa}%
  \BibitemOpen
  \bibfield  {author} {\bibinfo {author} {\bibfnamefont {F.}~\bibnamefont
  {Benini}}, \bibinfo {author} {\bibfnamefont {R.}~\bibnamefont {Eager}},
  \bibinfo {author} {\bibfnamefont {K.}~\bibnamefont {Hori}}, \ and\ \bibinfo
  {author} {\bibfnamefont {Y.}~\bibnamefont {Tachikawa}},\ }\href {\doibase
  10.1007/s11005-013-0673-y} {\bibfield  {journal} {\bibinfo  {journal} {Lett.
  Math. Phys.}\ }\textbf {\bibinfo {volume} {104}},\ \bibinfo {pages} {465}
  (\bibinfo {year} {2014})},\ \Eprint {http://arxiv.org/abs/1305.0533}
  {arXiv:1305.0533 [hep-th]} \BibitemShut {NoStop}%
%%CITATION = ARXIV:1305.0533;%%
\bibitem [{\citenamefont {Benini}\ \emph
  {et~al.}(2015{\natexlab{b}})\citenamefont {Benini}, \citenamefont {Eager},
  \citenamefont {Hori},\ and\ \citenamefont {Tachikawa}}]{Hori}%
  \BibitemOpen
  \bibfield  {author} {\bibinfo {author} {\bibfnamefont {F.}~\bibnamefont
  {Benini}}, \bibinfo {author} {\bibfnamefont {R.}~\bibnamefont {Eager}},
  \bibinfo {author} {\bibfnamefont {K.}~\bibnamefont {Hori}}, \ and\ \bibinfo
  {author} {\bibfnamefont {Y.}~\bibnamefont {Tachikawa}},\ }\href {\doibase
  10.1007/s00220-014-2210-y} {\bibfield  {journal} {\bibinfo  {journal}
  {Commun. Math. Phys.}\ }\textbf {\bibinfo {volume} {333}},\ \bibinfo {pages}
  {1241} (\bibinfo {year} {2015}{\natexlab{b}})},\ \Eprint
  {http://arxiv.org/abs/1308.4896} {arXiv:1308.4896 [hep-th]} \BibitemShut
  {NoStop}%
CITATION = ARXIV:1308.4896;
\bibitem [{\citenamefont {Imamura}\ and\ \citenamefont
  {Yokoyama}(2011)}]{Imamura}%
  \BibitemOpen
  \bibfield  {author} {\bibinfo {author} {\bibfnamefont {Y.}~\bibnamefont
  {Imamura}}\ and\ \bibinfo {author} {\bibfnamefont {S.}~\bibnamefont
  {Yokoyama}},\ }\href {\doibase 10.1007/JHEP04(2011)007} {\bibfield  {journal}
  {\bibinfo  {journal} {JHEP}\ }\textbf {\bibinfo {volume} {04}},\ \bibinfo
  {pages} {007} (\bibinfo {year} {2011})},\ \Eprint
  {http://arxiv.org/abs/1101.0557} {arXiv:1101.0557 [hep-th]} \BibitemShut
  {NoStop}%
%%CITATION = ARXIV:1101.0557;%%
\bibitem [{\citenamefont {Borokhov}(2004)}]{Borokhov}%
  \BibitemOpen
  \bibfield  {author} {\bibinfo {author} {\bibfnamefont {V.}~\bibnamefont
  {Borokhov}},\ }\href {\doibase 10.1088/1126-6708/2004/03/008} {\bibfield
  {journal} {\bibinfo  {journal} {JHEP}\ }\textbf {\bibinfo {volume} {03}},\
  \bibinfo {pages} {008} (\bibinfo {year} {2004})},\ \Eprint
  {http://arxiv.org/abs/hep-th/0310254} {arXiv:hep-th/0310254 [hep-th]}
  \BibitemShut {NoStop}%
%%CITATION = HEP-TH/0310254;%%
\bibitem [{\citenamefont {Cremonesi}(2015)}]{Cremonesi1}%
  \BibitemOpen
  \bibfield  {author} {\bibinfo {author} {\bibfnamefont {S.}~\bibnamefont
  {Cremonesi}},\ }\href {\doibase 10.1088/1751-8113/48/45/455401} {\bibfield
  {journal} {\bibinfo  {journal} {J. Phys.}\ }\textbf {\bibinfo {volume}
  {A48}},\ \bibinfo {pages} {455401} (\bibinfo {year} {2015})},\ \Eprint
  {http://arxiv.org/abs/1505.02409} {arXiv:1505.02409 [hep-th]} \BibitemShut
  {NoStop}%
%%CITATION = ARXIV:1505.02409;%%
\bibitem [{\citenamefont {Pasquetti}(2012)}]{Pasquetti}%
  \BibitemOpen
  \bibfield  {author} {\bibinfo {author} {\bibfnamefont {S.}~\bibnamefont
  {Pasquetti}},\ }\href {\doibase 10.1007/JHEP04(2012)120} {\bibfield
  {journal} {\bibinfo  {journal} {JHEP}\ }\textbf {\bibinfo {volume} {04}},\
  \bibinfo {pages} {120} (\bibinfo {year} {2012})},\ \Eprint
  {http://arxiv.org/abs/1111.6905} {arXiv:1111.6905 [hep-th]} \BibitemShut
  {NoStop}%
%%CITATION = ARXIV:1111.6905;%%
\bibitem [{\citenamefont {Beem}\ \emph {et~al.}(2014)\citenamefont {Beem},
  \citenamefont {Dimofte},\ and\ \citenamefont {Pasquetti}}]{Pasquetti2}%
  \BibitemOpen
  \bibfield  {author} {\bibinfo {author} {\bibfnamefont {C.}~\bibnamefont
  {Beem}}, \bibinfo {author} {\bibfnamefont {T.}~\bibnamefont {Dimofte}}, \
  and\ \bibinfo {author} {\bibfnamefont {S.}~\bibnamefont {Pasquetti}},\ }\href
  {\doibase 10.1007/JHEP12(2014)177} {\bibfield  {journal} {\bibinfo  {journal}
  {JHEP}\ }\textbf {\bibinfo {volume} {12}},\ \bibinfo {pages} {177} (\bibinfo
  {year} {2014})},\ \Eprint {http://arxiv.org/abs/1211.1986} {arXiv:1211.1986
  [hep-th]} \BibitemShut {NoStop}%
%%CITATION = ARXIV:1211.1986;%%
\bibitem [{\citenamefont {Gadde}\ \emph {et~al.}(2015)\citenamefont {Gadde},
  \citenamefont {Razamat},\ and\ \citenamefont {Willett}}]{Shlomo}%
  \BibitemOpen
  \bibfield  {author} {\bibinfo {author} {\bibfnamefont {A.}~\bibnamefont
  {Gadde}}, \bibinfo {author} {\bibfnamefont {S.~S.}\ \bibnamefont {Razamat}},
  \ and\ \bibinfo {author} {\bibfnamefont {B.}~\bibnamefont {Willett}},\ }\href
  {\doibase 10.1007/JHEP11(2015)163} {\bibfield  {journal} {\bibinfo  {journal}
  {JHEP}\ }\textbf {\bibinfo {volume} {11}},\ \bibinfo {pages} {163} (\bibinfo
  {year} {2015})},\ \Eprint {http://arxiv.org/abs/1506.08795} {arXiv:1506.08795
  [hep-th]} \BibitemShut {NoStop}%
%%CITATION = ARXIV:1506.08795;%%
\bibitem [{\citenamefont {Redlich}(1984{\natexlab{a}})}]{Redlich1}%
  \BibitemOpen
  \bibfield  {author} {\bibinfo {author} {\bibfnamefont {A.~N.}\ \bibnamefont
  {Redlich}},\ }\href {\doibase 10.1103/PhysRevLett.52.18} {\bibfield
  {journal} {\bibinfo  {journal} {Phys. Rev. Lett.}\ }\textbf {\bibinfo
  {volume} {52}},\ \bibinfo {pages} {18} (\bibinfo {year}
  {1984}{\natexlab{a}})}\BibitemShut {NoStop}%
%%CITATION = PRLTA,52,18;%%
\bibitem [{\citenamefont {Redlich}(1984{\natexlab{b}})}]{Redlich2}%
  \BibitemOpen
  \bibfield  {author} {\bibinfo {author} {\bibfnamefont {A.~N.}\ \bibnamefont
  {Redlich}},\ }\href {\doibase 10.1103/PhysRevD.29.2366} {\bibfield  {journal}
  {\bibinfo  {journal} {Phys. Rev.}\ }\textbf {\bibinfo {volume} {D29}},\
  \bibinfo {pages} {2366} (\bibinfo {year} {1984}{\natexlab{b}})}\BibitemShut
  {NoStop}%
%%CITATION = PHRVA,D29,2366;%%
\bibitem [{\citenamefont {Kapustin}\ and\ \citenamefont
  {Witten}(2007)}]{KapustinWitten}%
  \BibitemOpen
  \bibfield  {author} {\bibinfo {author} {\bibfnamefont {A.}~\bibnamefont
  {Kapustin}}\ and\ \bibinfo {author} {\bibfnamefont {E.}~\bibnamefont
  {Witten}},\ }\href {\doibase 10.4310/CNTP.2007.v1.n1.a1} {\bibfield
  {journal} {\bibinfo  {journal} {Commun. Num. Theor. Phys.}\ }\textbf
  {\bibinfo {volume} {1}},\ \bibinfo {pages} {1} (\bibinfo {year} {2007})},\
  \Eprint {http://arxiv.org/abs/hep-th/0604151} {arXiv:hep-th/0604151 [hep-th]}
  \BibitemShut {NoStop}%
%%CITATION = HEP-TH/0604151;%%
\bibitem [{\citenamefont {Yoshida}\ and\ \citenamefont
  {Sugiyama}(2014)}]{Yutaka}%
  \BibitemOpen
  \bibfield  {author} {\bibinfo {author} {\bibfnamefont {Y.}~\bibnamefont
  {Yoshida}}\ and\ \bibinfo {author} {\bibfnamefont {K.}~\bibnamefont
  {Sugiyama}},\ }\href@noop {} {\  (\bibinfo {year} {2014})},\ \Eprint
  {http://arxiv.org/abs/1409.6713} {arXiv:1409.6713 [hep-th]} \BibitemShut
  {NoStop}%
%%CITATION = ARXIV:1409.6713;%%
\bibitem [{\citenamefont {{Suzuki}}(2012)}]{Suzuki}%
  \BibitemOpen
  \bibfield  {author} {\bibinfo {author} {\bibfnamefont {M.}~\bibnamefont
  {{Suzuki}}},\ }\href@noop {} {\bibfield  {journal} {\bibinfo  {journal}
  {ArXiv e-prints}\ } (\bibinfo {year} {2012})},\ \Eprint
  {http://arxiv.org/abs/1211.2953} {arXiv:1211.2953 [math.CA]} \BibitemShut
  {NoStop}%
\bibitem [{\citenamefont {Marino}(2004)}]{Marino}%
  \BibitemOpen
  \bibfield  {author} {\bibinfo {author} {\bibfnamefont {M.}~\bibnamefont
  {Marino}}\ }(\bibinfo {year} {2004})\ \Eprint
  {http://arxiv.org/abs/hep-th/0410165} {arXiv:hep-th/0410165 [hep-th]}
  \BibitemShut {NoStop}%
%%CITATION = HEP-TH/0410165;%%
\bibitem [{\citenamefont {Mizoguchi}(2005)}]{Review}%
  \BibitemOpen
  \bibfield  {author} {\bibinfo {author} {\bibfnamefont {S.}~\bibnamefont
  {Mizoguchi}},\ }\href {\doibase 10.1016/j.nuclphysb.2005.03.035} {\bibfield
  {journal} {\bibinfo  {journal} {Nucl. Phys.}\ }\textbf {\bibinfo {volume}
  {B716}},\ \bibinfo {pages} {462} (\bibinfo {year} {2005})},\ \Eprint
  {http://arxiv.org/abs/hep-th/0411049} {arXiv:hep-th/0411049 [hep-th]}
  \BibitemShut {NoStop}%
%%CITATION = HEP-TH/0411049;%%
\bibitem [{\citenamefont {Benini}\ and\ \citenamefont {Zaffaroni}(2016)}]{BZ4}%
  \BibitemOpen
  \bibfield  {author} {\bibinfo {author} {\bibfnamefont {F.}~\bibnamefont
  {Benini}}\ and\ \bibinfo {author} {\bibfnamefont {A.}~\bibnamefont
  {Zaffaroni}},\ }\href@noop {} {\  (\bibinfo {year} {2016})},\ \Eprint
  {http://arxiv.org/abs/1605.06120} {arXiv:1605.06120 [hep-th]} \BibitemShut
  {NoStop}%
%%CITATION = ARXIV:1605.06120;%%
\bibitem [{\citenamefont {Gnecchi}\ \emph {et~al.}(2016)\citenamefont
  {Gnecchi}, \citenamefont {Gursoy}, \citenamefont {Papadoulaki},\ and\
  \citenamefont {Toldo}}]{Gnecchi}%
  \BibitemOpen
  \bibfield  {author} {\bibinfo {author} {\bibfnamefont {A.}~\bibnamefont
  {Gnecchi}}, \bibinfo {author} {\bibfnamefont {U.}~\bibnamefont {Gursoy}},
  \bibinfo {author} {\bibfnamefont {O.}~\bibnamefont {Papadoulaki}}, \ and\
  \bibinfo {author} {\bibfnamefont {C.}~\bibnamefont {Toldo}},\ }\href@noop {}
  {\  (\bibinfo {year} {2016})},\ \Eprint {http://arxiv.org/abs/1604.04221}
  {arXiv:1604.04221 [hep-th]} \BibitemShut {NoStop}%
%%CITATION = ARXIV:1604.04221;%%
\bibitem [{\citenamefont {Closset}\ \emph {et~al.}(2013)\citenamefont
  {Closset}, \citenamefont {Dumitrescu}, \citenamefont {Festuccia},\ and\
  \citenamefont {Komargodski}}]{Closset}%
  \BibitemOpen
  \bibfield  {author} {\bibinfo {author} {\bibfnamefont {C.}~\bibnamefont
  {Closset}}, \bibinfo {author} {\bibfnamefont {T.~T.}\ \bibnamefont
  {Dumitrescu}}, \bibinfo {author} {\bibfnamefont {G.}~\bibnamefont
  {Festuccia}}, \ and\ \bibinfo {author} {\bibfnamefont {Z.}~\bibnamefont
  {Komargodski}},\ }\href {\doibase 10.1007/JHEP05(2013)017} {\bibfield
  {journal} {\bibinfo  {journal} {JHEP}\ }\textbf {\bibinfo {volume} {05}},\
  \bibinfo {pages} {017} (\bibinfo {year} {2013})},\ \Eprint
  {http://arxiv.org/abs/1212.3388} {arXiv:1212.3388 [hep-th]} \BibitemShut
  {NoStop}%
%%CITATION = ARXIV:1212.3388;%%
\bibitem [{\citenamefont {Kapustin}\ \emph {et~al.}(2010)\citenamefont
  {Kapustin}, \citenamefont {Willett},\ and\ \citenamefont
  {Yaakov}}]{Kapustin}%
  \BibitemOpen
  \bibfield  {author} {\bibinfo {author} {\bibfnamefont {A.}~\bibnamefont
  {Kapustin}}, \bibinfo {author} {\bibfnamefont {B.}~\bibnamefont {Willett}}, \
  and\ \bibinfo {author} {\bibfnamefont {I.}~\bibnamefont {Yaakov}},\ }\href
  {\doibase 10.1007/JHEP03(2010)089} {\bibfield  {journal} {\bibinfo  {journal}
  {JHEP}\ }\textbf {\bibinfo {volume} {03}},\ \bibinfo {pages} {089} (\bibinfo
  {year} {2010})},\ \Eprint {http://arxiv.org/abs/0909.4559} {arXiv:0909.4559
  [hep-th]} \BibitemShut {NoStop}%
%%CITATION = ARXIV:0909.4559;%%
\bibitem [{\citenamefont {Alday}\ \emph {et~al.}(2013)\citenamefont {Alday},
  \citenamefont {Martelli}, \citenamefont {Richmond},\ and\ \citenamefont
  {Sparks}}]{Alday}%
  \BibitemOpen
  \bibfield  {author} {\bibinfo {author} {\bibfnamefont {L.~F.}\ \bibnamefont
  {Alday}}, \bibinfo {author} {\bibfnamefont {D.}~\bibnamefont {Martelli}},
  \bibinfo {author} {\bibfnamefont {P.}~\bibnamefont {Richmond}}, \ and\
  \bibinfo {author} {\bibfnamefont {J.}~\bibnamefont {Sparks}},\ }\href
  {\doibase 10.1007/JHEP10(2013)095} {\bibfield  {journal} {\bibinfo  {journal}
  {JHEP}\ }\textbf {\bibinfo {volume} {10}},\ \bibinfo {pages} {095} (\bibinfo
  {year} {2013})},\ \Eprint {http://arxiv.org/abs/1307.6848} {arXiv:1307.6848
  [hep-th]} \BibitemShut {NoStop}%
%%CITATION = ARXIV:1307.6848;%%
\bibitem [{\citenamefont {Peeters}(2007)}]{Cadabra}%
  \BibitemOpen
  \bibfield  {author} {\bibinfo {author} {\bibfnamefont {K.}~\bibnamefont
  {Peeters}},\ }\href@noop {} {\  (\bibinfo {year} {2007})},\ \Eprint
  {http://arxiv.org/abs/hep-th/0701238} {arXiv:hep-th/0701238 [HEP-TH]}
  \BibitemShut {NoStop}%
%%CITATION = HEP-TH/0701238;%%
\bibitem [{\citenamefont {Drukker}\ \emph {et~al.}(2014)\citenamefont
  {Drukker}, \citenamefont {Okuda},\ and\ \citenamefont {Passerini}}]{Drukker}%
  \BibitemOpen
  \bibfield  {author} {\bibinfo {author} {\bibfnamefont {N.}~\bibnamefont
  {Drukker}}, \bibinfo {author} {\bibfnamefont {T.}~\bibnamefont {Okuda}}, \
  and\ \bibinfo {author} {\bibfnamefont {F.}~\bibnamefont {Passerini}},\ }\href
  {\doibase 10.1007/JHEP07(2014)137} {\bibfield  {journal} {\bibinfo  {journal}
  {JHEP}\ }\textbf {\bibinfo {volume} {07}},\ \bibinfo {pages} {137} (\bibinfo
  {year} {2014})},\ \Eprint {http://arxiv.org/abs/1211.3409} {arXiv:1211.3409
  [hep-th]} \BibitemShut {NoStop}%
%%CITATION = ARXIV:1211.3409;%%
\bibitem [{\citenamefont {Hori}\ and\ \citenamefont {Romo}(2013)}]{Romo}%
  \BibitemOpen
  \bibfield  {author} {\bibinfo {author} {\bibfnamefont {K.}~\bibnamefont
  {Hori}}\ and\ \bibinfo {author} {\bibfnamefont {M.}~\bibnamefont {Romo}},\
  }\href@noop {} {\  (\bibinfo {year} {2013})},\ \Eprint
  {http://arxiv.org/abs/1308.2438} {arXiv:1308.2438 [hep-th]} \BibitemShut
  {NoStop}%
%%CITATION = ARXIV:1308.2438;%%
\bibitem [{\citenamefont {Wu}\ and\ \citenamefont {Yang}(1976)}]{WuYang}%
  \BibitemOpen
  \bibfield  {author} {\bibinfo {author} {\bibfnamefont {T.~T.}\ \bibnamefont
  {Wu}}\ and\ \bibinfo {author} {\bibfnamefont {C.~N.}\ \bibnamefont {Yang}},\
  }\href {\doibase 10.1016/0550-3213(76)90143-7} {\bibfield  {journal}
  {\bibinfo  {journal} {Nucl. Phys.}\ }\textbf {\bibinfo {volume} {B107}},\
  \bibinfo {pages} {365} (\bibinfo {year} {1976})}\BibitemShut {NoStop}%
%%CITATION = NUPHA,B107,365;%%
\bibitem [{\citenamefont {Benini}\ and\ \citenamefont
  {Cremonesi}(2015)}]{Cremonesi}%
  \BibitemOpen
  \bibfield  {author} {\bibinfo {author} {\bibfnamefont {F.}~\bibnamefont
  {Benini}}\ and\ \bibinfo {author} {\bibfnamefont {S.}~\bibnamefont
  {Cremonesi}},\ }\href {\doibase 10.1007/s00220-014-2112-z} {\bibfield
  {journal} {\bibinfo  {journal} {Commun. Math. Phys.}\ }\textbf {\bibinfo
  {volume} {334}},\ \bibinfo {pages} {1483} (\bibinfo {year} {2015})},\ \Eprint
  {http://arxiv.org/abs/1206.2356} {arXiv:1206.2356 [hep-th]} \BibitemShut
  {NoStop}%
%%CITATION = ARXIV:1206.2356;%%
\bibitem [{\citenamefont {Doroud}\ \emph {et~al.}(2013)\citenamefont {Doroud},
  \citenamefont {Gomis}, \citenamefont {Le~Floch},\ and\ \citenamefont
  {Lee}}]{Gomis}%
  \BibitemOpen
  \bibfield  {author} {\bibinfo {author} {\bibfnamefont {N.}~\bibnamefont
  {Doroud}}, \bibinfo {author} {\bibfnamefont {J.}~\bibnamefont {Gomis}},
  \bibinfo {author} {\bibfnamefont {B.}~\bibnamefont {Le~Floch}}, \ and\
  \bibinfo {author} {\bibfnamefont {S.}~\bibnamefont {Lee}},\ }\href {\doibase
  10.1007/JHEP05(2013)093} {\bibfield  {journal} {\bibinfo  {journal} {JHEP}\
  }\textbf {\bibinfo {volume} {05}},\ \bibinfo {pages} {093} (\bibinfo {year}
  {2013})},\ \Eprint {http://arxiv.org/abs/1206.2606} {arXiv:1206.2606
  [hep-th]} \BibitemShut {NoStop}%
%%CITATION = ARXIV:1206.2606;%%
\end{thebibliography}%
\end{document}